\documentclass[12pt]{iopart_mod}

\pdfoutput=1

\usepackage{iopams}  
\usepackage[utf8]{inputenc}
\usepackage{xcolor}
\usepackage{enumerate}
\usepackage{graphicx}

\usepackage{hyperref}
\hypersetup{
	colorlinks = true,
	linkcolor = black,
	citecolor = black,
	urlcolor = black,
	filecolor = black,
	linktoc = all,
}

\newcommand{\thetitle}{Algorithms for the explicit computation of Penrose diagrams}

\newcommand{\plotwidth}{1in}

\newcommand{\udl}{\tilde{u}}
\newcommand{\vdl}{\tilde{v}}
\newcommand{\ph}{\varphi}
\newcommand{\R}{\mathbb{R}}
\newcommand{\rstar}{r_{*}}
\newcommand{\orb}{\textrm{Orb}}

\newcommand{\del}{\nabla}
\newcommand{\cre}{C_{r,\epsilon}}
\newcommand{\hks}{h_{k_+, k_-}^{s_0}}

\newcommand{\xdl}{\tilde{x}}
\newcommand{\ydl}{\tilde{y}}
\newcommand{\cdl}{\tilde{c}}
\newcommand{\sgn}{\textrm{sgn}}

\newcommand{\ehat}{\hat{e}}
\newcommand{\G}[2]{\Gamma^{#1}_{ #2}}
\newcommand{\lra}{\leftrightarrow}
\newcommand{\vex}{\vec{x}}
\newcommand{\ddel}{\nabla}

\newcommand{\Ur}{U_0}   %{U_{V=V_0}}
\newcommand{\Vr}{V_0}    %{V_{U=U_0}}
\renewcommand{\Re}{\textrm{Re} \, }
\renewcommand{\Im}{\textrm{Im} \, }

\begin{document}

\title{\thetitle}

\author{J C Schindler, A Aguirre}

\address{Department of Physics, University of California Santa Cruz, Santa Cruz, CA, USA}

\ead{jcschind@ucsc.edu}
\vspace{10pt}
\begin{indented}
\item[]January 2018
\end{indented}

\begin{abstract}
An algorithm is given for explicitly computing Penrose diagrams for spacetimes of the form $ds^2 = -f(r)\, dt^2 + f(r)^{-1} \, dr^2 + r^2 \, d\Omega^2$. The resulting diagram coordinates are shown to extend the metric continuously and nondegenerately across an arbitrary number of horizons. The method is extended to include piecewise approximations to dynamically evolving spacetimes using a standard hypersurface junction procedure. Examples generated by an implementation of the algorithm are shown for standard and new cases. In the appendix, this algorithm is compared to existing methods.
\end{abstract}


\begin{indented}
\item[] \textit{Keywords:} general relativity, differential geometry, Penrose diagram, conformal diagram, null shell formalism
\item[] %\submitto{\CQG}
\end{indented}

%%%%%%%%%%%%%toc%%%%%%%%%%%%%%%
{\footnotesize \tableofcontents \listoffigures}
\renewcommand{\leftmark}{\thetitle}
\normalsize

\clearpage

%%%%%%%%%%%%%maintext%%%%%%%%%%%%%%%%%%%%%

\section{Introduction}
\label{sec:intro}

Visualizing the causal structure of curved spacetime is among the basic tasks of relativistic physics. A useful tool in this pursuit, the technique now known as Penrose diagram analysis, in which finite coordinate diagrams of conformally transformed spacetimes are used to visualize global structure, was first introduced by Penrose in 1964 \cite{penrose64,penrose65}. The same technique was soon implemented by Carter \cite{carter66}, who was first to provide such diagrams in a recognizably modern form. An important systematic analysis was later given by Walker in \mbox{1970 \cite{walker70}}. The significance of these techniques as a tool to study asympototic infinities in spacetime was quickly recognized \cite{hawking73}.

It is surprising, given the importance of Penrose diagrams, that one rarely sees a ``real" one. They are almost always hand-drawn --- in fact, it is rare even to find a computer-generated Penrose diagram of Minkowski space. There are some exceptions to this rule, including a number of especially nice diagrams due to Hamilton \cite{hamiltonbook}, and some others from Griffiths and Podolsk\`y \cite{griffiths09}. However, no general method for the numerical computation of diagrams across a broad and interesting class of metrics has been given. To do so is the goal of this article.

An algorithm will be given for constructing and numerically generating Penrose diagrams for spacetimes in two classes:
\begin{enumerate}[(A)]
\item Maximally extended completions of spacetimes which locally have the form \mbox{$ds^2 = -f(r) \, dt^2 + f(r)^{-1} \, dr^2 +r^2 \, d\Omega^2$}. (We refer to these spacetimes as \textit{strongly spherically symmetric} (SSS), see section \ref{sec:ssss}).
\item Piecewise-SSS spacetimes with null-shell junctions. These are constructed by joining pieces of spacetimes of class (A) across null shells of matter. These may have an arbitrary finite number of shells and piecewise regions.
\end{enumerate}

This is achieved by adopting a global contour integral definition of the tortoise coordinate (see section \ref{sec:alg1}), and making a careful choice of the function that squishes the local double-null coordinates (see sections \ref{sec:ssss} and \ref{sec:alg1}) into the global coordinate patch. The result is a \textit{global} double-null patch of ``Penrose" coordinates (see section \ref{sec:pen_di}) in which the metric is continuous and non-degenerate at the horizons.

The new techniques we describe are similar in most respects to those used by Carter, Walker, and others. Our technique differs, however, by achieving simultaneous global coordinates for an arbitrary number of blocks across an arbitrary number of horizons, being numerically computable with weak restrictions on the metric function $f(r)$, and yielding diagrams whose lines of constant radius take on an intuitive shape. For a detailed comparison of the new and existing methods, see \ref{sec:comp}.

There are a few reasons why Penrose diagrams have continued to be hand-drawn in the computer age, which is to say, why this algorithm has not been given sooner. Most importantly, the outline and qualitative appearance of Penrose diagrams for SSS spacetimes can be determined by the block diagram method of Walker \cite{walker70}. When the diagram is being constructed primarily for analysis at infinity, the interior structure is irrelevant, and so the qualitative block diagram method is sufficient. Moreover, most known Penrose diagrams represent either vacuum spacetimes, or spacetimes with a homogeneous distribution of matter, making interior analysis rather dull. In contrast to these historical precedents, we wish to study diagrams for spacetimes which have nontrivial matter distributions, and which are dynamically evolving in nontrivial ways. The detailed interior appearance of such diagrams is not obvious from the standard qualitative analysis.

In particular, a major motivation for this endeavor is the desire to produce a detailed Penrose diagram for the process of black hole formation and evaporation, such that the distribution and flow of matter can be clearly and explicitly tracked. Fortunately, since class (B) above includes piecewise approximations to many interesting dynamically evolving geometries (e.g. Vaidya metric \cite{griffiths09}, forming and evaporating Hayward black \mbox{hole \cite{hayward06}}, and stellar-collapse black hole models), the algorithms presented in this article will make that goal accessible.

An outline of the article is as follows.

Sections \ref{sec:pen_di}-\ref{sec:ssss} serve two purposes. First, to review the well-known theory of Penrose diagrams in general and as applied to strongly spherically symmetric spacetimes. And second, to establish a clear and modern formalism in which to state the results of later sections. We hope that this formalism helps distill the key features of standard Penrose diagram analysis, and that these sections might be used as a pedagogical introduction to the subject for students with a strong background in differential geometry.

Sections \ref{sec:alg1}-\ref{sec:alg2} present new techniques for the practical construction of Penrose diagrams, while Section \ref{sec:imp} describes an implementation of these techniques, and gives examples generated by the implementation. \ref{sec:comp} gives a detailed comparison between existing methods and the new methods.

Some of the appendices may be of general interest. Readers interested in the symmetry of manifolds may enjoy \ref{sec:spherical}, which shows how spherical symmetry about a particular rotation axis can be defined, even when the symmetry axis is not itself a part of the manifold. \ref{sec:unitconventions} describes a useful unit convention for tensor components in relativity. And \ref{sec:moreprops} collects a variety of useful geometric information about SSS spacetimes, including discussions of their trapped surfaces, physical singularities, and energy condition violations. The other appendices cover details particular to the text.

For those wishing to quickly see the practical algorithms we employ, the most direct route is to read section \ref{subsec:enumalg} followed by \ref{subsec:junctalg}. The examples in figures \ref{fig:example1} -- \ref{fig:example2}, and the comparison to other methods in \ref{sec:comp}, should also be consulted. Additionally, scanning \mbox{figure \ref{fig:sss_ill}} will help clarify the SSS spacetime terminology. This quick path through the paper is mostly, but not entirely, self-contained.

\section{Penrose diagrams in two and more dimensions}

\label{sec:pen_di}

We begin by reviewing the general theory of Penrose diagrams, and establishing a formalism commensurate with standard practices.

In general, the term ``Penrose diagram" refers to a broad class of spacetime diagrams from which the causal structure of a spacetime can be easily read off. In particular, a Penrose diagram should make evident (i) the lightcone structure, and (ii) the causal structure of conformal infinity (defined below). This is typically achieved by covering a two-dimensional slice of a spacetime with a finite patch of double-null coordinates. Although most authors need not bother to have a rigorous definition of Penrose diagram in mind, it is possible to give a precise definition which is in line with typical use. We do so now for the case of two dimensions, and then discuss the generalization to higher dimensions.

\begin{figure}[t]
\centerline{
{
\def\arraystretch{0.3}
\begin{tabular}{|c|c|}
\hline & \\
\includegraphics[scale=1]{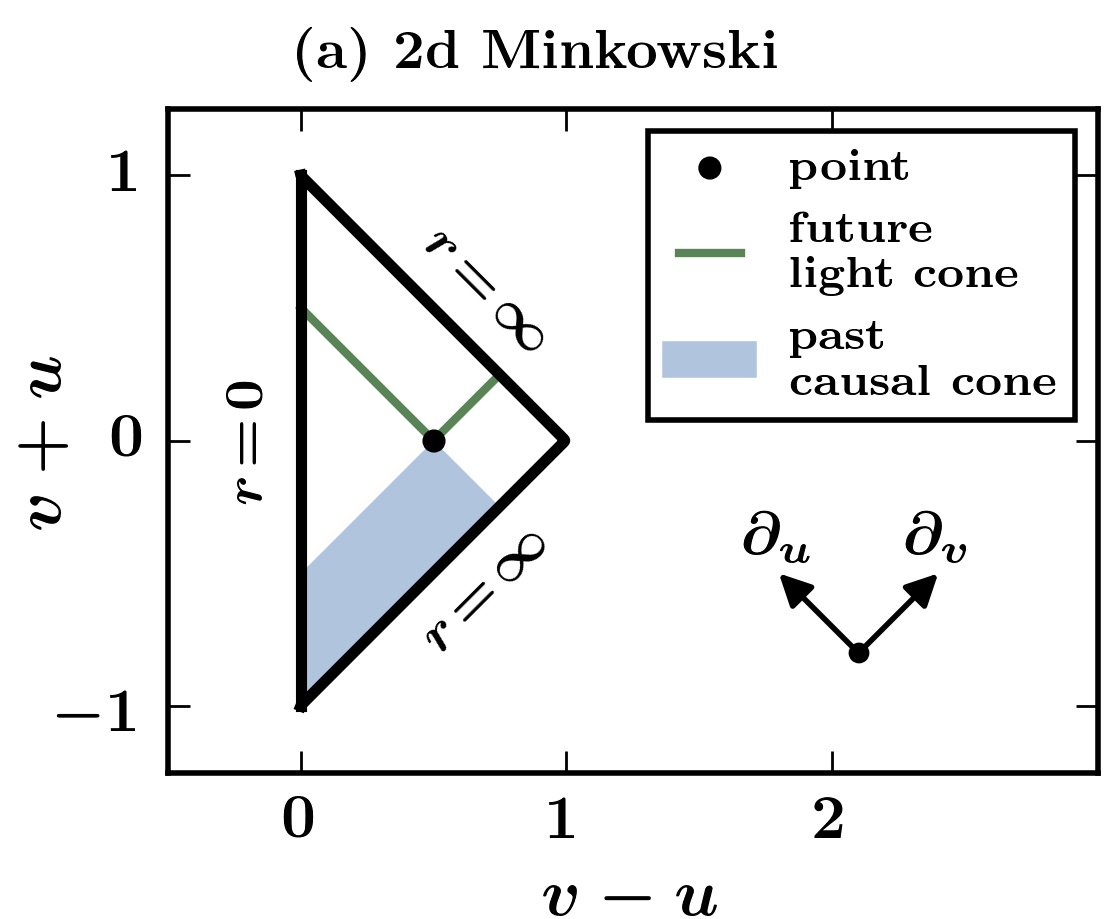}
&
\includegraphics[scale=1]{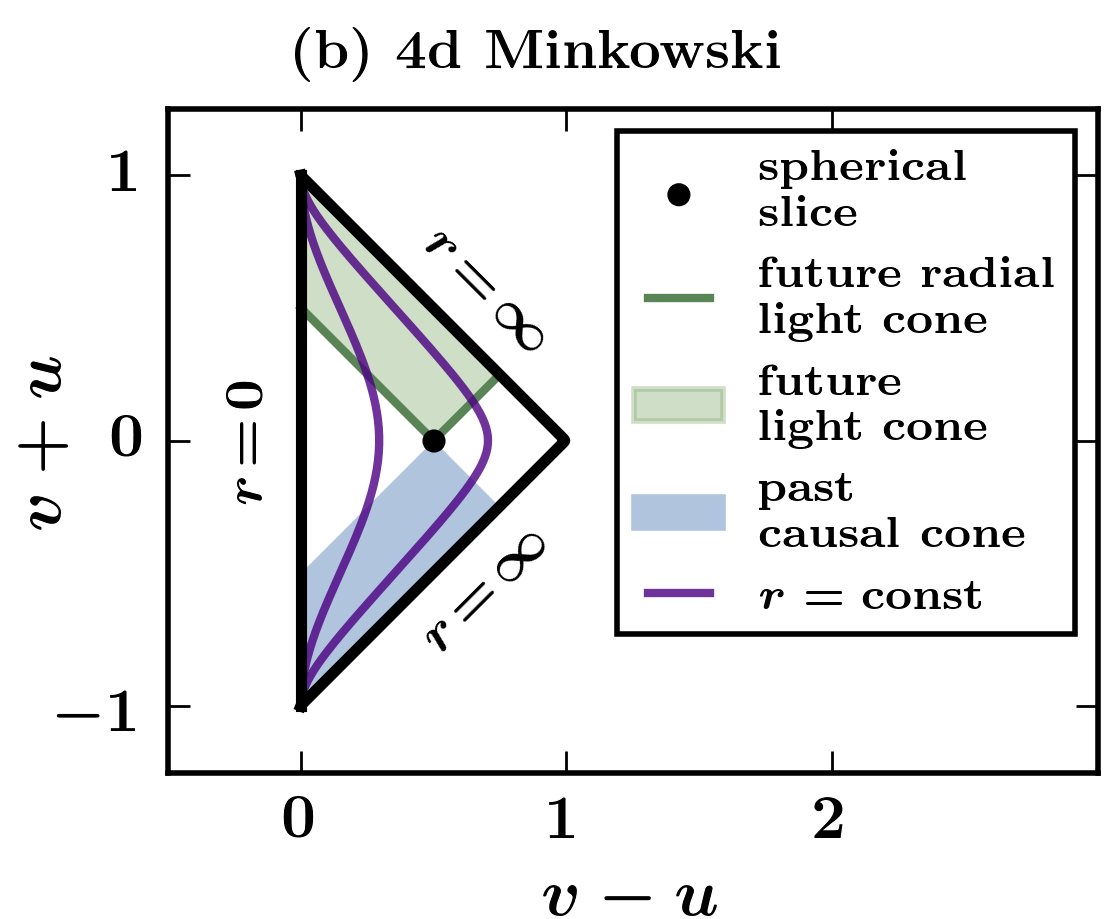}
\\[1mm]
\hline
\end{tabular}
}
}
\caption[Minkowski space]{ \label{fig:mink_basic}
(Color online). Penrose diagrams for flat spacetimes in \mbox{(a) two} and \mbox{(b) four} dimensions. See text at the end of section \ref{subsec:penmored} for details.
}
\end{figure}

\subsection{Rigorous definition in two dimensions}
Consider a two dimensional spacetime $M$, and an open set $U \subset M$. Let $\ph : U \to \R^2$ be  a chart on $M$, with coordinates  $\ph(p)=(u,v)$ for $p\in U$. Let $\overline{U}$ denote the closure of $U$, and let $\ph(U)$ denote the image of $U$ in $\R^2$. Then $\ph$ may be called a \textit{Penrose chart} if it satisfies three conditions: (I) $\overline{U}=M$; (II) there exists a compact $V \subset \R^2$ such that $\ph(U) \subset V$; (III) in coordinates $(u,v)$ the metric takes the form $ ds^2 = - \, g(u,v) \, du \, dv $, such that $g(u,v)>0$ and $\partial_u$, $\partial_v$ are both future-directed. When $\ph$ is a Penrose chart the coordinates $(u,v)$ may be called \textit{Penrose coordinates}, and the boundary of the closure of $\ph(U)$ is called the \textit{conformal boundary} of $M$. Any plot of $M$ in Penrose coordinates is called a \textit{Penrose diagram}.

Condition (I) ensures%
\footnote{%
Sometimes $M$ has a periodic structure, in which case this condition can be weakened. Suppose $M$ consists of a periodic arrangement of regions isometric to $N$. Then it suffices to require of a Penrose chart only that (I') $\overline{U}=N$, so long as we specify how the regions $N$ are connected. This information is, of course, equivalent to knowing the global causal structure. Standard examples of the periodic case are the maximally extended Reissner-Nordstrom and Kerr (on axis) spacetimes \cite{hawking73}.%
}
that the diagram includes all of $M$, while allowing some points to be left out to avoid technical difficulties (such as polar coordinate singularities) associated with attempting to cover all of $M$ in a single chart. Condition (II) ensures that the coordinate patch is finite, which allows the entirety of $M$ to be represented in a finite diagram, and allows analysis of the conformal boundary. And condition (III) ensures that the Penrose coordinates are double-null, making it trivial to identify lightcones and causal cones in the coordinates. Indeed, for any parameterized null curve $\dot{u} \dot{v} = 0$, which implies that the curve follows lines of constant $u$ or $v$ (where dot represents a derivative with respect to curve parameter). This determines the lightcone at each point. The causal interior of the lightcone may then be determined by the condition $\dot{u} \dot{v} > 0$ for timelike curves. Because of these restrictions on the lightcones, a Penrose diagram is typically plotted with x-axis $(v-u)$ and y-axis $(v+u)$. In such a case the lightcone is formed by rays at $45^{\circ}$ angles to the axes, with the top wedge being the future causal cone and the bottom wedge being the past causal cone (see Figure \ref{fig:mink_basic}).

The conformal boundary $B$ of $M$ under $\ph$, as defined above, plays a key role in understanding global causal structure because it allows the analysis of causal structure at ``infinity". The existence of a nonempty $B$ under a Penrose chart $\ph$ is guaranteed by condition (II) above. In general, points $b \in B$ may be one of several types: (i) $b$ may represent points at ``infinity"; (ii) if $M$ is incomplete at a curvature singularity, $b$ may represent the singularity; (iii) if $M$ is incomplete without curvature singularity, $b$ may represent a boundary where ``missing" parts of $M$ are simply left out. When working in more dimensions (see below), there is an additional possibility that (iv) $b$ represents a coordinate boundary of the projection (e.g. $r=0$). In practice it is usually easy to distinguish between the various possibilities, and to identify the boundary set $\mathcal{I} \subset B$ representing infinity. This set $\mathcal{I}$ is called \textit{conformal infinity}.  Our definition of conformal boundary is similar to that originally set forth by Penrose \cite{penrose65}. The more common definition in terms of terminal indecomposable sets \cite{Geroch72} is more general but less easily applicable when a Penrose chart exists.

In two dimensions, the Penrose chart $\ph$ describes a conformal isometry of $U$ into a subset of Minkowski space, when the coordinate space is equipped with metric $ds^2 = -\, du \, dv$. This allows the geometry of the boundary to be studied in the conformal Minkowski space, and is the reason for the term ``conformal" boundary. Since conformal isometries preserve causal structure, studying $\mathcal{I}$ in the conformal space determines the causal structure at infinity in $M$. Determining this structure is one of the main goals of constructing a Penrose diagram for $M$.

Practically speaking, the effort of constructing the Penrose diagram comes in two parts: (i) obtaining local double null coordinates; and (ii) manipulating the local patches to achieve a global double null coordinate system in which the metric is well-behaved (in the sense of condition (III)). Once (ii) has been achieved, it is trivial to (if necessary) squish and flip the global coordinates so as to attain Penrose coordinates. In sections \ref{sec:ssss} - \ref{sec:alg1} we will show for a certain class of spacetimes that (i) is trivial, and describe a method for resolving (ii). This method results immediately in Penrose coordinates.

\subsection{Generalization to higher dimensions}
\label{subsec:penmored}
How does this definition extend to higher dimesions? In the case of spherical symmetry, the theory goes through nearly unchanged. In this section, let $M$ have $D=2+n$ spacetime dimensions.

When spherical symmetry is present, a Penrose chart should be defined analogously to the two dimensional case, with the modification that for $p\in U$, $\ph(p)= (u,v,\Omega)$ with \mbox{$ds^2 = - \, g(u,v) \, du \, dv + r(u,v)^2 \; d\Omega^2$}. Here $\Omega$ represents a collection of angular coordinates, and $d\Omega^2$ the metric of an $n$-sphere. The diagram is then constructed by defining the \textit{projective Penrose chart} $\tilde{\ph}$ by $\tilde{\ph}(p)=(u,v)$ and the \textit{projective metric} $d\tilde{s}^2 = - \, g(u,v) \, du \, dv$. In this way, one essentially creates a Penrose diagram of the two dimensional spacetime transverse to the angular directions. Each point on the diagram represents a sphere of areal radius $r(u,v)$. It is important not to discard the radial information, as only by retaining the function $r(u,v)$ can the geometry at each point of the diagram be specified.

But even in case of spherical symmetry, the theory is slightly modified. The interpretation of the conformal space is no longer strict, since the coordinate Minkowski space is only conformal to $d\tilde{s}^2$ after projection into two dimensions. The projection into two dimensions also has the effect that the appearance of lightcones in the diagram is qualitatively altered. Null curves in $D$ dimensions  obey \mbox{$\dot{u} \dot{v} = g(u,v)^{-1} \, r^2 \, \dot{\Omega}^2 \; \geq 0$}, while timelike curves obey \mbox{$\dot{u} \dot{v} > g(u,v)^{-1} \, r^2 \, \dot{\Omega}^2 \; \geq 0$}. Thus, two-dimensional null curves in the confomal Minkowski space now represent only the radial null curves in $M$, while $D$-dimensional null curves with angular momentum in $M$ appear timelike in the conformal space. The $D$-dimensional lightcones of $M$, therefore, fill the interior of the two-dimensional causal cones in the conformal space. Despite these several modifications to the interpretation of the conformal space, the conformal method for studying infinity remains useful, and the name conformal boundary is retained.

For most of our purposes it will be convenient to deal with spherically symmetric shells and particles constrained to move in the $(u,v)$ plane. It is therefore useful to identify the \textit{radial lightcones}, defined by $\dot{u} \dot{v} = 0$, which are the effective lightcones for such objects. The $D$-dimensional radial lightcones are equivalent to the two-dimensional conformal lightcones. \textit{Radial causal cones} can be defined similarly. This concludes the extension from two dimensions to spherical symmetry in $D$ dimensions.

In some cases the spherically symmetric formalism can be generalized further. Let $A$ be a two-dimensional Lorentzian manifold with line element $dA^2$, and let $h(a)^2$ be a positive real scalar function on $A$. Let $B$ be an $n$-dimensional homogeneous Riemannian manifold with line element $dB^2$.  If there exist such an $A$ and $B$ for which a dense open submanifold $U \subset M$ is isometric to the product $A \times B$ with metric $ds^2 = dA^2 + h(a)^2 \, dB^2$, then a formalism directly analogous to that for spherical symmetry can be used (although if $B$ is not compact, condition (II) should be modified to apply only to the projective chart $\tilde{\ph}$, and one should be careful in interpretation). This includes cases of planar, hyperbolic, and spherical symmetry, among others. The above condition implies that $M$ has at least $n$ spacelike Killing vector fields, and is essentially equivalent to the condition that $M$ can be acted on by an $n$-dimensional group of spacelike isometries.

When $M$ lacks sufficient symmetry for analogous methods to be applied, by having nontrivial geometry in more than two dimensions, one must resort to piecing together the structure by observing various two-dimensional projections. This case is less common due to its complexity, and the theory of projection diagrams due to Chru\'{s}ciel \etal should be consulted \cite{chrusciel12}.

Figure \ref{fig:mink_basic} illustrates the basic features of a Penrose diagram in two and four dimensions using the simple case of flat spacetime. In this example, the two-dimensional spacetime is defined by $ds^2 = -dt^2+dr^2$ on the coordinate patch $r\in(0,\infty)$ and $t\in(-\infty,\infty)$, while the four-dimensional spacetime is defined by $ds^2 = -dt^2+dr^2 + r^2 \, d\Omega^2$ on the same coordinate patch. In both cases, the Penrose coordinates are given by 
$u = \pi^{-1} \tan^{-1} \left( t - r \right)$
and
$v = \pi^{-1} \tan^{-1} \left( t + r \right)$.
The resulting metrics are 
$ds^2=- \pi^2 \sec^2 (\pi u) \, \sec^2 (\pi v) \, du \, dv$
and 
$ds^2=- \pi^2 \sec^2 (\pi u) \, \sec^2 (\pi v) \, du \, dv + r^2 \, d\Omega^2$, where, in four dimensions, the areal radius at each point is given by 
$r = r(u,v) = (\tan \pi v - \tan \pi u)/2$.
Note that the two-dimensional example, which is half of a two-dimensional Minkowski space, is incomplete at $r=0$ by construction so as to more closely parallel the four-dimensional case.

\section{A note about units}
\label{sec:units}
A detailed review of our unit conventions and their justification is given in \mbox{\ref{sec:unitconventions}}. In short: \textit{all coordinates, parameters, tensor components, and lengths appearing in the article are unitless}. Proper units are restored by establishing an overall length scale, which can be propagated through all quantities. For details, please see the appendix.

To make the convention as clear as possible, consider the example 
\begin{equation}
ds^2 = -(1-R/r) \, dt^2 + (1-R/r)^{-1} \, dr^2 + r^2 \; d\Omega^2 \; .
\end{equation} 
The radius $r$, the coordinates $(t,r,\Omega)$, the parameter $R$, the one-forms $(dt,dr,d\Omega)$, and the line element $ds^2$ should all be regarded as unitless. To relate this to a physical metric, one would establish an overall length scale $l$. Then the physical coordinates $(lt,lr,l\Omega)$ and physical line element $d\bar{s}^2=l^2 ds^2$ would all have units of length, and other quantities would inherit units as appropriate.

\section{Strongly spherically symmetric spacetimes and their maximal extensions}
\label{sec:ssss}
The class of spacetimes with metric of the form
\begin{equation} \label{eqn:sssm}
ds^2 = -f(r) \, dt^2 + f(r)^{-1} \, dr^2 + r^2 \, d\Omega^2
\end{equation}
(where $d\Omega^2$ signifies the metric on a unit $n$-sphere) is of great historical and practical significance: common examples include the Schwarzschild, Reissner-Nordstrom, de Sitter, Anti de Sitter, and Minkowski solutions of GR, among numerous others. Strangely, given their ubiquity, this class of spacetimes lacks a standard name. We introduce some new  terminology, and review some properties of these spaces, below.

In particular, we will see that every spacetime which locally has the metric (\ref{eqn:sssm}) can be isometrically embedded into a spacetime of a larger class, which we will call the ``strongly spherically symmetric" (SSS) spacetimes. In this section we develop a detailed geometrical description of such spacetimes.

Historically, our ``strongly spherically symmetric" spacetimes have sometimes been called ``static spherically symmetric" \cite{graves60}. But when $f(r)\leq 0$ they are not static (do not have a timelike Killing vector field), so the term is not apt. The new name seems more fitting: the symmetry is ``strong" in the sense that, in addition to the spherical symmetries, there exists a Killing vector field normal to the angular directions, which allows the metric components to be expressed as functions of the radius only.

\begin{figure}[t]
\centering
\fbox{
\includegraphics[scale=1.05]{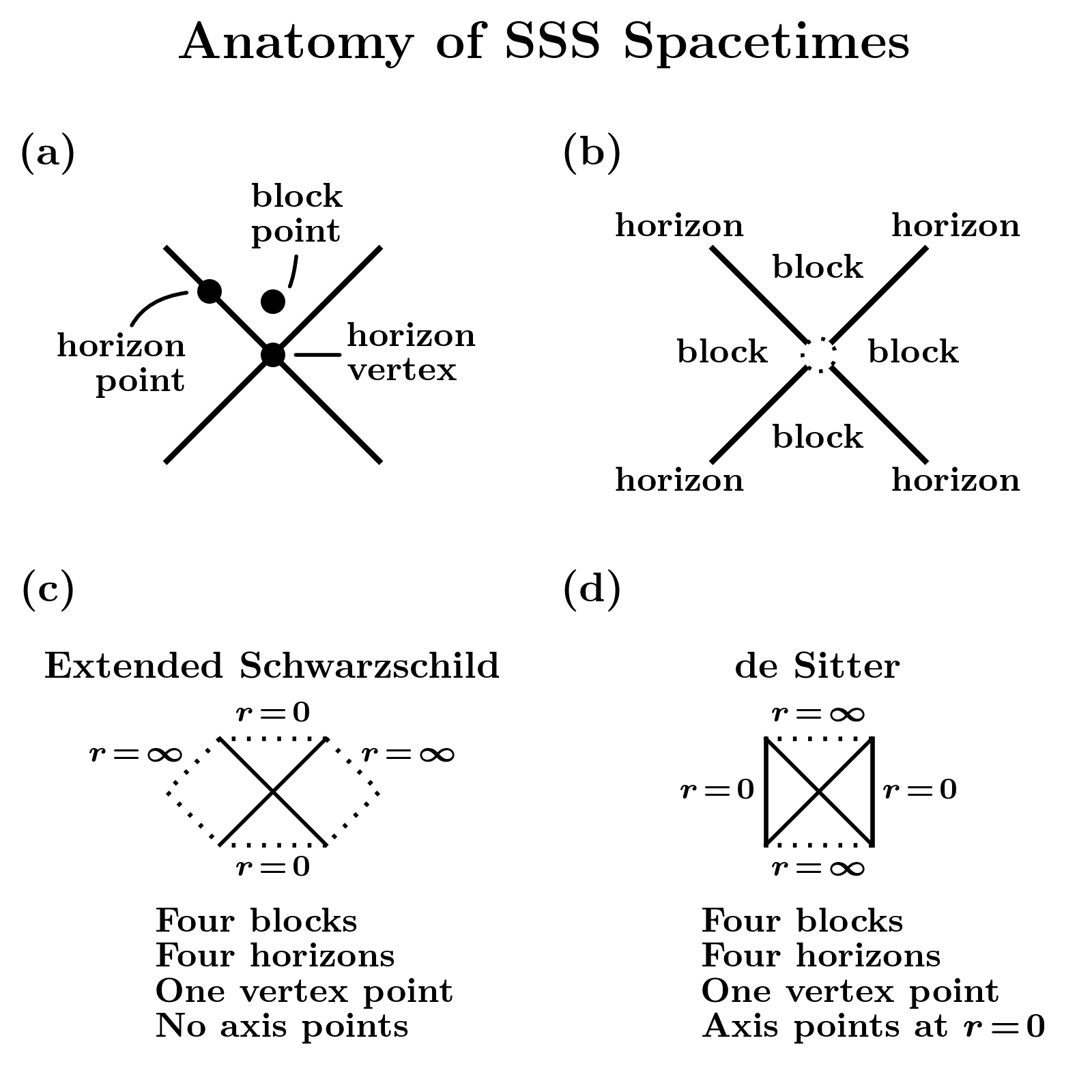} 
}
\caption[Anatomy of SSS spacetimes]{ \label{fig:sss_ill}
Illustration of the anatomy of SSS spacetimes. (a) The classification of points is given in section \ref{subsec:sss}. (b) The definition of blocks and horizons is given in section \ref{subsec:horblock}. In this image the vertex point is omitted to clarify that four disconnected horizons are present. (c,d) Block diagrams are discussed in section \ref{subsec:block}. In these two block diagrams conformal boundary points which are not in $M$ are dotted (which is not our standard convention). The extended Schwarzschild solution has metric function  $f(r)=1-R/r$, and has no axis points since the singularity at $r=0$ is excluded from the spacetime. In de Sitter space, with metric function $f(r)=1-(r/l)^2$, every point at $r=0$ is an axis point. In all cases, radii are measured by defining orbits of a rotation group, as discussed in section \ref{subsec:spherical}.
}
\end{figure}

\subsection{Spacetimes with spherical symmetry about a fixed origin}
\label{subsec:spherical}
To properly describe strong spherical symmetry requires the concept of spherical symmetry about a fixed origin. Since manifolds need not contain their symmetry axes, this requires a little bit of finagling. In \ref{sec:spherical} it is shown that for any spherically symmetric spacetime $M$, one can specify the origin of spherical symmetry by selecting a particular algebra $\sigma$ of Killing vector fields satisfying certain assumptions. This choice determines spherical orbits of a rotation group; the curvature of these orbits is used to define the radius at each point.

Once the ``origin of symmetry" has been fixed by choosing $\sigma$, the areal radius $r=r_\sigma(p)$ is an intrinsic property of each point $p \in M$, independent of coordinate system. $M$ is foliated by spheres (lying tangent to $\sigma$) with intrinsic metric $ds^2=r_\sigma(p)^2 \, d\Omega^2$, except on the axis of symmetry where $r_\sigma(p)=0$ by definition. Everywhere except on the axis, there is a local coordinate system respecting this foliation, in terms of which the metric is (\ref{eqn:spheremetric}).

In our treatment of strong spherical symmetry, we will assume that $M$ is spherically symmetric, and that the origin of symmetry has been fixed by selecting a particular $\sigma$. This ensures that local SSS patches all have symmetry about the same origin.

\subsection{Coordinate naming conventions}
\label{subsec:coordnames}

Up to now, our convention has been to denote Penrose coordinates by $(u,v)$. Hereafter, Penrose coordinates will usually be denoted by $(\udl,\vdl)$, while coordinates $(u,v)$ will be reserved for the usual local double-null coordinates with metric (\ref{eqn:dnmetric}) below. Coordinates $(t,r)$ refer of course to standard Schwarzschild-like coordinates with metric (\ref{eqn:sssm}). And coordinates $(w,r)$ will denote the Eddington-Finklestein (EF) coordinates with the metric (\ref{eqn:efmet}) below, such that the EF time $w$ may be either advanced or retarded depending on a parameter $\epsilon=\pm 1$ and a choice of time-orientation. In either case, the vector field $\partial_w$ is locally equivalent (up to a global normalization) to $\partial_t$ wherever the latter is defined. For reasons that will become clear below, we will define strong spherical symmetry in terms of the EF coordinate system $(w,r)$, and its associated Killing vector field $\partial_w$.

At times it is convenient to refer ambiguously to one of the double-null coordinates $(u,v)$, without specifying which one. When this is necessary we will utilize the placeholders $(x,y)$, where it is understood that either $(x,y) \equiv (u,v)$ or $(y,x) \equiv (u,v)$. The same placeholder convention extends to $(\udl,\vdl)$.

\subsection{Spacetimes with strong spherical symmetry}
\label{subsec:sss}

Suppose $M$ is a spacetime of dimension $D=2+n$, and that $M$ is $n$-spherically symmetric with axis fixed by $\sigma$ (see section \ref{subsec:spherical}). Then $M$ will be said to have \textit{strong spherical symmetry (about axis $\sigma$) with metric \mbox{function $f(r)$}} if every open set $U \subset M$ has an open subset $V\subset U$ isometric to%
\footnote{%
Here and later we abuse terminology by omitting the full statement ``isometric to an open subset of $\mathbb{R}^D$ on which the metric $ds^2$ is defined'', which should be obvious from context. This is equivalent to saying there exists a coordinate patch on $M$ in which the metric takes this form.%
}
\begin{equation} \label{eqn:efmet}
ds^2 = - f(r) \, dw^2 - 2 \epsilon \, dw \, dr + r^2 \, d\Omega^2 
\hspace{10mm} (\epsilon = \pm1) \; ,
\end{equation}
in coordinates $(w,r,\Omega)$, such that $r(p)=r_\sigma(p)$ and $\sigma$ lies tangent to $\Omega$. Those last bits ensure that all points have a strong spherical symmetry about the same origin. In practice we always define $\sigma$ in terms of this metric, so these technicalities become trivial. Assumptions on the function $f(r)$ are given below, near the end of this section.

The definition implies that the set
\begin{equation}
X = \left\lbrace p \in M  \, | \, p \textrm{  has a neighborhood isometric to (\ref{eqn:efmet})}   \right\rbrace 
\end{equation}
is open and dense in $M$. Using the Eddington-Finkelstein (EF) form (\ref{eqn:efmet}) of the metric allows $X$ to contain points where $f(r)=0$, and implies that $M$ is locally isometric to $ds^2 = -f(r) \, dt^2 + f(r)^{-1} \, dr^2 + r^2 \, d\Omega^2$  at points in $X$ where $f(r)\neq 0$. 

The spherical symmetry is called ``strong" due to the presence of a Killing vector field normal to the angular directions, which allows the metric components to be expressed as functions of the radius only. Indeed, by definition, the metric is independent of coordinate $w$, and so $\partial_w$ is a local Killing vector field. It is easily seen from the metric that  $\partial_w$ is null wherever $f(r)=0$, and that $\partial_w$ is always normal to the angular directions. 

In general, SSS spacetimes can contain four distinct types of points:
\begin{enumerate}
\item points not in $X$ at which $r=0$ (\textit{axis points})
\item points in $X$ at which $f(r)\neq 0$ (\textit{block points})
\item points in $X$ at which $f(r)=0$ (\textit{horizon points})
\item points not in $X$ at which $r\neq0$ (\textit{horizon vertices})
\end{enumerate} 
The reason for these names will become clear later. The isometry associated with flow along $\partial_w$ preserves this classification, and has the horizon vertices as fixed points. Moreover, every point in $X$ has a neighborhood where $\partial_w$ is an everywhere-nonzero Killing vector field tangent to lines of constant radius. For an illustration of the different types of points, see figure \ref{fig:sss_ill}.

It is now appropriate to explain the need for fixing the symmetry axis. Without doing so, the presence of additional symmetries can make it impossible to geometrically distinguish points in $M$. Take, for example, the de Sitter spacetime, with metric function $f(r)=1-(r/l)^2$. Since de Sitter is homogeneous, every point is geometrically indistinguishable. In particular, every point may be described as lying on a cosmological horizon. From the point of view where we consider de Sitter an SSS spacetime, the additional freedom to choose the origin of symmetry is superfluous (though by no means unimportant). By fixing the origin, we allow points and regions to be classified as above. Additionally, fixing the origin allows us to refer to the radius of a point without reference to any particular system of coordinates.

We typically describe a particular strongly spherically symmetric spacetime by specifying its metric function $f(r)$ on the interval $r \in (0,\infty)$. For simplicity, we assume that $f(r)$ always has the following properties: (I) $f(r)$ is continuous and once differentiable; (II) $f(r)$ has a finite number $N$ of zeroes; (III) $f(r)$ is analytic at its zeroes; (IV) all zeroes of $f(r)$ are isolated and simple (linear); and (V) $\lim_{\,r\to 0} f(r) \neq 0$. 

The assumption that $f(r)$ is analytic at its zeroes allows the use of a concise contour integral definition of the tortoise coordinate (see section \ref{subsec:globtort}); since $f(r)$ need not be analytic globally, the assumption is fairly weak. Note that $f(0)$ need not be defined. And moreover, the assumption that $f(r)$ does not approach zero in the $r \to 0$ limit is not strictly necessary; it conveniently avoids the treatment of certain edge cases, but can be relaxed with no serious consequences.

It is useful to define a consistent notation for critical values of the radius, in order to partition the radial coordinate into intervals separated by horizon radii. The endpoints of these intervals occur at $r=0$, at the $N$ horizon locations where $f(r)=0$, and at $r=\infty$. We therefore let $r_0=0$, denote the zeroes of $f(r)$ by $r_i$ for $i=1,2,\ldots,N$, and let $r_{N+1}=\infty$. Then the radius values $r_j$, for $j=0,1,\ldots,N$, partition the radial coordinate into intervals $I_j=(r_j,r_{j+1})$, and in each interval $I_j$ the sign of $f(r)$ is constant. In this context we reserve the subscript $i$ to refer only to the zeroes of $f(r)$, while expressions with subscript $j$ additionally include $r_0$ and $r_{N+1}$.

In what follows, a parameter $k_j$ will control coordinate transformations in the vicinity of each critical value $r_j$ of the radius. Near each zero $r_i$ of $f(r)$ it must have a particular value, equal to the slope of $f(r)$ at $r_i$, in order for the metric to extend continuously across the corresponding horizon. Where no horizon matching is needed, however, it may be set to zero, yielding a simplified coordinate transformation. In accordance with these requirements, we define $k_0=0$ at $r_0=0$, $k_i=f'(r_i)$ at the $N$ zeroes of $f(r)$, and $k_{N+1}=0$ at $r_{N+1}=\infty$. This ensures matching at all horizons, while providing the simplest possible transformations near $r=0$ and $r=\infty$, where no matching is needed.

Specifying the metric function determines the local structure of $M$, by insisting that (\ref{eqn:efmet}) holds for the fixed function $f(r)$. Some ambiguity remains in the global structure. We will see in section \ref{subsec:block} below that in fact such an $M$ might be any subset of a maximally extended $M'$ corresponding to $f(r)$. This is the class of spacetimes intended by (A) in section \ref{sec:intro}.

For the remainder of this section, let $M$ denote a strongly spherically symmetric spacetime with metric function $f(r)$.

\subsection{Horizons and blocks}
\label{subsec:horblock}

It is useful to think of $M$ as being partitioned into ``blocks" separated by ``horizons".

First let us mention the horizons. A connected hypersurface consisting of horizon points in $M$, on which $r=\textrm{const}$ and $f(r)=0$, is called a \textit{horizon}. Correspondingly, the values $r_i$ where $f(r)=0$ may be called \textit{horizon radii}. Every horizon is a null hypersurface, being normal to the (locally) null vector $\partial_w$. Moreover, horizons are always Killing horizons, and often trapping horizons (see \cite{wald01,senovilla11} for definitions). The Killing horizon property is immediate, since the Killing vector field $\partial_w$ is null at a horizon. The trapping property we will return to shortly. Now let us move on to the blocks.

A \textit{block} is a region of $M$, consisting of a connected set of block points, which corresponds to a single interval $I_j$ over which the metric function is nonzero. Each block can be covered by the metric (\ref{eqn:sssm}) defined on a coordinate patch $r\in I_j$ and $t \in (-\infty,\infty)$. Note that this metric may approach a coordinate singularity at the boundaries of the patch. There is a one-to-one correspondence between the intervals $I_j$ and the \textit{types} of blocks in $M$. However, $M$ may contain many blocks of the same type, each corresponding to the same interval $I_j$. We will often label blocks by their corresponding interval, which indicates their type.

\begin{figure}[t]
\centerline{
{
\def\arraystretch{0.3}
\begin{tabular}{|c|c|c|}
\hline & & \\
\includegraphics[scale=1]{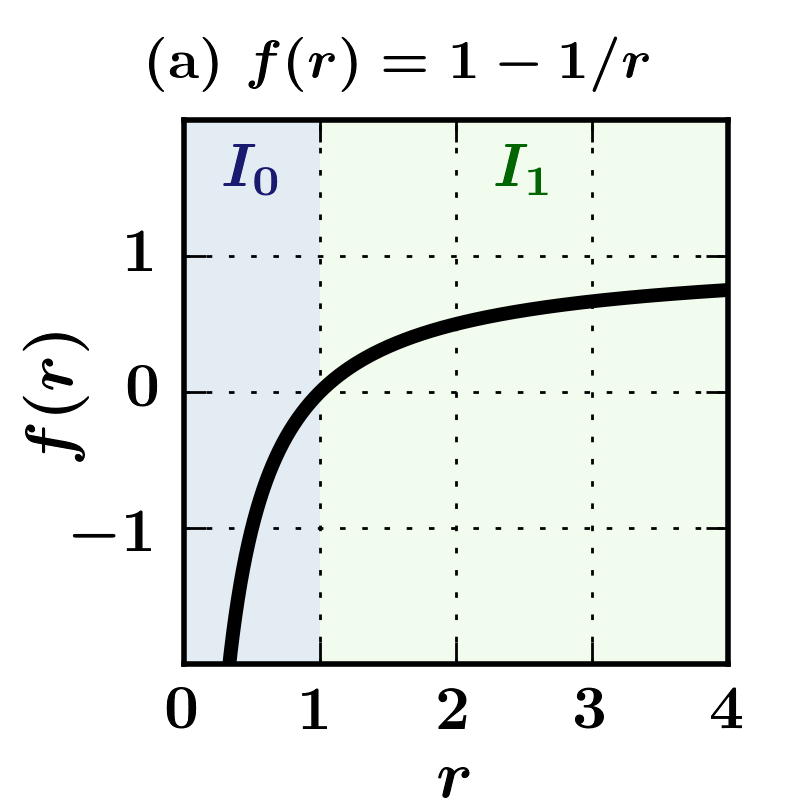}
&
\includegraphics[scale=1]{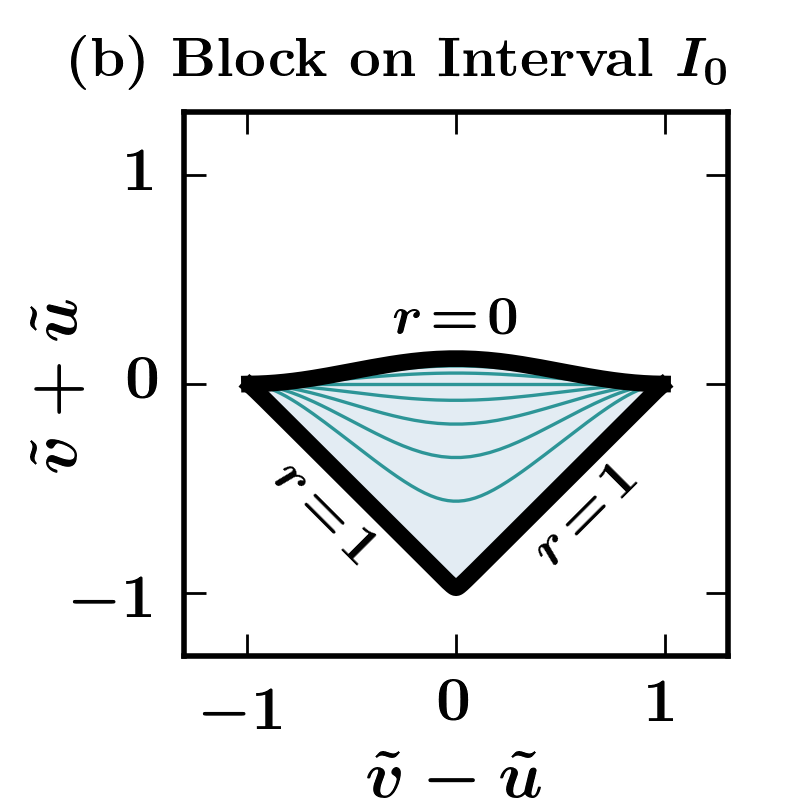}
&
\includegraphics[scale=1]{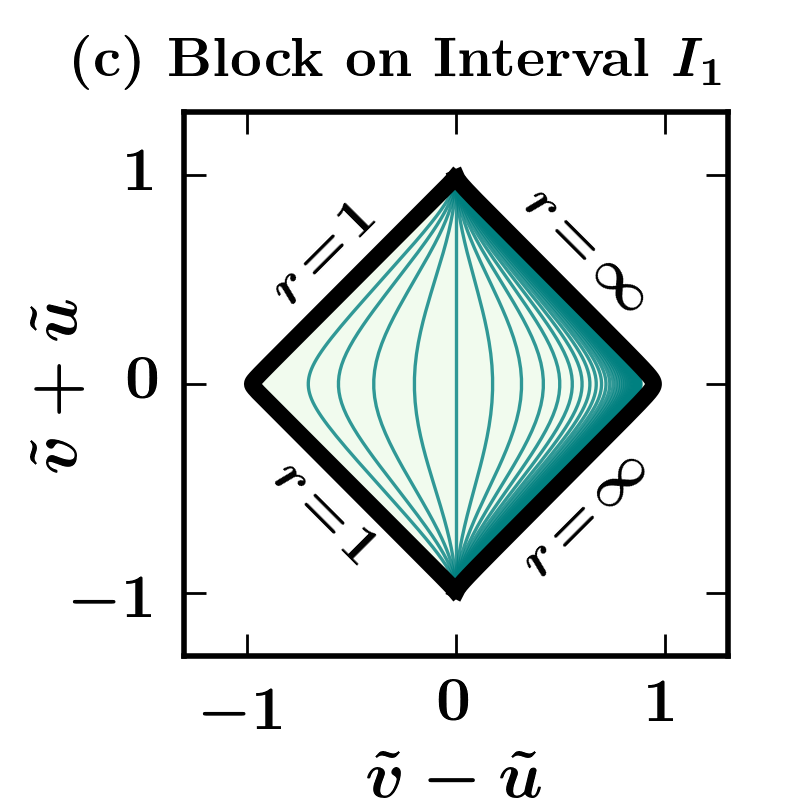}
\\[1mm]
\hline
\end{tabular}
}
}
\caption[Individual SSS blocks]{ \label{fig:indyblocks}
(Color online). Penrose diagrams for some individual blocks of an SSS spacetime with $f(r)=1-1/r$. We have employed the methods of secion \ref{subsec:horblock}, with integration points 
$a_1=0.5$ and $a_2=1.5$, 
integration constants 
$c_0=c_1=0$, 
and squishing functions 
$\udl_0(u)= - \udl_1(u) = - \pi^{-1} \tan^{-1}(u)$ and
$\vdl_0(v)= \vdl_1(v) = \pi^{-1} \tan^{-1}(v)$,
in regions corresponding to the intervals $I_0=(0,1)$ and $I_1=(1,\infty)$. 
Lines of constant areal radius $r=const$ (teal) are depicted in the diagram at intervals $dr=0.1$, and the radius at any diagram point can be determined numerically by (\ref{eqn:rudlvdl}). Heavy black lines denote the conformal boundary of each block. The causal shapes are triangle for $I_0$ and diamond for $I_1$, and the block $I_0$ is a trapped region containing future-trapped surfaces. Each block is bordered on one side by two horizons (and a horizon vertex) at $r=1$. Although the blocks can in principle be joined continuously at the matching horizon, this individual-block method doesn't give a way to do so.
}
\end{figure}

To construct Penrose coordinates for a single block is straightforward.  To begin, choose an arbitrary point $a \in I_j$. Define the \textit{tortoise coordinate} $\rstar$ and \textit{tortoise function} $F(r)$ by
\begin{equation} \label{eqn:tortoise1}
\rstar = F(r) = \int_a^r \frac{dr'}{f(r')} \; .
\end{equation}
The tortoise function obeys $dF/dr = f(r)^{-1}$, and is monotonic over $I_j$ since $f(r)$ is continuous and nonzero there.  Thus $F(r)$ is invertible on $I_j$, and we denote the inverse function by $r = F^{-1} (\rstar)$. The arbitrary choice of $a \in I_j$ amounts to an arbitrary additive constant in $F(r)$, which we later absorb into a choice of double null coordinates. The range of $\rstar$ depends on the behavior of $f(r)$ near the endpoints of $I_j$; the value becomes infinite in magnitude near each simple zero of $f(r)$. 

Next a set of double null coordinates for the block, with a parameter $c\in\R\;$ absorbing the tortoise function's free additive constant, is defined by
\begin{equation} \label{eqn:uvdef}
 u = t - \rstar + c  \, ,
 \qquad
 v = t + \rstar - c  \, ,
\end{equation}
in terms of which the metric becomes 
\begin{equation} \label{eqn:dnmetric}
ds^2 = -f(r)\,du\,dv+ r^2 \, d\Omega^2 \; ,
\end{equation}
with $r = F^{-1} \left( (v-u)/2 + c  \right)$.

Finally, one chooses two invertible monotonic functions 
$\udl(u)$ and $\vdl(v)$, 
called the \textit{squishing functions}, each with domain $\R$ and finite range, such that 
\begin{equation}
f(r) \, \frac{du}{d\udl} \,  \frac{dv}{d\vdl} > 0 \qquad 
\textrm{for} \qquad 
(r \in I_j),
\end{equation}
and such that $\partial_{\udl}$ and $\partial_{\vdl}$ are both future directed. 

The resulting metric reads 
\begin{equation}
ds^2 = -f(r) \frac{du}{d\udl} \frac{dv}{d\vdl}\,d\udl\,d\vdl+ r^2 \, d\Omega^2 \; ,
\end{equation}
where
\begin{equation} \label{eqn:rudlvdl}
r=r(\udl,\vdl)=F^{-1}\left( \frac{v(\vdl)-u(\udl)}{2} + c \right),
\end{equation}
and $(\udl,\vdl,\Omega)$ are Penrose coordinates for the block.

Any block bounded by simple zeroes of $f(r)$ admits the full range of coordinates $u\in(-\infty,\infty)$ and $v\in(-\infty,\infty)$, and therefore covers a diamond in the Penrose diagram. Blocks admitting a smaller range of $(u,v)$ lie inside this diamond (when the same squishing functions are used).

The freedom involved in obtaining Penrose coordinates according to the above process is less than it may at first seem. First of all, note that the squishing functions should be chosen for simplicity and convenience. Once some squishing functions have been chosen, there is always the possibility to rescale by monotonic increasing functions in the $u$ and $v$ direction. This basically amounts to a freedom to mess up a nice-looking block, without making any structurally meaningful changes. Therefore let us focus on the choice of $\rstar,u,v$. As seen already, the tortoise function $F(r)$ implicitly contains an arbitrary constant, which was absorbed into $u,v$ by the free constant $c$. In light of this fact and the freedoms to translate $u,v,t$ without disturbing the metric, one is tempted to write 
$(u-u_0) = (t-t_0) - (\rstar - c)$ 
and 
$(v-v_0) = (t-t_0) + (\rstar - c)$. 
These may be rearranged, however, to yield 
$u = (t-t'_0) - (\rstar - c')$ and $ v = (t-t'_0) + (\rstar - c')$.
But translations of $t$, being isometries of the block which preserve the range of $t$ values (since $t\in(-\infty,\infty)\,$), are entirely nonphysical and have no effect on the appearance of Penrose diagrams. The equations (\ref{eqn:uvdef}) are therefore sufficiently general to exhibit all relevant freedoms in the process. Like with the squishing functions, changing the parameter $c$ does not cause any important changes to the diagram; it simply alters the appearance, and may be chosen for convenience. Although $c$ is free in each block individually, we will see in section \ref{sec:alg1} that in order to construct global Penrose coordinates for many blocks, the values of $c$ in each block must be carefully coordinated, and only a global additive constant remains.

Having now constructed Penrose coordinates for each block, let us use them to investigate the properties of these blocks more deeply.

\begin{figure}[t]
\centering
\fbox{
\includegraphics[scale=1]{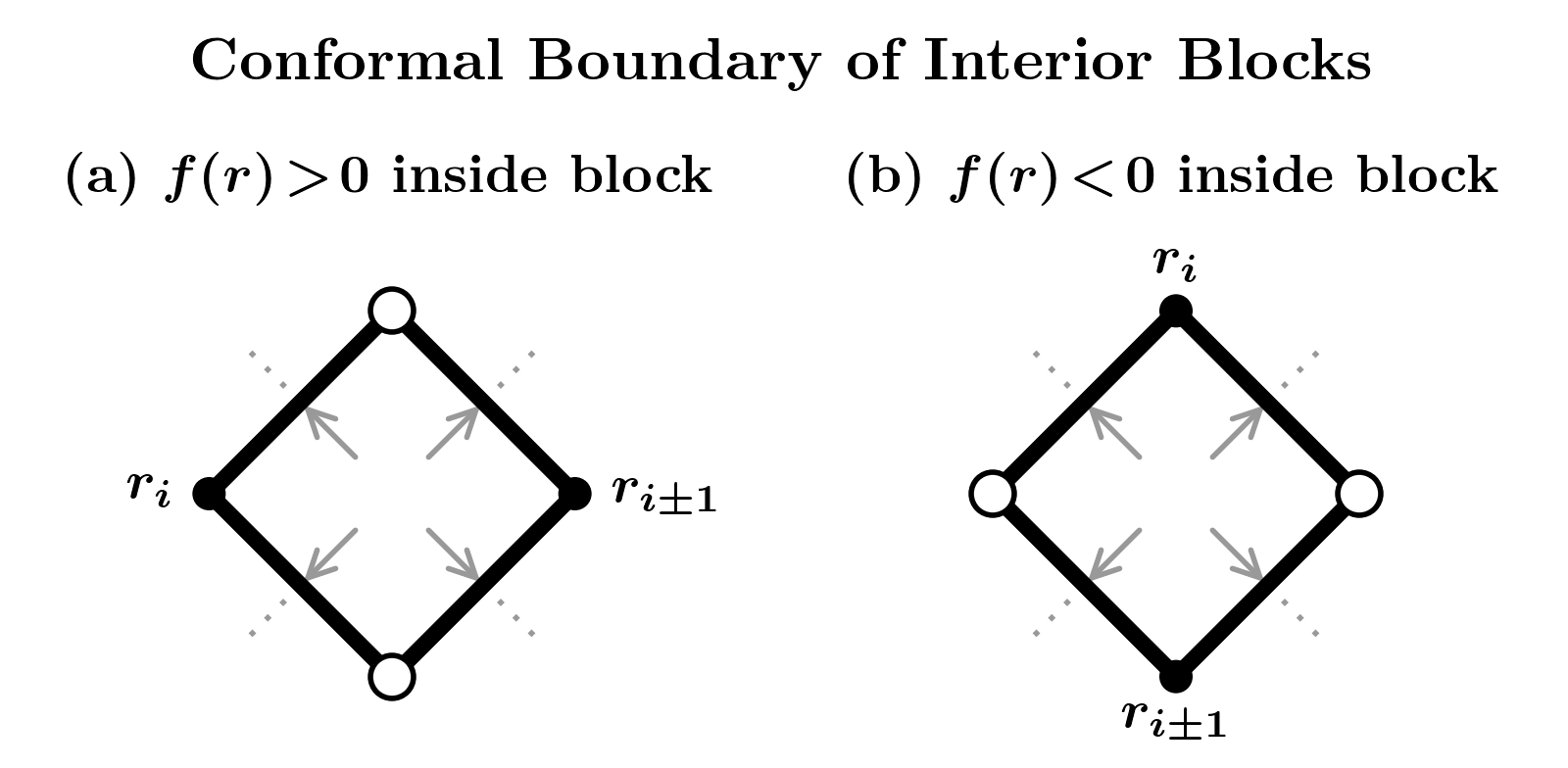}  
}
\caption[Conformal boundary of interior blocks]
{ \label{fig:interior-block}
The conformal boundary of a block corresponding to interval \mbox{$I_j=(r_j,r_{j+1})$} can be decomposed into boundaries corresponding to $r_j$ and $r_{j+1}$. Additionally, there are two points (empty circles) on the boundary which do not correspond to any point in $M$, and do not have a well defined radius. Each illustration above shows an ``interior" block, which is bounded by horizon radii (zeroes of $f(r)$) on both sides. For such a block, the pieces of the boundary associated with $r_j$ and $r_{j+1}$ each correspond to two horizons (bold lines) joined at a horizon vertex (solid circles). All interior blocks have four classes of incomplete radial null geodesics (gray arrows), each exiting the block through a different horizon. These classes can be classified as past or future directed and normal to $\partial_u$ ($\nwarrow$) or $\partial_v$ ($\nearrow$).
}
\end{figure}

The appearance of a block in the diagram is largely determined by the limits of the function $F(r)$ on the interval $I_j$. In general, each block is bounded by conformal boundaries corresponding to $r_j$ and $r_{j+1}$. When $|F(r_j)|$ is finite, the block fails to fill its diamond near $r_j$, and the corresponding boundary is either timelike or spacelike. When $|F(r_j)|$ is infinite, the block fills a corner of the diamond, and the corresponding conformal boundary is null: it consists of two null horizons joined at a vertex (see figure \ref{fig:interior-block}). This makes sense, since whenever $r_j$ is a horizon radius, the line of constant $r_j$ is necessarily null, and the value $|F(r_j)|$ is necessarily infinite.

As determined by the above-stated dependence on $F(r)$, each block has a \textit{causal shape} corresponding to its shape in the Penrose diagram. There are three possibilities; we denote them ``diamond", ``triangle", and ``slug". When the conformal boundaries at $r_j$ and $r_{j+1}$ both are null, the shape is \textit{diamond}. When one is null and the other either timelike or spacelike, the shape is \textit{triangle}. When either both are timelike, or both are spacelike, the shape is \textit{slug}. All blocks except the first and last necessarily have a diamond shape, and slugs are only possible when $f(r)$ has no zeroes. A block's causal shape is an intrinsic property of the type of block; it is the same in any Penrose diagram of the block, and for all blocks corresponding to the same $I_j$. Indeed, the tortoise function alone determines causal shape. The orientation of a block in the diagram, however, depends on the sign of $f(r)$ in the block, and on the block's time-orientation.

Every non-slug block, on its own, is extendible (geodesically incomplete without singularity) at its horizons. Indeed, to each horizon radius bounding $I_j$ are associated two classes of incomplete null geodesics: either future-directed and past-directed, or left-going and right-going. These correspond to null rays exiting the two horizons making up the block's conformal boundary at that radius. It follows that each block bounded by two horizon radii (these are necessarily diamonds) has four such classes of incomplete null geodesics, while each block bounded by one horizon radius (these may be triangles or diamonds) has two such classes. Counting classes of extendible null geodesics suggests how many neighbors a block can have.

We now return to the question of trapping. In \ref{sec:moreprops}, it is shown that a sphere of constant $(r,t)$ is a trapped surface if and only if $f(r)<0$. It follows that within a block with $f(r)<0$, every point intersects a trapped surface. In light of this fact, blocks on which $f(r)<0$ may sometimes be referred to as \textit{trapped blocks}, and a union of trapped blocks is a \textit{trapped region} (more generally, trapped regions are open sets on which every point intersects a trapped surface \cite{senovilla11b}). Trapped regions are important: an appealingly pragmatic definition of nonsingular black hole is ``a future-trapped region terminating in a region of extreme density". There are various technical notions describing the horizon associated with a trapped region. Suffice it to say that under certain conditions, the boundary of the trapped region can be called a trapping horizon and/or apparent horizon \cite{senovilla11}.

Figure \ref{fig:indyblocks} exhibits Penrose diagrams for some individual blocks of an SSS spacetime with $f(r)=1-1/r$. Note that there remains a freedom to invert the blocks by $(\udl,\vdl) \to (-\udl,-\vdl)$, since no physical criteria has been given to establish a time-orientation. Parameters for the diagram construction are given in the caption. 

Figure \ref{fig:interior-block} illustrates the conformal boundary structure of blocks  bounded by two horizon radii.

\subsection{Ingoing/outgoing regions, Kruskal quad-blocks, and horizon vertices}
\label{subsec:inout}

\begin{figure}[t]
\centerline{
{
\def\arraystretch{0.3}
\begin{tabular}{|ccc|}
\hline &  & \\
\includegraphics[scale=1]{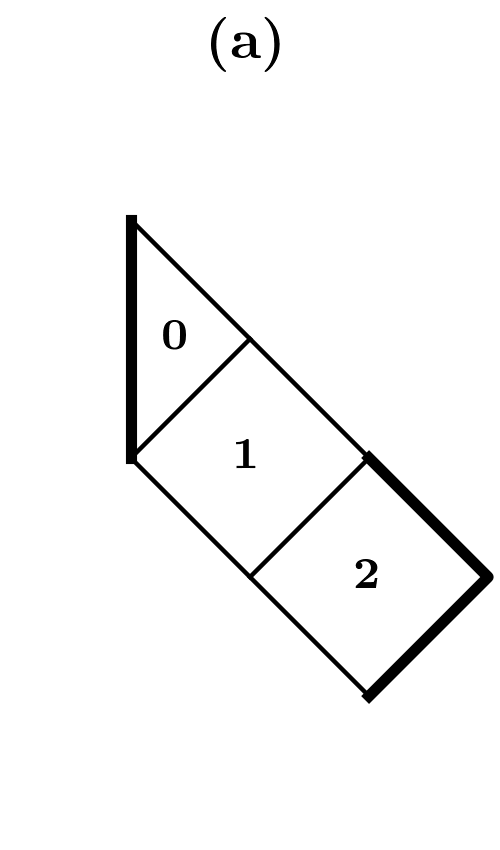}
&
\includegraphics[scale=1]{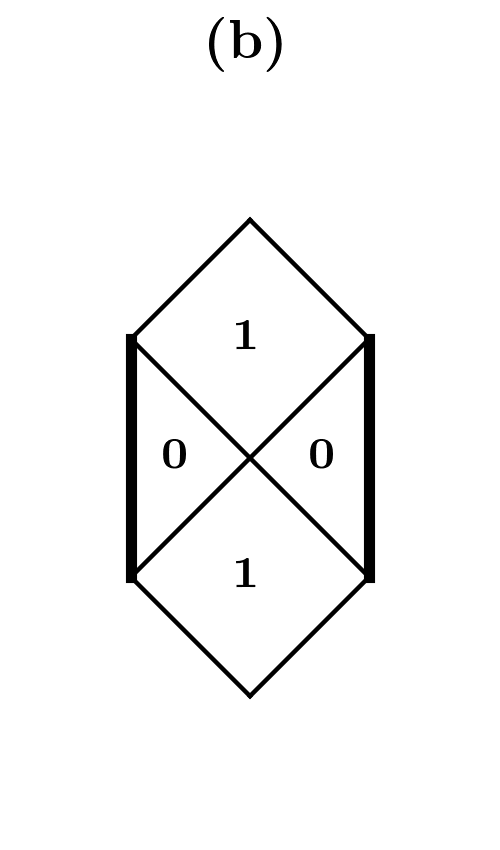}
&
\includegraphics[scale=1]{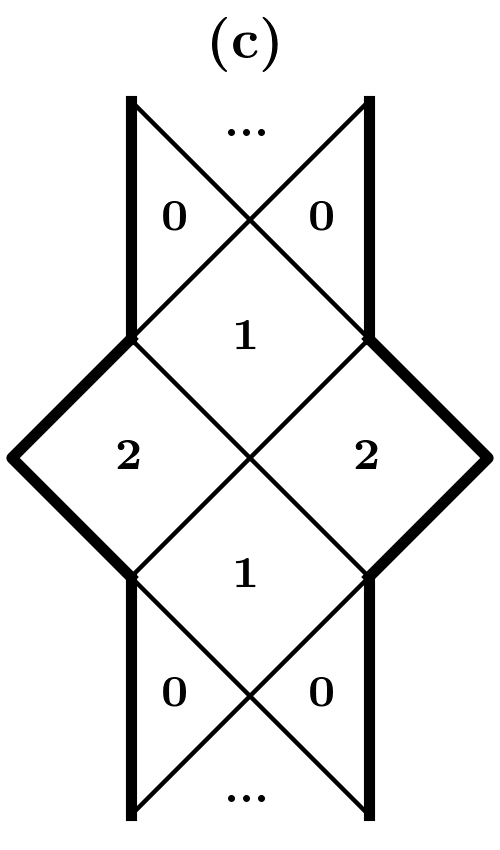}
\\[1mm]
\hline
\end{tabular}
}
}
\caption[Block diagrams]{ \label{fig:blockdiagram}
Block diagrams for an SSS spacetime with metric function \mbox{$f(r)=1 - 2 r^2 / (1+r^3)$}. The metric function has two zeroes. Shown are (a) an ingoing EF region, (b) a ``quad-block" region, and (c) a piece of a maximal extension. Bold lines indicate conformal boundaries where $r=0$ or $r=\infty$. Thin lines represent horizons at $r=1$ and $r\approx 1.62$. Each block is labelled according to its interval $I_j$.
}
\end{figure}

Having observed that $M$ is built from blocks, the next step is to see how these blocks can be joined together. There are two useful constructions that make this clear: Eddington-Finklestein (EF) regions, and Kruskal ``quad-block" regions. As this section will demonstrate, each of these units highlights an important aspect of how blocks may be joined. In particular, the EF regions show how a chain of blocks can be linked to cover the entire range of radii. Meanwhile, the Kruskal ``quad-block" regions show how four blocks can be joined at a horizon radius by four horizons and a vertex. These constructions demonstrate that each block admits either zero, two, or four neighbors. These regions can be visualized using the block diagrams (see section \ref{subsec:block}) of figure \ref{fig:blockdiagram}.

First, we introduce the Eddington-Finklestein regions (figure \ref{fig:blockdiagram}a). A \textit{full Eddington-Finklestein region} (\textit{EF region}) is a subset of $M$ which has the metric (\ref{eqn:efmet}) everywhere on a coordinate patch $r\in(0,\infty)$ and $w\in(-\infty,\infty)$. Note that there is no coordinate singularity in this patch. When an EF region is time-oriented such that $-\partial_r$ is future (past) directed, it is called an \textit{ingoing EF region} (\textit{outgoing EF region}). EF regions are the smallest regions of $M$ containing a point at every radius, and thereby are the smallest regions exhibiting the global function $f(r)$. They contain one of each type of block for $M$, with the blocks joined naturally across the horizons. The connectivity of the blocks is exactly the connectivity of the intervals $I_j$. Nonetheless, EF regions are, in general, extendible.

Every block of $M$ can be isometrically embedded into two distinct EF regions, corresponding to the choice $\epsilon=\pm 1$ in (\ref{eqn:efmet}), by using the transformation $w=t \mp \rstar$ (see section \ref{subsec:horblock} for definition of $\rstar$) to obtain the coordinates $(w,r)$ from the $(t,r)$ in (\ref{eqn:sssm}). The difference between these two regions is that different sets of incomplete null geodesics are extended. These two instances exhaust the block's extendible null geodesics, indicating that each block naturally has two neighbors for each horizon-radius bounding $I_j$. Within a given EF region, half of these possible neighbors are realized.

\begin{figure}[t]
\centerline{
{
\def\arraystretch{0.3}
\begin{tabular}{|c|c|}
\hline &  \\
\includegraphics[scale=1]{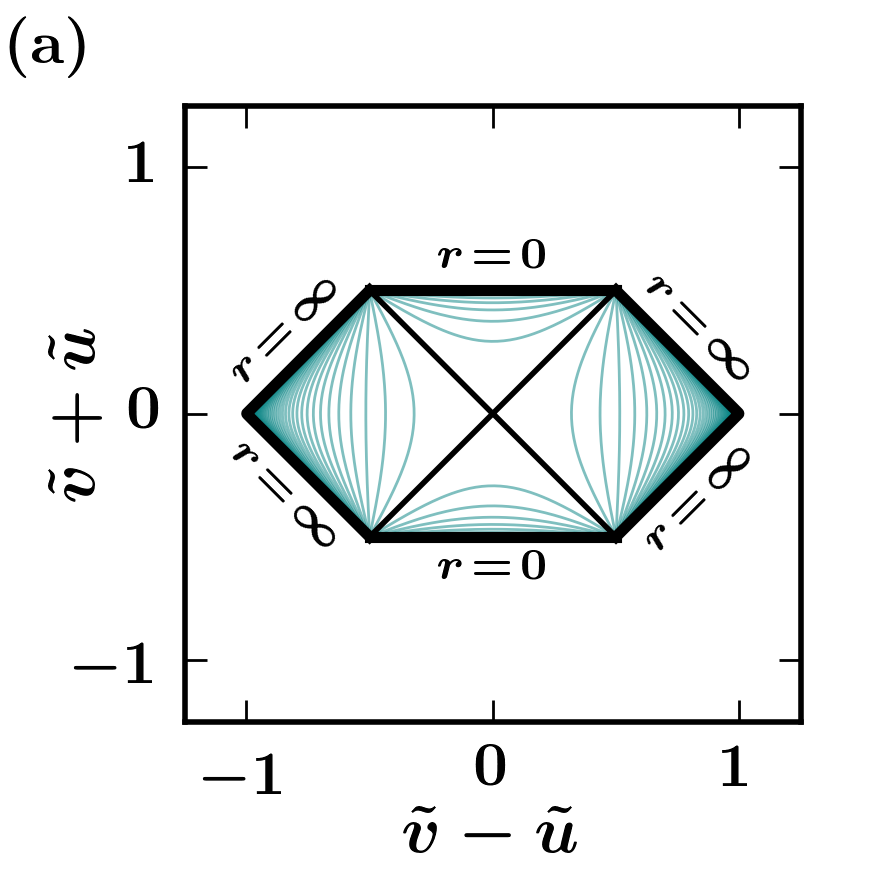}
&
\includegraphics[scale=1]{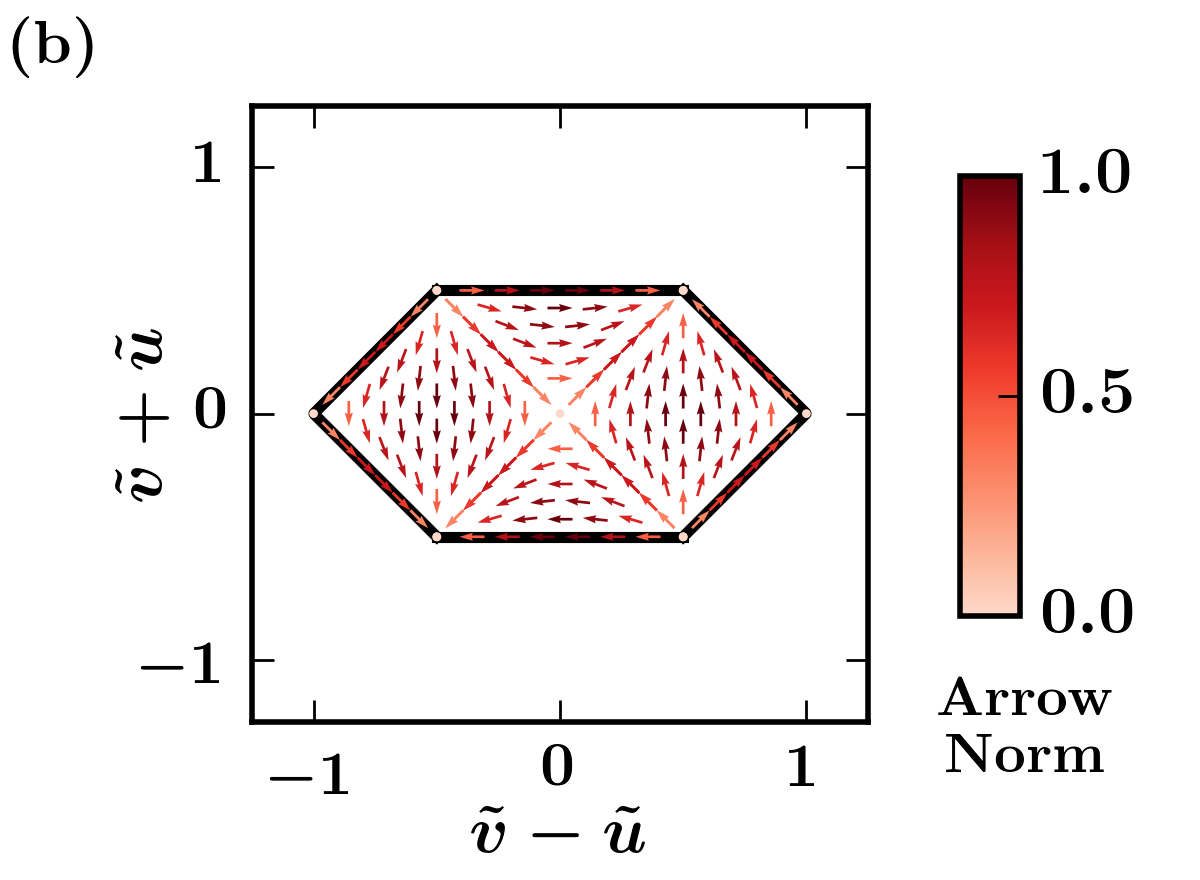}
\\[1mm]
\hline
\end{tabular}
}
}
\caption[Kruskal quad-block]{ \label{fig:kruskal}
(Color online). Penrose diagram for the classic Kruskal extension of a Schwarzschild black hole, with metric function $f(r)=1 - 1/r$, and tortoise function $\hat{F}(r) = r + \ln|r-1|$, according to the methods of section \ref{subsec:inout}. Panel (a) shows lines of constant radius (teal) with spacing $dr=0.1$. Interior black lines in (a) represent the horizons and vertex at $r=1$. Panel (b) shows a global Killing vector field which is locally $\partial_w$ (also locally $\partial_t$, see section \ref{subsec:coordnames}), pushed forward into Penrose coordinates. In (b), the depicted unit vectors (arrows) must be multiplied by a scale factor (arrow norm color scale) to obtain the components of the Killing vector field. Note that the Killing vector field lies everywhere tangent to lines of constant radius. 
    Apparent reflection symmetry within blocks in the Kruskal method is related to a curious identity: the function $\tan^{-1} ( e^{x}) - \pi/4$ is odd. 
     Since the Schwarzschild metric function has just one zero, the extended Schwarzschild spacetime has only one quad-block. Other SSS spacetimes may have many such regions, in which case the Kruskal extension method does not achieve a global coordinate system for all blocks simultaneously.
}
\end{figure}

Next we will introduce the Kruskal quad-blocks (figure \ref{fig:blockdiagram}b). The classic example of this structure is the Kruskal diagram for extended Schwarzschild spacetime, with metric function $f(r)=1-1/r$. In this case the Kruskal coordinates are usually constructed by defining $(u,v)$ using the tortoise function $\hat{F}(r)= r + \ln|r-1|$. Exponential transformations are then applied to $(u,v)$ to attain the Kruskal coordinates, and the resulting metric is shown to be nondegenerate at the horizons. Because this tortoise function is defined analytically, it is easy to miss its key property: that $\hat{F}(r) - \ln|r-1|  \to 1$ as $r \to 1$ from both the left and right. This did not have to be the case, as different constants could be added on either side of the discontinuity at $r=1$ without messing up the derivative. This would be quite unnatural to do when $\hat{F}(r)$ is defined analytically, but is perfectly natural when it is defined by a definite integral (e.g. equation (\ref{eqn:tortoise1}) above, which would in this case be applied on each side of $r=1$). The limit obtained by subtracting out the logarithmic infinity may take any value (a global constant is allowed), but if the left and right limits were different, the metric would be undefined at the horizons.

The issues associated with defining a well-matched tortoise function are described in detail in section \ref{subsec:globtort} below. In the present case, it suffices to take a simple generalization of the above observations:  to construct a Kruskal quad-block centered at $r_i$ in the general case, a tortoise function $\hat{F}(r)$ must be defined spanning both intervals $I_{i-1}$ and $I_{i}$,  such that $\lim_{r\to r_i} ( \hat{F}(r) - \frac{1}{k_i}\,\ln|r-r_i| ) = c$, where $k_i=f'(r_i)$, for some $c\in R$. This function exists and is unique due to our assumptions (section \ref{subsec:sss}) on $f(r)$, and the indeterminate form $|f(r)| \, e^{-k_i \hat{F}(r)}$ is analytic at $r=r_i$ with the limit $|k_i| \, e^{-k_i c}$ (see section \ref{subsec:globtort}). The desired definition may now proceed.

A \textit{Kruskal quad-block region} is centered on a horizon vertex at $r=r_i$ where $f(r_i)=0$. It contains two types of blocks, corresponding to $I_{i-1}$ and $I_{i}$, and the tortoise function must be defined as in the preceding paragraph above. A \textit{Kruskal quad-block centered at $r_i$} has the metric $ds^2 = 4 \, k_i^{-2} \, |f(r)| \, e^{-k_i \hat{F}(r)} \,d\hat{u} d\hat{v} + r^2 \, d\Omega^2$, in coordinates $(\hat{u},\hat{v},\Omega)$, where $|\hat{u}\hat{v}| = e^{k_i \hat{F}(r)}$. The metric is defined on a coordinate patch $ min < \hat{u}\hat{v}  < max  $, where the max and min values depend on $f(r)$ and $r_i$. The patch is always nonempty and includes $\hat{u}\hat{v}=0$. The point $\hat{u}=\hat{v}=0$ is a horizon vertex; it has no neighborhood isometric to (\ref{eqn:efmet}) because if it did there would be a smooth everywhere-nonzero vector field tangent to lines of constant radius in that neighborhood. The rest of $\hat{u}\hat{v}=0$ consists of horizon points at $r=r_i$. In case of the traditional Kruskal region with $f(r)=1-R/r$ and $\hat{F}(r)=r+R\ln|r/R-1|$, one immediately obtains $4 \, k_i^{-2} \, |f(r)| \, e^{-k_i \hat{F}(r)} = 4 R^3 r^{-1} e^{-r/R}$, which is the usual form of the prefactor in the traditional Kruskal metric.

Each quadrant of the Kruskal region is isometric to a single block of $M$. To make this evident, take the transformations $\hat{u} = \pm e^{-k_i u /2}$ and $\hat{v} = \pm e^{k_i v /2}$, followed by $u=t-\hat{F}(r)$ and  $v=t+\hat{F}(r)$, yielding the usual block metric (\ref{eqn:sssm}). Since every open set in the Kruskal region contains an open patch of an individual quadrant, this also proves that the region has strong spherical symmetry. In these coordinates the action of the Killing vector field $\partial_w$ looks similar to a boost, and has the vertex as a fixed point. Penrose coordinates for the Kruskal region can be obtained by applying  arbitrary squishing functions. Figure \ref{fig:kruskal} shows a Penrose diagram for one Kruskal region. 

Having exhibited the EF and Kruskal regions, the possible arrangements of joined blocks should be qualitatively clear. These arrangements are shown in figure \ref{fig:blockdiagram}. Figure \ref{fig:blockdiagram}a and \ref{fig:blockdiagram}b show the EF and Kruskal quad-block building blocks. Figure \ref{fig:blockdiagram}c shows how these can be combined to form larger structures. In  figure \ref{fig:blockdiagram}c, each block is a part of two EF regions, and one or two quad-block regions.

\subsection{Block diagrams and maximal extension}
\label{subsec:block}

Since we already know how each individual block looks in a Penrose diagram, and we now know how the blocks can be connected, it makes sense to draw a schematic \textit{block diagram} of $M$. In such a diagram each block is drawn with the appropriate causal shape and orientation, and connected blocks are drawn sharing the relevant horizon. In this way, the global causal structure and topology of $M$ can be accurately presented without need for a global Penrose diagram. Examples of block diagrams are given in figure \ref{fig:blockdiagram}.

Although a block diagram shares the qualitative appearance of its corresponding Penrose diagram, it lacks a global coordinate system. This has no effect on its usefulness as a tool for studying the global causal structure and topology. It does, however, stop us from accurately identifying the lines of constant radius, or plotting scalar functions of spacetime (e.g. the local density or local WEC inequality violation) on the digaram. For detailed analysis of dynamically evolving piecewise-SSS spacetimes, we need these more advanced tools at our disposal.

To wrap up discussion of the global structure of SSS spacetimes, let us introduce the concept of maximal extension. Technically speaking, a (connected) manifold $M'$ is \textit{maximally extended} if it can't be isometrically embedded into a proper subset of another (connected) manifold $M''$ of the same number of dimensions \cite{hawking73}. In practice, it is often true that a spacetime is maximally extended when all geodesics are either complete or approach a physical singularity. In the context of SSS spacetimes, what we usually mean is that the block diagram for $M$ leaves no open horizons. Block diagrams allow maximal extensions of SSS spacetimes to be described pictorially.

For any metric function $f(r)$, there is a maximally extended $M'$ with the metric function $f(r)$ everywhere. The structure of such an an $M'$ depends on the number $N$ of zeroes of the metric function. For $N=0$ every block is already maximally extended on its own. For $N=1$, the maximal extension is unique and consists of a single quad-block region. For $N=2$, maximal extensions may have a finite or infinite number of blocks, with a simple periodic structure similar to that seen in \mbox{figure \ref{fig:blockdiagram}c}. A finite number of blocks is possible (but only in the case $f(0)\!<\!0$) because of a topological ambiguity: torus-like boundary conditions can be allowed so long as closed timelike curves are avoided. Therefore maximal extensions in the case $N=2$ are not always unique, but do always have a unique simply-connected cover. For $N>2$, maximal extensions are constructed from an infinite chain of quad-block regions, and are not easily represented in a two dimensional block diagram. These necessarily have an infinite number of blocks, and suffer a similar topological ambiguity as the $N=2$ case. When analyzing maximal extensions in the case $N>2$, it is easiest to represent spacetime by a lattice of horizon vertices, rather than by a set of blocks. However, that method is pursued no further here.

\subsection{Boundary points with undefined radius}
\label{subsec:undefined_rad}

\begin{figure}[t]
\centering
{
\begin{tabular}{cc}
{\footnotesize \textbf{(a)}} & {\footnotesize \textbf{(b)}}
\\
\includegraphics[scale=1, trim={0 5mm 0 5mm}, clip]
{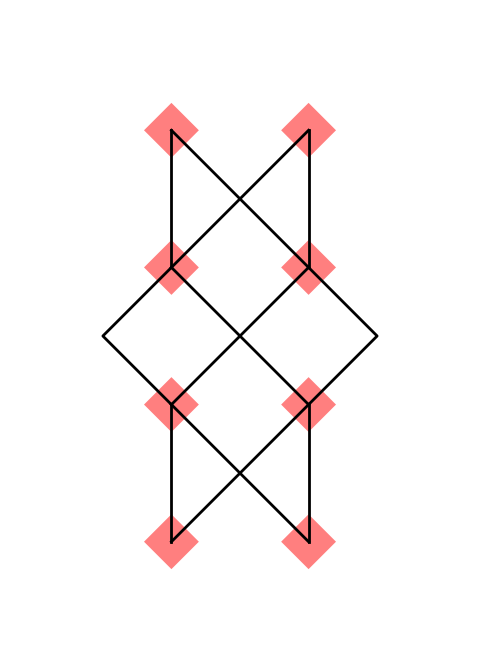}
&
\includegraphics[scale=1, trim={0 5mm 0 5mm}, clip]
{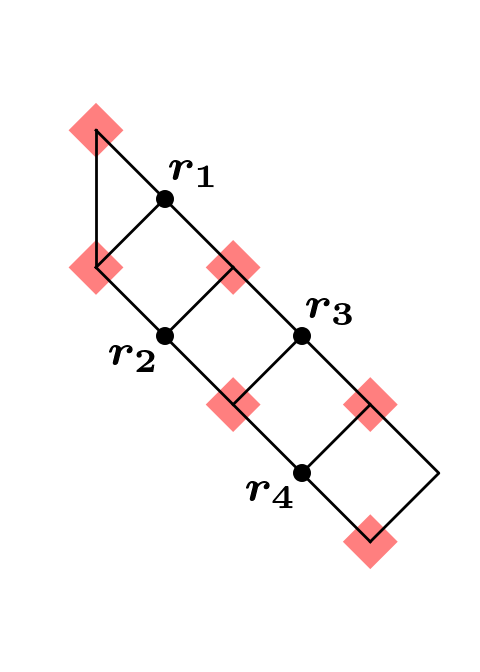}%
\end{tabular}

}
\caption[Boundary points with undefined radius]
{ \label{fig:undefined_radius}
(Color online). 
Conformal boundary points where $t \to \pm \infty$ are highlighted in red. At each highlighted boundary point, no radius can be defined. Panel (a) shows a maximally extended region with $N=2$ horizon radii, and panel (b) shows an EF region with $N=4$ horizon radii. In panel (b), horizon vertices are dotted (solid black circles) and labeled by radius. Due to a fundamental limitation, the metric will be discontinuous in some (arbitrarily small) neighborhood of the undefined radius points, in any Penrose diagram of an SSS spacetime. See section \ref{subsec:undefined_rad}.
}
\end{figure}

Points where $t \to \pm \infty$ on the conformal boundary of an SSS spacetime have a slight defect: their radius can't be well defined. This is obvious, since the same $t \to \pm \infty$ boundary point can be approached along lines of constant radius for a continuum of $r$ values. For this reason, it is best to think of these points as representing an entire continuum of boundary points, each corresponding to a unique radius. Nonetheless, this continuum is represented in Penrose diagrams by a single point. Figure \ref{fig:undefined_radius} highlights these defective points in two examples. 

The undefined radius points impose a fundamental, but minor, limitation on the possibility of constructing global Penrose coordinates for SSS spacetimes: in any Penrose diagram of an SSS spacetime, each undefined radius point on the conformal boundary has a neighborhood in which the metric may be discontinuous (continuity occurs as a special case but not in general). The affected neighborhoods may be made arbitrarily small by choice of how the diagram is constructed. The red patches in figure \ref{fig:undefined_radius} are meant to schematically represent the affected neighborhoods.

Since the neighborhoods where the metric is discontinuous can be made arbitrarily small, the problem imposed by these points is inconsequential. For any subset of $M$, or any worldline contained in $M$, the neighborhoods in which the metric is discontinuous can be ``moved far enough out to infinity" so as not to effect the physical problem. In our construction of Penrose coordinates in section \ref{sec:alg1}, a parameter $s_0$ will directly control the size of the affected neighborhoods, such that the neighborhoods become arbitrarily small as $s_0 \to \infty$.

Fundamentally, the issue is caused by the fact that a single undefined radius point on the conformal boundary may correspond to many different horizon radius values, in many different blocks, at the same time. The technical problem with this may be understood by observing the point between $r_2$ and $r_4$ on the bottom edge of \mbox{figure \ref{fig:undefined_radius}b}. Heuristically speaking, the metric will only be continuous at this point if the same coordinate transformation is applied on both sides. But on either side individually, a transformation must be carefully selected to make the metric continuous at the appropriate horizon (supposing more blocks are added to continue the region through the relevant horizons). Using the same transformation on both sides is, in general, incompatible with choosing the correct transformation on both sides. But since one only needs to be picky about transformations in a small neighborhood of the horizons, the problem can be pushed out into the corner.

The issue of undefined radius points is of a fundamental nature, and is not an artifact of any particular method for constructing diagrams. When there are more than two horizon radii (i.e. $N>2$ zeroes of $f(r)$), the problem cannot be avoided. On the other hand, in cases where $N=0$ or $N=1$, there are not enough horizon radii to force a discontinuity in the metric. In the case $N=2$ there are enough horizons to cause a problem, but the discontinuity can be removed by a special choice of transformations. However, the procedure to do so is rather unnatural. Instead, in the method of \mbox{section \ref{sec:alg1}} we will leave the discontinuities, in order to use more natural transformations, and to keep the treatment unified.

\subsection{True Penrose diagrams}

We have seen that $M$ can be partitioned into blocks, each with a well-defined causal shape. Moreover, we have seen how these blocks can be joined together across horizons in a regular way, and collections of blocks can be represented in a schematic block diagram. Lacking, so far, is a method for joining an arbitrary number of blocks, explicitly, in double-null coordinates.

This goal can be achieved in two key steps:
\begin{enumerate}
\item taking a global contour integral definition of the tortoise function; and
\item using a particular form of the squishing functions for each block.
\end{enumerate}
Applying these steps results immediately in Penrose coordinates for $M$, in which the metric extends continuously and nondegenerately across all horizons. This will be the subject of section \ref{sec:alg1}.

\section{Explicitly computing Penrose diagrams for SSS spacetimes}
\label{sec:alg1}

This section provides the recipe for explicitly constructing Penrose coordinates for SSS spacetimes. First, a global tortoise function is defined using a contour integral. Then a useful choice of the squishing function for each block is defined. Finally, the algorithm for generating Penrose coordinates is stated in full, and the resulting metric is shown to be nondegenerate.

We continue to assume that $f(r)$ satisfies the criteria laid out at the end of section \ref{subsec:sss}. In particular $f(r)$ has simple zeroes at $r=r_i$ (for $i=1,2,\ldots,N$) such that $f'(r_i)\neq 0$, and $f(r)$ is analytic at each of its zeroes. We denote $r_0=0$ and $r_{N+1}=\infty$, and the intervals $I_j=(r_j,r_j+1)$ are defined for $j=0,1,\ldots,N$. The parameters $k_j$ are defined by $k_0=k_{N+1}=0$ where no matching is needed, and by $k_i=f'(r_i)$ at the horizons.

\subsection{Global tortoise function}
\label{subsec:globtort}

\begin{figure}[b]
\centering
\includegraphics[scale=1]{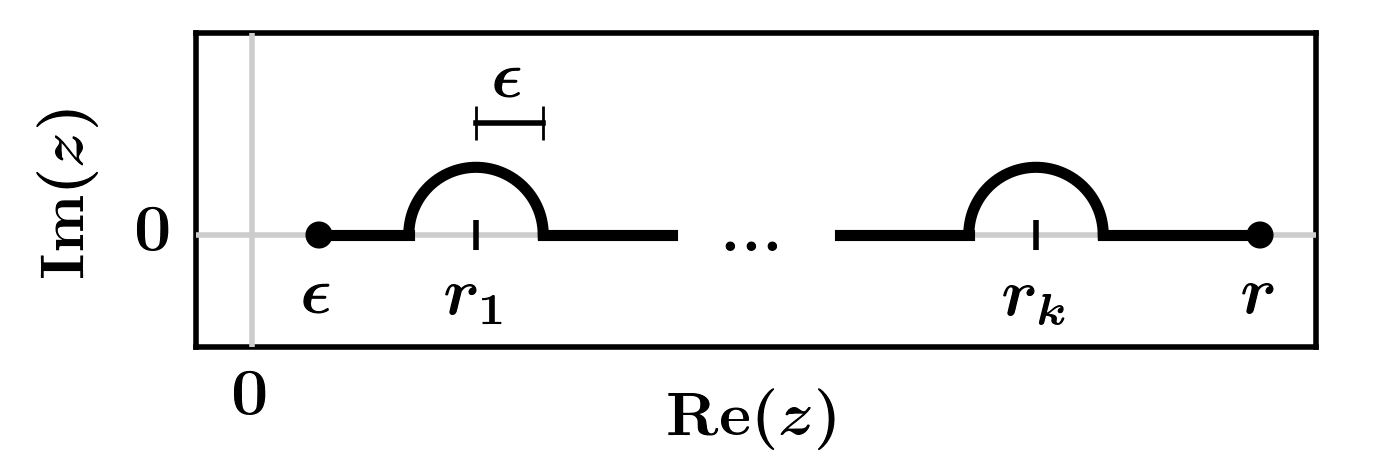} 
\caption[Tortoise function contour]{ \label{fig:contour}
The contour $\cre$ used to define the global tortoise function. The contour avoids each pole of $f(r)^{-1}$ using a semicircle of radius $\epsilon$ in the upper half plane.
}
\end{figure}

A tortoise function is by definition an antiderivative of $f(r)^{-1}$. This leaves an arbitrary integration constant in each block. These must be coordinated such that, when the logarithmic infinities at the zeroes of $f(r)$ are subtracted out, what remains is continuous. We adopt a definition for such a global function which is both analytically useful and numerically approximable. The only free parameter in this definition is a global additive constant, which has been absorbed into the definitions of $(u,v)$.

Let $\cre$ denote a contour in the complex plane which begins at $z=\epsilon$, ends at $z=r$, and follows the real axis except for avoiding zeroes of $f(r)$ using semicircles of radius $\epsilon$ in the upper half plane. Since $f(r)$ is assumed analytic at its zeroes, $f(z)$ is well-defined on the contour for sufficiently small $\epsilon$. This contour is depicted in figure \ref{fig:contour}.

\begin{figure}[t]
\centerline{
{
\def\arraystretch{0.3}
\begin{tabular}{|c|c|c|}
\hline & & \\
\includegraphics[scale=1]{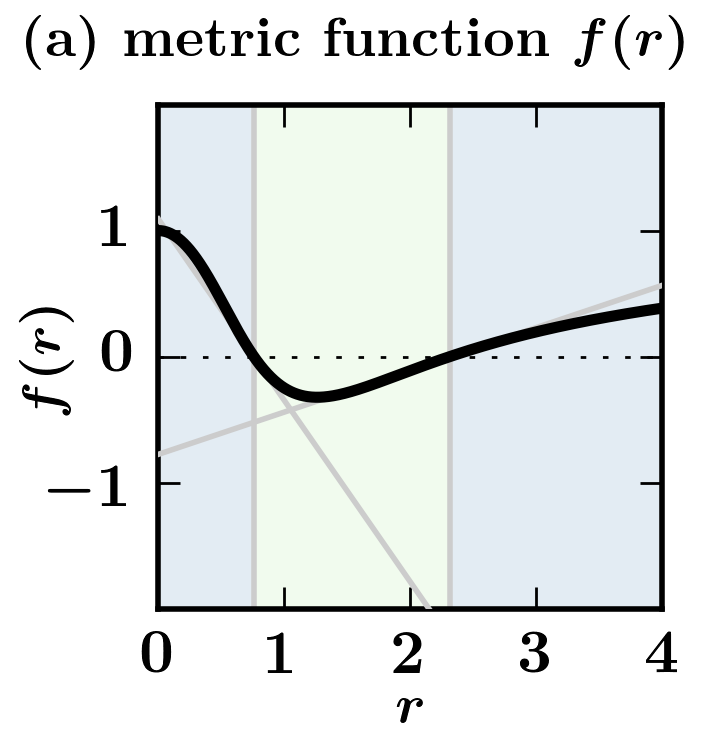}
&
\includegraphics[scale=1]{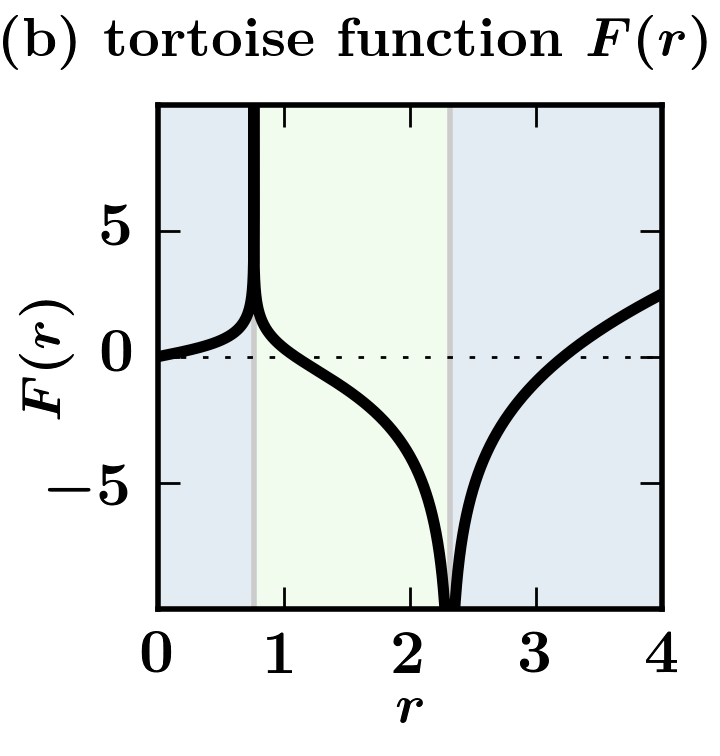}
&
\includegraphics[scale=1]{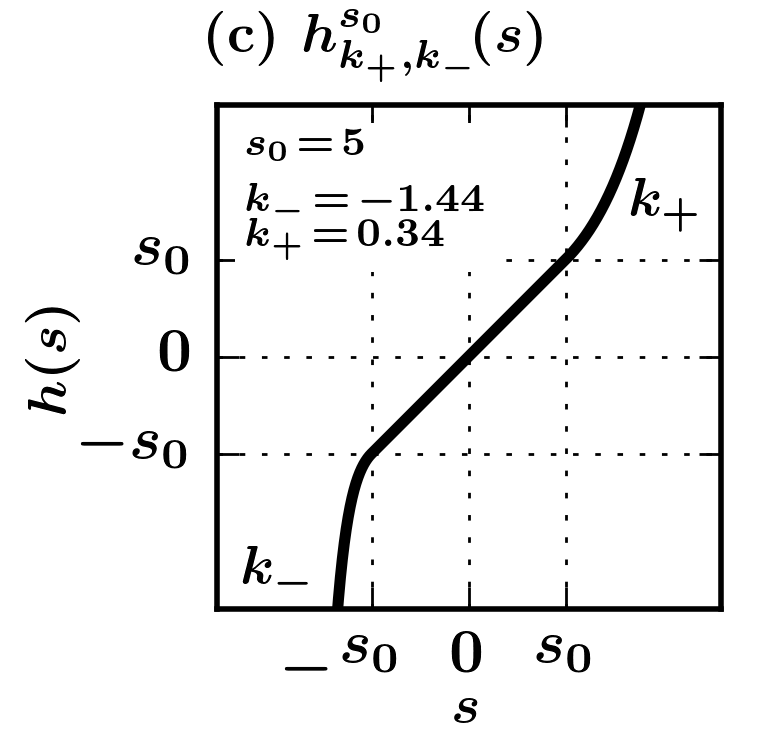}
\\[1mm]
\hline
\end{tabular}
}
}
\caption[Tortoise function and pre-squishing function]{ \label{fig:tortsquish}
(Color online).
     Functions used to construct the Penrose diagram for an SSS spacetime with metric function $f(r) = 1 - 2.5 \, r^2 / (1+r^3)$. 
     Panel (a) shows the metric function $f(r)$.  Zeroes of $f(r)$ (vertical gray lines), which define horizon radii, are located at $r_1 \approx 0.758$ and $r_2 \approx 2.313$. Tangent lines (gray) to $f(r)$ at its zeroes have slopes $k_1 \approx -1.44$ and $k_2 \approx 0.34$ respectively. 
     Panel (b) shows the corresponding global tortoise function $F(r)$, as defined by section \ref{subsec:globtort}. Note that near the ``stronger" horizon with greater $|k_i|$, the logarithmic infinity of $F(r)$ appears ``tighter".
     \mbox{Panel (c)} shows the pre-squishing function $h(s)$, as defined in section \ref{subsec:dnpen}, corresponding to the interior interval $I_1=(r_1,r_2)$ (light green), with the linear window parameter set to $s_0=5$. The exponential growth parameters are $k_+ = k_2 \geq 0$ and $k_- = k_1 \leq 0$, as appropriate for this interval. Note that positive (negative) values of $s$ are associated with the positive (negative) parameter $k_+$ ($k_-$), and that the linear segment where $h(s)=s$ connects once-differentiably to the exponential tails.
}
\end{figure}

The \textit{global tortoise coordinate $\rstar$} and \textit{global tortoise function $F(r)$} are defined by
\begin{equation} \label{eqn:globtort}
\rstar = F(r) = \textrm{Re} \; \lim_{\epsilon\to 0} \; \int_{\cre} \frac{dz'}{f(z')} \; .
\end{equation}
Clearly this definition obeys the defining relation $F'(r)=f(r)^{-1}$.  Within each interval $I_j$ the tortoise function is monotonic and invertible, and we denote the inverse functions by $r = F^{-1}_j(\rstar)$. The semi-circular contours each contribute the purely imaginary value $i\pi/k_i$  in the small $\epsilon$ limit. Thus  $F(r_i+\epsilon') \approx F(r_i-\epsilon')$ for sufficiently small $\epsilon'$.

It follows from the above considerations that (\ref{eqn:globtort}) is approximated by
\begin{equation} \label{eqn:tortapprox1}
F(r) = \int_\epsilon^r \! \frac{dr'}{f(r')} \hspace{4cm} (r \in I_0)
\end{equation}
\begin{equation} \label{eqn:tortapprox2}
F(r) \approx F(r_j - \epsilon) + \int_{r_j+\epsilon}^r \frac{dr'}{f(r')} \hspace{1.45cm} (r \in I_j)  \; (j>0)
\end{equation}
with a small parameter $\epsilon>0$. This form can be computed on a dense spacing of points by numerical integration. The forward and inverse functions can then both be approximated by linear interpolation. For improved numerical precision, it may be best to implement these formulae by integrating from some less extremal point in the interval, and matching boundary terms with additive constants. An example of the global tortoise function thus defined is given in figure \ref{fig:tortsquish}.

\begin{figure}[t]
\centerline{
{
\def\arraystretch{0.3}
\begin{tabular}{|c|c|}
\hline & \\
\includegraphics[scale=1]{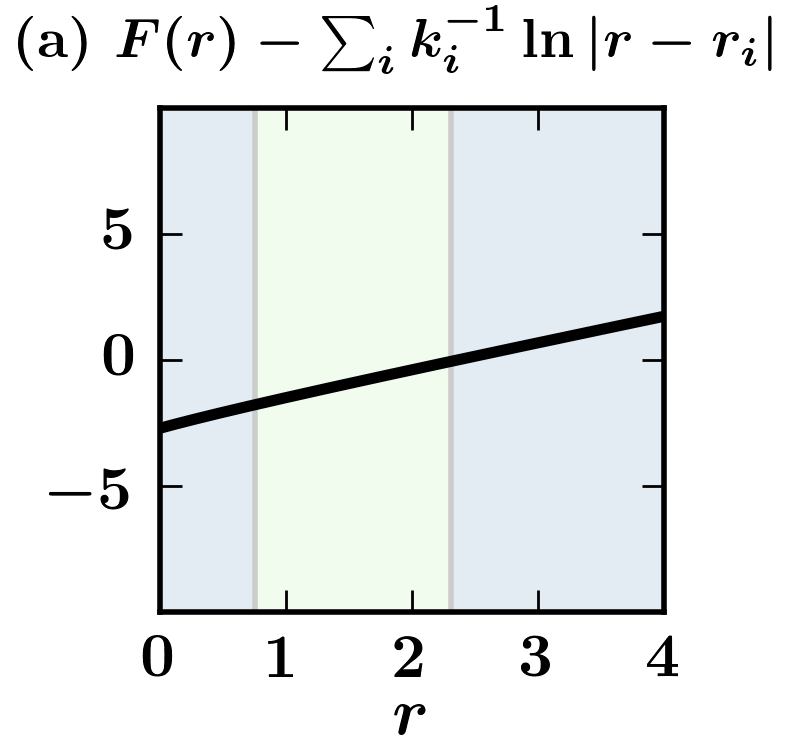}
&
\includegraphics[scale=1]{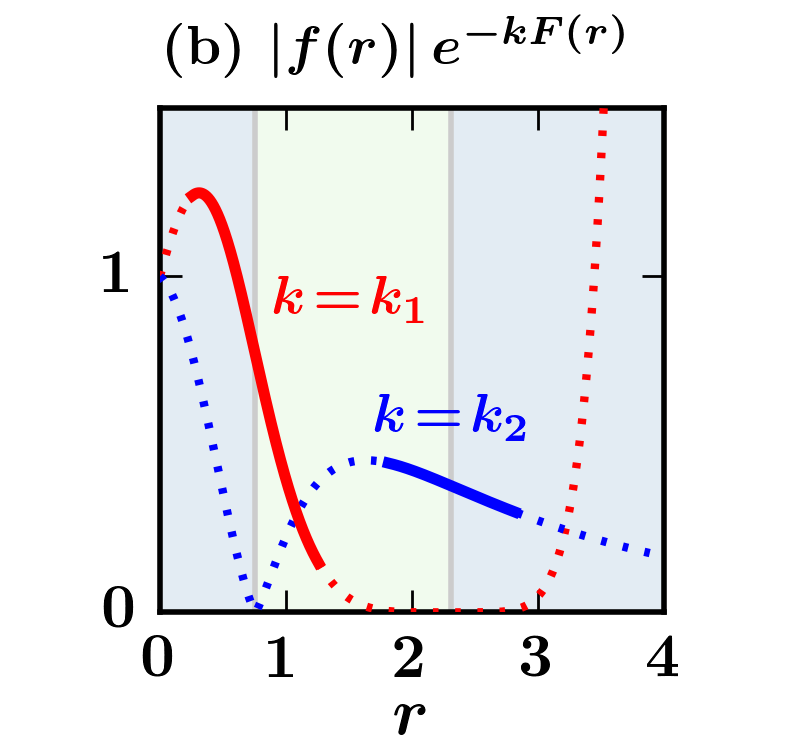}
\\[1mm]
\hline
\end{tabular}
}
}
\caption[Continuity properties of the global tortoise function]{ \label{fig:continuity}
(Color online).
     Properties of the global tortoise function $F(r)$, for the same metric function $f(r) = 1 - 2.5 \, r^2 / (1+r^3)$ shown in figure \ref{fig:tortsquish} above. (a) When the appropriate logarithmic infinities are subtracted out, a continuous function remains. (b) The function $|f(r)| \, e^{-k_i F(r)}$, which appears in the Penrose coordinate metric (\ref{eqn:penmetric}), is continuous and analytic at $r=r_i$. See section \ref{subsec:globtort} for details.
}
\end{figure}

By assumption, $f(r)$ is analytic and linear at each of its zeroes. This fact, in tandem with the definition (\ref{eqn:globtort}), gives the global tortoise function two key properties. Most importantly, the expression
\begin{equation} \label{eqn:analytic}
|f(r)| \; e^{- k_i F(r)} 
\end{equation}
is analytic and strictly positive in a neighborhood of $r=r_i$. This fact is essential for establishing that the metric is nondegenerate and continuous at the horizons; we will find that (\ref{eqn:analytic}) appears explicitly in the Penrose coordinate metric (\ref{eqn:penmetric}).
And secondly, the expression
\begin{equation} \label{eqn:cont}
F(r) - \textstyle\frac{1}{k_i} \, \ln|r-r_i|
\end{equation}
is analytic in a neighborhood of $r=r_i$. This result is of mainly conceptual importance, and relates closely to the first. These two properties can be understood by simple heuristic arguments in the real domain, and are also proved more rigorously in \mbox{\ref{subsec:tortproof}}; they are demonstrated by example in figure \ref{fig:continuity}.

Having defined the global tortoise function and determined its properties, the transformation to global Penrose coordinates can be undertaken.

\subsection{Double-null and Penrose coordinates}
\label{subsec:dnpen}

The global tortoise function makes it possible to define Penrose coordinates     $(\udl,\vdl)$ in which the metric is non-degenerate at the horizons. But to do so still requires that the coordinates be defined just right. 

The approach given here differs in philosophy from the more well-known Kruskal method. There, the blocks are joined first and then all jointly squished. Here, instead, each block is squished individually, and the squished blocks are placed next to each other. This type of technique was pioneered by \mbox{Carter \cite{carter66}}. The transformation in each block relies on an exquisite balance between arctangent compression and exponential expansion; when done right, a seemingly miraculous cancellation at the block boundaries renders the metric continuous.

First, the transformation to double-null block coordinates $(u,v)$ proceeds in the standard way, utilizing the global tortoise function. In particular,
\begin{equation}
u=t-F(r)+c \; ,
\qquad
v=t+F(r)-c \; .
\end{equation}
The global parameter $c \in \R$ absorbs the global integration constant left in the tortoise function, and must be the same in every block. For the same reasons explained in section \ref{subsec:horblock}, the definition of $(u,v)$ should have only the parameter $c$, and cannot be made more general by including other parameters.

The transformation from block to Penrose coordinates is then defined by
\begin{equation}
\tan  \, \pi \, (\udl - \cdl_u) =  \epsilon_u \,  h(u/2) \; ,
\qquad
\tan \, \pi \, (\vdl - \cdl_v) =  -\epsilon_v \,  h(-v/2) \; .
\end{equation}
The constants $\cdl_u \in \R$ and $\cdl_v \in \R$ locate the center of the block. The constants $\epsilon_u = \pm1$ and $\epsilon_v = \pm1$ determine the block's orientation, and must obey
\begin{equation}
\epsilon_u \, \epsilon_v = \sgn(f).
\end{equation} 
The function $h$ must be chosen carefully in each block, and is defined below. For notational purposes, $h$ will be written as $h(s)$, with $s$ a generic argument of no physical significance. The function $h(s)$ is referred to as the \textit{pre-squishing function}, since it is applied to block coordinates before arctangent squishing functions are applied.

All that remains is to define the function $h(s)$ in each block. There is some amount of freedom in this definition, as discussed in more detail below. We take the definition
\begin{equation} \label{eqn:hks}
h(s) = \hks(s) = \left\lbrace
\begin{tabular}{lll}
$-s_0 + H_{k_-}(s+s_0)$, & & $s <  - s_0$ \\
$s$, & & $|s| \leq s_0$ \\
$s_0 + H_{k_+}(s-s_0)$, & & $s > s_0$
\end{tabular}
\right. \; ,
\end{equation}
where
\begin{equation} \label{eqn:bigH}
H_k(s) = \left\lbrace
\begin{tabular}{ll}
$(e^{ks}-1)/k$ & $k \neq 0$ \\
$s$ & $k = 0$
\end{tabular}
\right. \; ,
\end{equation}
with $s_0 \geq 0$ a parameter of the diagram construction, and with $k_+ \in \R$ and $k_- \in \R$ parameters determined by the metric function in the block. In particular, for a block on radius interval $I_j=(r_j,r_{j+1})$, the parameters are
\begin{equation}
k_- = \min (k_j,k_{j+1}),
\qquad
k_+ = \max (k_j,k_{j+1}),
\end{equation}
where $k_0=k_{N+1}=0$ at $r_0=0$ and $r_{N+1}=\infty$, and $k_i = f'(r_i)$ at horizons (see second half of section \ref{subsec:sss}). Since by this prescription $k_- \leq 0 \leq k_+$, it is always true that $h(s)$ is continuous, differentiable, monotonic increasing, and unbounded as $|s|\to\infty$. Figure \ref{fig:tortsquish}c shows an example of $h(s)$ for a particular block.

The parameters $k_{\pm}$ control the exponential behavior at the tails of $h(s)$. This behavior is essential to obtaining a continuous metric at the horizons. The setup is such that near each horizon radius $r_i$, the corresponding slope $k_i$ controls the exponential transformation. Near $r=0$ and $r=\infty$ there is no need to match the metric; recalling that, by definition, $k_0=0$ at $r_0=0$ and $k_{N+1}=0$ at $r=\infty$ (end of section \ref{subsec:sss}), one can see that $h(s)$ remains linear at these locations. Although the definitions ensure that $h(s)$ is everywhere continuous and once-differentiable, large values of $|k_\pm|$ cause a rapid exponential turn-on that appears similar to (but is not) a kink in the first derivative.

Whereas $k_\pm$ control the form of the exponential regions of $h(s)$, the parameter $s_0\geq 0$ controls the location of the exponential regions in the diagram. As $s_0 \to \infty$, the exponential turn-ons are pushed arbitrarily far up against the horizons and into the corners of the diagram. As discussed in section \ref{subsec:undefined_rad}, as $|t|\to\infty$ there must always be a neighborhood where the metric may be mismatched. The parameter $s_0$ controls the size of these neighborhoods: the metric is never discontinuous unless both $|u/2|>s_0$ and $|v/2|>s_0$. We typically choose $s_0$ such that the exponential regions are visible but not overwhelming in the diagram.

There is some amount of freedom to choose an $h(s)$. Only the monotonicity and asymptotic behavior at $s \to \pm \infty$ are absolutely essential. But choosing the function poorly can lead to difficult calculations and very ugly diagrams. The choice above is based on several additional criteria: (i) the function is continuous and once differentiable; (ii) lines of constant radius in the resulting diagram look natural; (iii) the function has a closed-form inverse; (iv) metric calculations are relatively simple; and (v) issues near the points $|t| \to \infty$ are minimized.

Thus have the Penrose coordinates been constructed. The resulting metric is given in section \ref{subsec:enumalg}. It remains only to show how the blocks should be matched up such that the Penrose metric is continuous and non-degenerate at horizons.

\begin{figure}[t]
\centering
\fbox{
\includegraphics[scale=1]{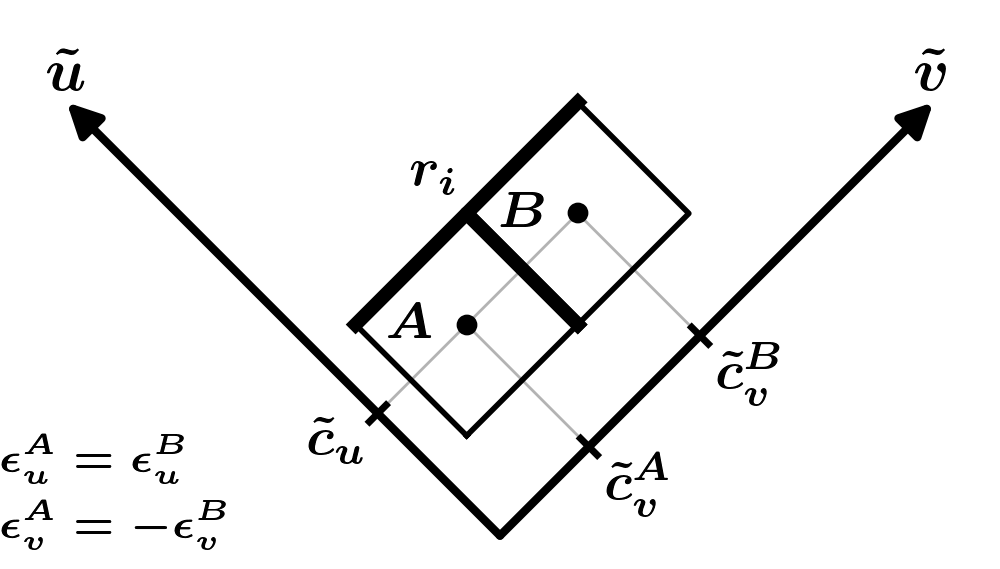} 
}
\caption[Joining blocks in Penrose coordinates]{ 
\label{fig:joining}
Illustration of the setup for joining neighboring blocks in the Penrose coordinates. See section \ref{subsec:joinpen}. In the case depicted here, $x \equiv u$ and $y \equiv v$, so that the blocks are joined on a line of constant $\vdl$. The bold boundary indicates horizons at the joining radius $r=r_i$.
}
\end{figure}

\subsection{Joining blocks in Penrose coordinates}
\label{subsec:joinpen}

Two blocks can be joined along a horizon at $r=r_i$ if and only if the two blocks correspond to $I_{i-1}$ and $I_{i}$. Let two such blocks be denoted $B_A$ and $B_B$, respectively. The necessary parameter constraints for matching are determined by simple considerations. The setup is depicted in figure \ref{fig:joining}.

In order that the blocks be squarely aligned, it is necessary that either \mbox{$\cdl_u^{\,B}=\cdl_u^{\,A}$} or \mbox{$\cdl_v^{\,B}=\cdl_v^{\,A}$}. For the horizons to intersect, the pair which is not equal must obey $|\cdl_y^{\,B}-\cdl_y^{\,A}|=1$ (where $y$ represents either $u$ or $v$). We can therefore say that the blocks are shifted in the coordinate $y$, and aligned in the other coordinate $x$. The matched horizon then traces out a line of constant $\tilde{y}$ in the diagram. In order that the orientation of this horizon be preserved, it is necessary that 
$\epsilon_{x}^{\,B} = \epsilon_{x}^{\,A}$. But since the metric function switches sign in subsequent intervals, one obtains the further requirement $\epsilon_u^{\,B} \epsilon_v^{\,B} = - \epsilon_u^{\,A} \epsilon_v^{\,A}$. This implies that $\epsilon_{y}^{\,B} = -\epsilon_{y}^{\,A}$ along the shifted coordinate. The only remaining parameter freedom is the sign of the above translation. In order to avoid introducing a time-orientation, this can be handled by simply stating that the matched horizons must be at an equal radius. This condition can be easily checked.

These requirements may be summarized as follows. Let $B_A$ and $B_B$ be blocks corresponding to $I_{i-1}$ and $I_{i}$. The two blocks may be joined either along a line of constant $\udl$, or along a line of constant $\vdl$. In order to treat both cases simultaneously, fix the symbols $(x,y)$ to mean either $(u,v)$ or $(v,u)$ (see section \ref{subsec:coordnames}). The blocks will be joined along a line of constant $\ydl$.  Then the blocks are properly matched at $r=r_i$ if:
\begin{enumerate}[(i)]
\item $\cdl_x^{\,A} = \cdl_x^{\,B}$ and $\epsilon_x^{\,A} = \epsilon_x^{\,B}$ ;
\item $\epsilon_y^{\,A} = - \epsilon_y^{\,B}$ ;
\item $|\cdl_y^{\,A} - \cdl_y^{\,B}| = 1$ ;
\item both intersecting horizons have the same radius $r=r_i$ .
\end{enumerate}
For a fixed orientation of $B_A$, this yields exactly two ways to attach $B_B$, corresponding to the choice of $(x,y)$. In \ref{subsec:proofan}, it will be shown that these conditions suffice to ensure non-degeneracy of the metric.

\subsection{Enumeration of the algorithm}
\label{subsec:enumalg}
All the ingredients are in place. In this section, we concisely summarize and enumerate the algorithm that has been presented for generating global Penrose coordinates $(\udl,\vdl)$ for an SSS spacetime with metric function $f(r)$.

Suppose $M$ is an SSS spacetime with metric locally of the form 
\begin{equation}
ds^2 = -f(r) \, dt^2 + f(r)^{-1} \, dr^2 + r^2 \, d\Omega^2 \; .
\end{equation}
Assume the metric function $f(r)$ is once-differentiable, has a finite number $N$ of zeroes on the interval $r \in (0,\infty)$, and that $\lim_{\, r \to 0} f(r) \neq 0$. Further, assume that $f(r)$ is analytic at each of its zeroes, although it may not be analytic globally, and that each zero is isolated and simple (linear).

The zeroes of $f(r)$ are denoted $r_i$ for $i=1,2,\ldots,N$, and at each zero there is a slope $k_i = f'(r_i) \neq 0$. Additionally, denote $r_0 = 0$ and $r_{N+1} = \infty$, along with $k_0 = k_{N+1} = 0$. Then the radius values $r_j$, for $j=0,1,\ldots,N$, partition the radial coordinate into intervals $I_j=(r_j,r_{j+1})$, and in each interval $I_j$ the sign of $f(r)$ is constant. Note that unlike $k_i$ at the zeroes of $f(r)$, the values $k_0$ and $k_{N+1}$ do not correspond to slopes, but are zero by fiat. This is because at $r_0$ and $r_{N+1}$, no horizon matching is needed.

First one must choose several global parameters associated with choices in the diagram construction. These control the appearance of the diagram. Additionally, the tortoise function $F(r)$, which determines the transformation into double null coordinates, must be defined globally in the correct way.
\begin{enumerate}[\hspace{2mm} (i)]
\item Choose the global parameters $c \in \R$ and $s_0 \geq 0$.
\item Define the global tortoise function $F(r)$ by (\ref{eqn:globtort}) as approximated by \mbox{(\ref{eqn:tortapprox1}--\ref{eqn:tortapprox2})}.
\end{enumerate}
The parameter $c$ absorbs an integration constant in the tortoise function and acts as a translation in the double null block coordinates. The parameter $s_0$ determines the location of exponential turn-ons in the piecewise function $h(s)$ below. Increasing $s_0$ pushes certain details into the corners of each block.

A block is specified by its interval $I_j=(r_j,r_{j+1})$. For each block in the diagram, appropriate block parameters must be chosen for the transformation to Penrose coordinates. For each block:
\begin{enumerate}[\hspace{2mm} (i)] 
\setcounter{enumi}{2}
\item Let $k_- = \min(k_j,k_{j+1}) \leq 0$. \\
Let $k_+ = \max(k_j,k_{j+1}) \geq 0$. \\
(Recall that $k_i = f'(r_i)$ for $1 \leq i \leq N$ and $k_0=k_{N+1}=0$).
\item Choose block parameters $\cdl_u,\cdl_v \in \R $ and $\epsilon_u,\epsilon_v \in \pm 1$. \\
Ensure these are chosen such that $ \epsilon_u \, \epsilon_v = \sgn(f)$.
\end{enumerate}
These control the location and orientation of a block.

Having set the parameters in each block, the transformation to double null coordinates $(u,v)$, then to Penrose coordinates $(\udl,\vdl)$, can be defined. For \mbox{each block:}
\begin{enumerate}[\hspace{2mm} (i)] 
\setcounter{enumi}{4}
\item Let $h(s)=\hks(s)$ as defined by (\ref{eqn:hks}).
\item Let $u=t-F(r)+c $. \\
Let $v=t+F(r)-c$. 
\item Let $\tan  \, \pi \, (\udl - \cdl_u) =  \epsilon_u \,  h(u/2) $. \\
Let $ \tan \, \pi \, (\vdl - \cdl_v) =  -\epsilon_v \,  h(-v/2)$.
\end{enumerate}
The coordinates $(\udl,\vdl)$ are now Penrose coordinates for each block by the definition of section \ref{sec:pen_di}. In order to ensure that the Penrose coordinates are global, parameters in each block must be compared.
\begin{enumerate}[\hspace{2mm} (i)] 
\setcounter{enumi}{7} 
\item Check that all adjacent blocks are properly matched according to the criteria listed at the end of \mbox{section \ref{subsec:joinpen}}.
\end{enumerate}
These conditions ensure that all blocks overlap only on the appropriate horizons, and that relative block orientations are correct.

The resulting metric in Penrose coordinates $(\udl,\vdl)$ is implicitly defined in terms of $(u,v)$ and $r$ by
\begin{equation} \label{eqn:penmetric}
ds^2 = -  \left( \frac{4 \pi^2}{e^{-kc}}  \right) \frac{|f(r)|}{e^{k F(r)} } \;\, G_u(u,k) \; G_v(v,k) \;  d\udl \, d\vdl + r^2 \; d\Omega^2
\end{equation}
where we have introduced a free parameter $k\in \R$, and
\begin{equation}
\label{eqn:Gu}
G_u(u,k) = e^{-ku/2} \; \frac{1+h(u/2)^2}{ h'(u/2)},
\end{equation}
\begin{equation}
\label{eqn:Gv}
G_v(v,k) = e^{kv/2} \; \frac{1+h(-v/2)^2}{ h'(-v/2)} \; .
\end{equation}
The areal radius is given in each block by
\begin{equation}
r = r(\udl,\vdl) = F_j^{-1} \left(\frac{v(\vdl)-u(\udl)}{2} + c \right) \; .
\end{equation}
The metric (\ref{eqn:penmetric}) is independent of the free parameter $k$, which cancels out entirely when $G_u$ and $G_v$ are substituted back into the metric. Thus (\ref{eqn:penmetric}) holds for every $k \in \R$. The purpose of introducing artificial dependence on $k$ is to make it easy to evaluate limits at each $r \to r_j$ by setting $k=k_j$. This trick works  on both sides of every $r_i$ (i.e. approaching $r_i$ from blocks on $I_i$ or $I_{i+1}$), due to the piecewise definition of $h(s)$ in each block. Note that with the appropriate substitution for $k$, the explicit function of radius appearing in the metric is equivalent to (\ref{eqn:analytic}).

In the above form, all metric coefficients of $-d\udl d\vdl$ and of $d\Omega^2$ are explicitly non-negative everywhere, and explicitly positive at all block points. In can be shown that these coefficients are also positive at horizon and horizon vertex points. In \ref{subsec:proofan} it is shown that this metric is continuous and nonzero everywhere, except along the polar coordinate singularity at $r=0$, and except possibly in arbitrarily small corners of the diagram where $|u/2|>s_0$, $|v/2|>s_0$, and $|t|\to \infty$ (as discussed in section \ref{subsec:undefined_rad}). Thus, the metric extends continuously and without degeneracy across all horizons and horizon vertices.

This form of the metric depends upon the Penrose coordinates only implicitly, through its dependence on $(u,v)$ and $r$. This leaves the freedom to translate and flip blocks in the Penrose coordinates without a meaningful disturbance of the metric. Once the diagram is set and the blocks are matched, arbitrary monotonic functions $\udl'(\udl)$ and $\vdl'(\vdl)$ can be applied without compromising the Penrose coordinates; this usually just makes the diagram look worse, but is necessary in section \ref{sec:alg2}.

\section{Extension to piecewise-SSS spacetimes with null-shell junctions}
\label{sec:alg2}

In the astrophysical universe, spacetime is not strongly spherically symmetric -- it is constantly, dynamically, evolving. Fortunately, many dynamical phenomena can be approximately modeled by spacetimes which are piecewise-SSS. Basic examples include stellar radiation, stellar collapse to a black hole, and the emission of Hawking radiation from a black hole.

The class of spacetimes treated in this section, which we refer to as \textit{piecewise-SSS spacetimes with null-shell junctions}, is defined by having a finite number of SSS components, each joined together along null junction hypersurfaces. The junction hypersurfaces correspond to thin null shells of matter. These junctions may have either the geometry of a single null shell (figure \ref{fig:corner}b), or of two null shells colliding at a corner (figure \ref{fig:corner}a). The full metric and other tensors are defined distributionally on the piecewise spacetime.

In this section we state the procedure necessary to construct piecewise-SSS spacetimes, making use of the Penrose coordinates previously identified. This immediately yields Penrose coordinates for the piecewise spacetime. First, we present the procedure for shell and corner junctions under the minimal requirement that junction shells have well-defined intrinsic geometry. Then, we determine the matter content of junction shells and discuss the conservation of energy.

This section is largely informed by the classic exposition of Barrab\`es and \mbox{Israel \cite{israel91}}, which is the standard reference for the analysis of thin shells and piecewise-defined spacetimes.

\subsection{Geometry of a radial null slice in SSS spacetime}
\label{subsec:geonull}

\begin{figure}[t]
\centering
\fbox{
\includegraphics[scale=1]{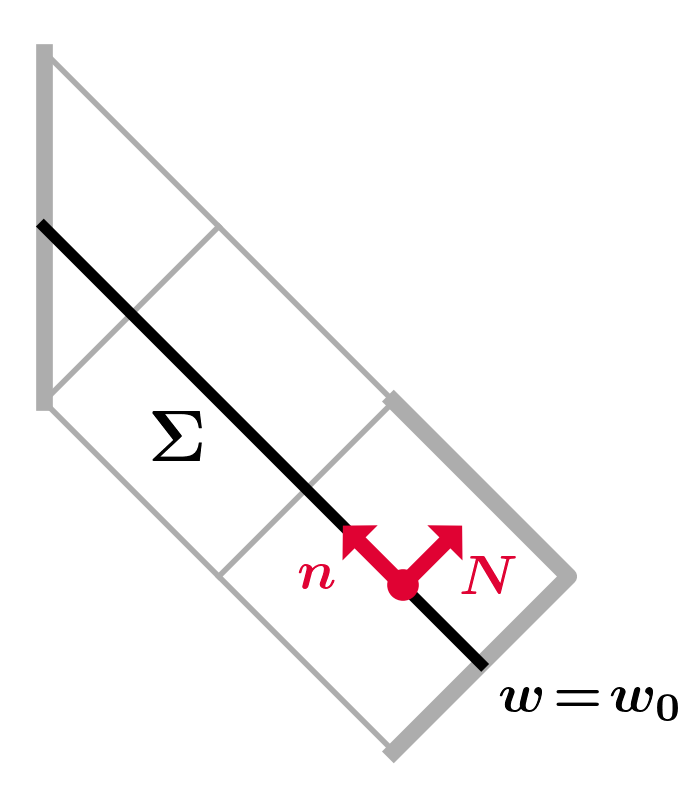} 
}
\caption[Radial null slice in SSS spacetime]{ \label{fig:nullslice}
(Color online). Illustration of a radial null slice in SSS spacetime. The hypersurface $\Sigma$ (black) lies on a surface of constant $w=w_0$. The vector $n^a$ (red) is both normal and tangent to $\Sigma$, while the vector $N^a$ (red) is transverse to $\Sigma$.
}
\end{figure}

Consider an EF patch of an SSS spacetime, covered by the coordinates and \mbox{metric (\ref{eqn:efmet})}. As usual, $\epsilon=\pm 1$ is the parameter appearing in the EF metric. Assume, without loss of generality, that $\partial_w$ is future-directed. (This is general because $w\to-w$ with $\epsilon\to-\epsilon$ is an isometry of the EF patch.) With this convention, it follows that
\begin{eqnarray}
\epsilon =1 \quad \Longleftrightarrow \quad \partial_r \hspace{10pt}\textrm{future-directed,}
\\
\epsilon =-1 \quad \Longleftrightarrow \quad \partial_r \hspace{10pt}\textrm{past-directed,}
\end{eqnarray}
and that $(\epsilon \, \partial_r)$ is always future-directed. This convention applies throughout section \ref{sec:alg2}.

With this setup, a radial null slice $\Sigma$ of the patch is defined as a hypersurface of constant $w=w_0$, as depicted in figure \ref{fig:nullslice}. Such a hypersurface is parameterized by the coordinates $x^i=(r,\Omega)$, and obtains the induced metric
\begin{equation}
ds^2 = r^2 \; d\Omega^2
\end{equation}
in the coordinate basis. 

A normal vector $n^a$ and transverse null vector $N^a$ to $\Sigma$ can be defined (in the abstract index notation for spacetime tensors) by
\begin{equation}
n^a = \epsilon \; (\partial_r)^a \; ,   
\qquad   
N^a = (\partial_w)^a - \textstyle\frac{1}{2} \; \epsilon \, f(r) \, (\partial_r)^a \; ,
\end{equation}
such that $n_a n^a = N_a N^a = 0$ and $n_a N^a = -1$. Note that $n^a$ and $N^a$ are both future-directed. As usual for a null hypersurface, the normal vector is tangent to $\Sigma$, and is a degenerate vector of $T_p \Sigma$ in the induced metric. This implies, as usual, that $\Sigma$ has a geometrical dimension $D-2$ one less than its topological dimension $D-1$.

Before moving on, it worth noting that the normal and transverse vectors to $\Sigma$ are eigenvectors of the stress tensor on the EF patch. At block points in the patch, the stress tensor ${T^a}_b = (1/8\pi) \, {G^a}_b$ is given in an ortho-normalized Schwarzschild coordinate basis by (\ref{eqn:stress}) of the appendix. This stress tensor has two degenerate eigenspaces. Using this fact, one can note without calculation that, at block points,
\begin{equation}
{T^a}_b \, n^b = -\rho \, n^a ,   \qquad    {T^a}_b \, N^b = -\rho \, N^a  \; ,
\end{equation}
since both $n^a$ and $N^a$ lie entirely in the $(\partial_t,\partial_r)$ plane. Direct calculation in the EF coordinates reveals that these same eigenvalue relations hold also at horizon points, and thus everywhere throughout the EF patch.

\subsection{Algorithm for implementing corner and shell junctions}
\label{subsec:junctalg}

Four SSS (or piecewise-SSS) spacetimes can be joined together at a corner, which represents a pair of colliding null shells. The basic restriction for such a junction is that every point have a well-defined radius, so that the induced metric on the junction hypersurfaces is well-defined. Here we present a method for attaining a junction under this minimal condition. Later, more detailed conditions for energy conservation are given. For the sake of simplicity, in this treatment the corner junction point is always located at a block point. Although it is possible to locate the corner junction at a horizon or vertex point, to do so requires a separate treatment. Keep in mind that, despite the similarity in their schematic illustrations, a corner junction point is \textit{not} the same thing as a horizon vertex. Once the algorithm for corner junctions is established, the procedure for shell junctions follows as a special case.

\begin{figure}[t]
\centerline{
\def\arraystretch{0.3}
\begin{tabular}{|c|c|c|}
\hline
 & & \\
\includegraphics[scale=1]{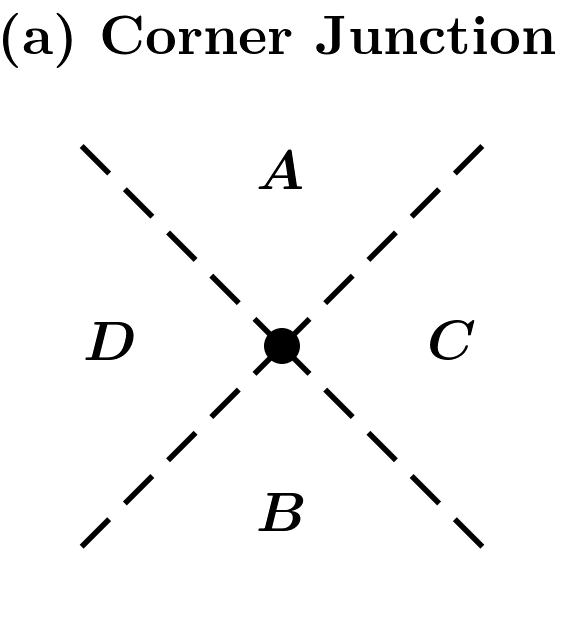}
&
\includegraphics[scale=1]{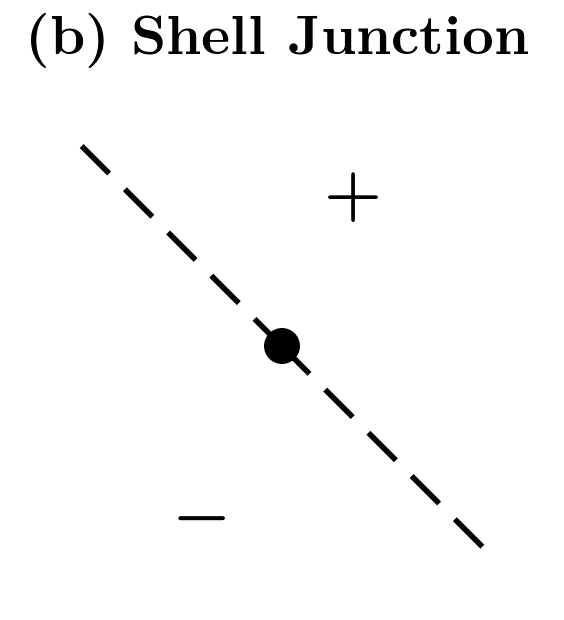}
&
\includegraphics[scale=1]{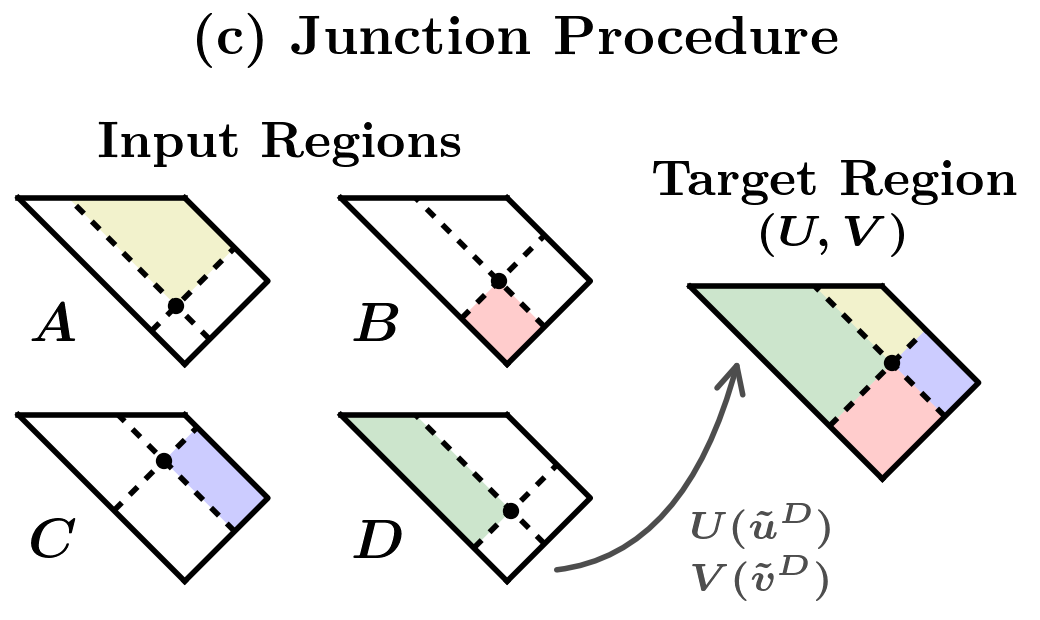} 
\\
\hline
\end{tabular}
}
\caption[Piecewise junction schematic]{ 
\label{fig:corner}
(Color online).
Schematic depiction of corner and shell junctions. \mbox{(a) The} collision of two null shells at a corner point separates four SSS regions at a piecewise junction. In each of the four regions, the metric function $f(r)$ may be different. The corner point (black circle) has a well-defined radius, $r=r_0$, called the ``corner radius". Note that, despite the similar schematic appearance, a corner point is \textit{not} the same thing as a horizon vertex. (b) A null shell separates a past region ($-$) from a future region ($+$) at a piecewise junction. The shell junction is a special case of the more general corner junction. (c) Illustration of the procedure for creating properly matched junctions. Coordinates in the target region are denoted $(U,V)$, and each input region has its own Penrose coordinate system $(\udl,\vdl)$. The appropriate subset of each input region is mapped into the target space by null transformations $U(\udl)$ and $V(\vdl)$. These mappings must be self-consistent, such that every point on the junction hypersurfaces (dashed lines) has a well-defined radius. Compare to figure \ref{fig:example2} for an implemented example.
}
\end{figure}

Consider four SSS (or piecewise-SSS) regions, labelled $A,B,C,D$, and called the \textit{input regions}, to be combined into a piecewise unit. Let each input region have Penrose coordinates $(\udl,\vdl,\Omega)$. The input regions will be mapped into a \textit{target space} representing the joint spacetime, as in figure \ref{fig:corner}. The coordinates of the target space we call $(U,V,\Omega)$. For each input region, the transformations $U(\udl)$ and $V(\vdl)$ into the target space must be specified.

The procedure by which the input regions are joined can be stated simply: each input region must be sliced along null junction surfaces in the $u$ and $v$ directions, then stretched and shifted in the $u$ and $v$ directions such that the radii at junction surfaces are all matched up. This procedure is illustrated schematically in figure \ref{fig:corner}c. A precise formulation of the procedure is as follows.

First, one must choose null slices in each input region to act as the junction hypersurfaces. This amounts to choosing values $\udl_0$ and $\vdl_0$ in each (the values in each region may be distinct). Junction slices are then located at $\udl=\udl_0$ and $\vdl=\vdl_0$, with the corner point located at the intersection $(\udl_0,\vdl_0)$. The values must be chosen consistently, such that the corner points for each input region have the same radius $r_0 = r(\udl_0,\vdl_0)$. We call this shared radius the \textit{corner radius}.

With the junction hypersurfaces consistently defined in the input regions, one is free to define an arbitrary radial parameterization of the junction hypersurfaces in target coordinates. This is done by specifying two arbitrary monotonic functions $\Ur(r)$ and $\Vr(r)$. Each input region will be mapped into the target region so that the radii match these functions on the junction surfaces. The corner point will therefore attain the coordinates $(U_0(r_0),V_0(r_0))$ in target coordinates.

Having achieved this setup, the transformation of each input region into target space coordinates is given by
\begin{equation} 
\label{eqn:pentotarget}
U(\udl) = \Ur \Big( r(\udl,\vdl_0) \Big), 
\qquad 
V(\vdl)=\Vr \Big( r( \udl_0,\vdl) \Big).
\end{equation}
When junction slices cover less than the entire domain of $\udl,\vdl$ values, the above transformations may be extended by an arbitrary monotonic extrapolation. The transformations ensure that the radii $r(U,V_0(r_0))$ and $r(U_0(r_0),V)$ are well-defined regardless of the region of evaluation. Moreover, a simple chain rule calculation leveraging the results of sections \ref{subsec:globtort} and \ref{subsec:dnpen} reveals that when the input regions are SSS, the junction slice radii $r(\udl,\vdl_0)$ and $r(\udl_0,\vdl)$ are monotonic functions. This implies that whenever $\Ur(r)$ and $\Vr(r)$ are monotonic, the target-space metric components for each input region are regular and nonzero. This ensures that the transformations (\ref{eqn:pentotarget}) will yield Penrose coordinates for the joint spacetime. Furthermore, it follows from the chain rule that $r(U,V_0(r_0))$ and $r(U_0(r_0),V)$ are monotonic functions in the target coordinates. Thus even when the input regions are themselves piecewise-SSS, the above procedure yields Penrose coordinates in the target space.

The full metric of the joint spacetime is defined piecewise in terms of the input region target-space metrics; it can be written distributionally using Heaviside \mbox{$\Theta$-functions}. The arrangement is depicted in figure \ref{fig:corner}a. With this setup, the induced metric on the junction hypersurfaces is $ds^2 = r^2 \; d\Omega^2$. Since the radii at these hypersurfaces has been properly matched, the geometry of the joint spacetime is well-defined and non-degenerate. The coordinates $(U,V,\Omega)$ are Penrose coordinates, and our theory of corner junctions is complete.

A shell junction can be regarded as a special case of the corner junction, in which two pairs of the input spacetimes are identical. In such a case, all the above considerations remain valid. Additionally, for a shell, radius matching in either the $U$ or $V$ direction becomes trivial (depending on the direction of the shell), so that one of the transformations in (\ref{eqn:pentotarget}) can be replaced by $U(\udl)=\udl$ or $V(\vdl)=\vdl$, or some other monotonic function, if desired.

This procedure achieves our goal of constructing a composite spacetime under the minimal junction condition.

\subsection{Matter content at shell junctions}
\label{subsec:mattercont}

In order to analyze the matter content at a shell junction generated by the above algorithm, it is easiest to work in local EF coordinates of the type set up in \mbox{section \ref{subsec:geonull}}, rather than in the Penrose coordinates of the previous section. Suppose we focus on a local patch $M_0$, which is separated into a past region $M_-$ and future region $M_+$ by the null junction hypersurface $\Sigma$, with the metric functions $f_{\pm}(r)$ in the two regions. Each of $M_{\pm}$ can be expressed in terms of EF coordinates, with metric of the form (\ref{eqn:efmet}), such that $\Sigma$ lies on a line of $w=const$ in each region, and such that $\epsilon_+=\epsilon_-$. It is therefore possible to choose a joint coordinate system $(w,r,\Omega)$ on $M_0$, such that $\Sigma$ is defined by the level set $\Phi=w=0$, and such that the metric is (\ref{eqn:efmet}) with metric function $f(r)=f_{\pm}(r)$ in the appropriate regions. In accordance with section \ref{subsec:geonull} and the requirements of \cite{israel91}, the metric parameter $\epsilon=\pm 1$ indicates the future-/past- directedness of $\partial_r$, the normal and transverse vectors $n^a$ and $N^a$ are future-directed, and the level set function $\Phi=w$ increases toward the future.

In order to conveniently express the stress tensor, let us define the mass function $m(r)$ by $f(r)= 1 - 2\, m(r)/r$, and define the mass jump $[m(r)]=m_+(r)-m_-(r)$. Note that no restriction on $m(r)$ is implied --- it is simply a useful way to write the metric function. Then, in the joint EF coordinate system described above, the stress tensor associated with the junction shell may be read off from \cite[(s. II, IV)]{israel91}. In the abstract index notation for spacetime tensors (as opposed to the notation convention of \cite{israel91}, which uses latin indices for hypersurface coordinates), it reads
\begin{equation}
T^{ab}_{\Sigma} =  \sigma \, n^a n^b \; \delta(w) , 
\qquad 
\sigma = (-\epsilon) \, \frac{[m(r)]}{4\pi r^2} \; ,
\end{equation}
where $n^a = \epsilon \, (\partial_r)^a$ is both normal and tangent to the shell, the metric parameter $\epsilon=\pm 1$ indicates an outgoing ($+$) or \mbox{ingoing ($-$)} shell (see sec \ref{subsec:geonull}), and $\delta(w)$ is the Dirac \mbox{$\delta$-distribution}. 

The coefficient $\sigma$ may be thought of as the surface energy density of the shell, up to an arbitrary normalization factor associated with the null vector $n^a$. A more physical quantity is the surface energy density relative to an observer with future-directed timelike tangent vector $t^a$, which is given by $\sigma_t =  (n_a t^a)^2 \;  \sigma$. Evidently the sign of $\sigma$ is physically meaningful: timelike observers measure a positive energy density if and only if $\sigma>0$. Indeed, one can show that when $\sigma<0$ the null, weak, and dominant energy conditions are violated, and the energy flux vector (relative to a future-directed timelike observer) is past-directed null. It is therefore sensible to say that shells with $\sigma<0$ have \textit{negative mass}, while shells with $\sigma>0$ have \textit{positive mass}. The sign of $\sigma$ is a local property, and in principle (in physically unusual cases) a single shell may have positive and negative mass at different points.

\begin{figure}[t]
\centerline{
\def\arraystretch{0.3}
\begin{tabular}{|cc|}
\hline
\multicolumn{2}{|c|}{} \\[3pt]
\multicolumn{2}{|c|}{\textbf{ \small Schwarzschild Junction: Positive Mass Shell Scenarios}} \\[2pt]
\multicolumn{2}{|c|}{} \\
\hline
 &  \\[1pt]
\includegraphics[scale=1]{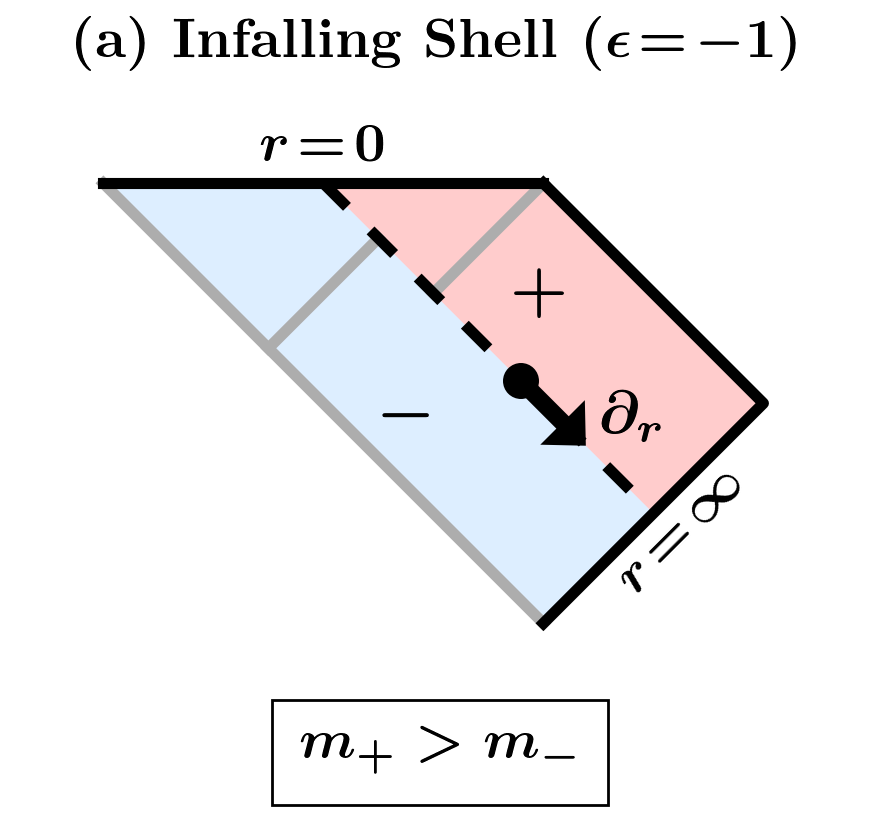} 
 &
\includegraphics[scale=1]{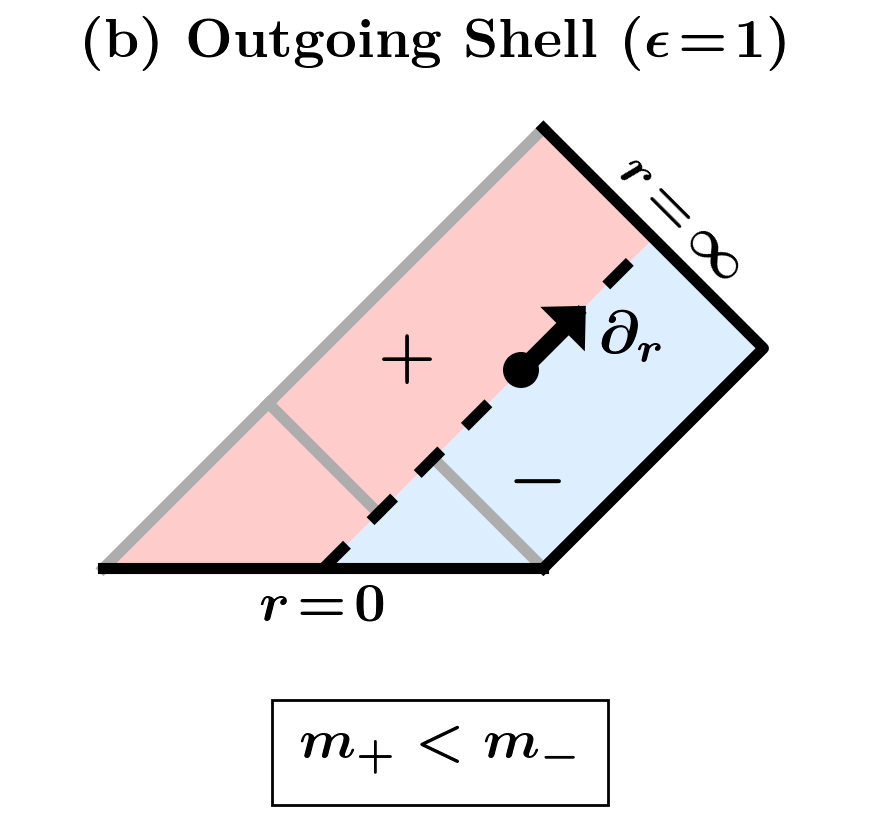} 
 \\[2mm]
\hline
\end{tabular}
}
\caption[Positive mass Schwarzchild junctions]{
\label{fig:posmass}
(Color online). 
Illustration of the possible scenarios for joining two Schwarzschild spacetimes across a positive-mass shell (i.e. a shell with $\sigma > 0$, see \mbox{section \ref{subsec:mattercont}}). The junction shell $\Sigma$ (dashed black line) separates the joint spacetime into future (red fill, labelled ``$+$") and past (light-blue fill, labelled ``$-$") regions. The parameter $\epsilon=\pm 1$ is determined by the future-/past- directedness of $\partial_r$, while the normal vector $n^a = \epsilon \, (\partial_r)^a$ (not shown) is always future-directed. As expected, an infalling (outgoing) positive-mass shell necessarily increases (decreases) the mass $m_+$ of the future region. This change is reflected by a shift in the horizons (gray lines) at $r=2m$ in each region.
}
\end{figure}

For the junction of two patches of Schwarzschild spacetime, the mass jump is a constant value $\Delta m$; the positive mass shell scenarios ($\sigma>0$) for this case are depicted schematically in figure \ref{fig:posmass}. If the inequalities in the figure were reversed, the junction would yield a shell with negative mass.

\subsection{Energy conservation at shell and corner junctions}
Local conservation of energy and momentum in General Relativity is expressed by the relation $\ddel_a \, T^{ab}=0$. For any smooth metric, the contracted Bianchi identities provide $\ddel_a \, G^{ab}=0$, ensuring energy conservation by way of Einstein's equation. When the metric is not differentiable, the standard derivation of the Bianchi identities does not hold up. So the question remains: for the piecewise case, does local energy conservation hold everywhere as a distributional identity?

For the case of shell junctions, it has been shown by Barrab\`es and Israel \mbox{\cite[(eqns. A10 - A13)]{israel91}} that $\ddel_a \, G^{ab}=0$ does indeed hold as a distributional identity. It is therefore true that, without any further constraints, the joint spacetime produced by our junction algorithm is automatically energy conserving at all shell junctions. In context of our junction algorithm, every junction is locally a shell junction except at the corner points $(U_0(r_0),V_0(r_0))$ in the target coordinates. So the only remaining question is that of energy conservation at the corner points. 

At corner points, the question of energy conservation is slightly more complicated, but there is nonetheless a well-established theory \cite{israel91,dray1985,redmount1985,barrabes2003}. In order that conservation hold at a corner point with radius $r_0$, the metric functions of the four input regions must satisfy the DTR (Dray - 't Hooft - Redmount) relation \cite{barrabes2003}
\begin{equation}
f_A(r_0) \, f_B(r_0)  = f_C(r_0) \, f_D(r_0) \; ,
\end{equation}
with the region labels defined by figure \ref{fig:corner}a. This formula encodes a relativistic version of conservation of mass in the shell collision. Note that when shell junctions are regarded as a special case of corner junctions, the DTR relation is satisfied trivially. It is well known that the DTR relation is necessary for energy to be conserved, and general consensus of the standard treatments of shell collision (cited above) suggests that the relation is also sufficient. However, the authors are not aware of an explicit proof that the distributional equation $\ddel_a \, G^{ab}=0$ holds at corner points when DTR is satisfied.

\section{Implementation and examples}
\label{sec:imp}
The methods described in this article have been implemented in a package called \texttt{xhorizon} for Python 2.7. Source code for the implementation is available from the authors.

\begin{figure}[ht]
\centering
\renewcommand{\plotwidth}{2.5in}
\begin{tabular}{cc}
\includegraphics[width=\plotwidth]{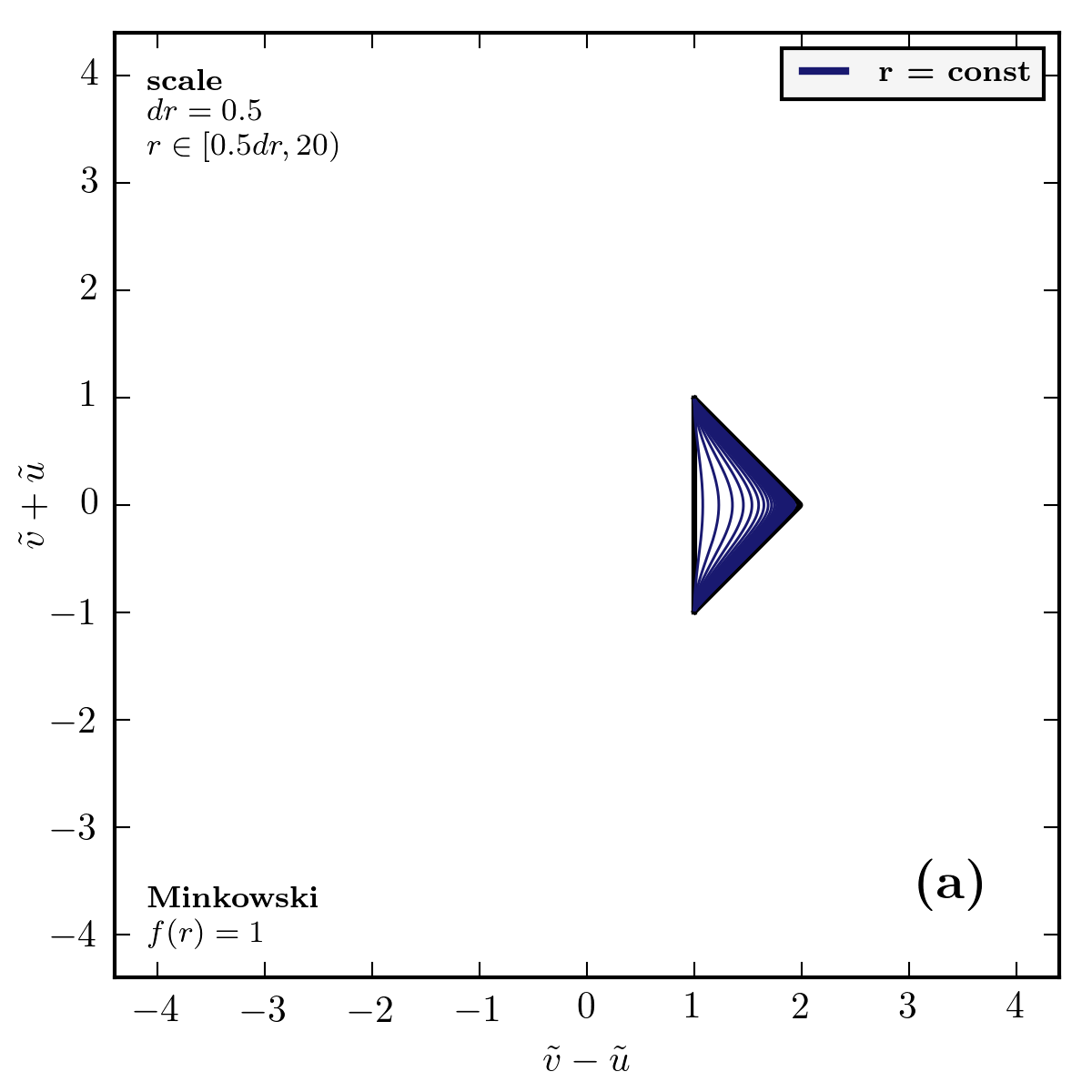} 
&
\includegraphics[width=\plotwidth]{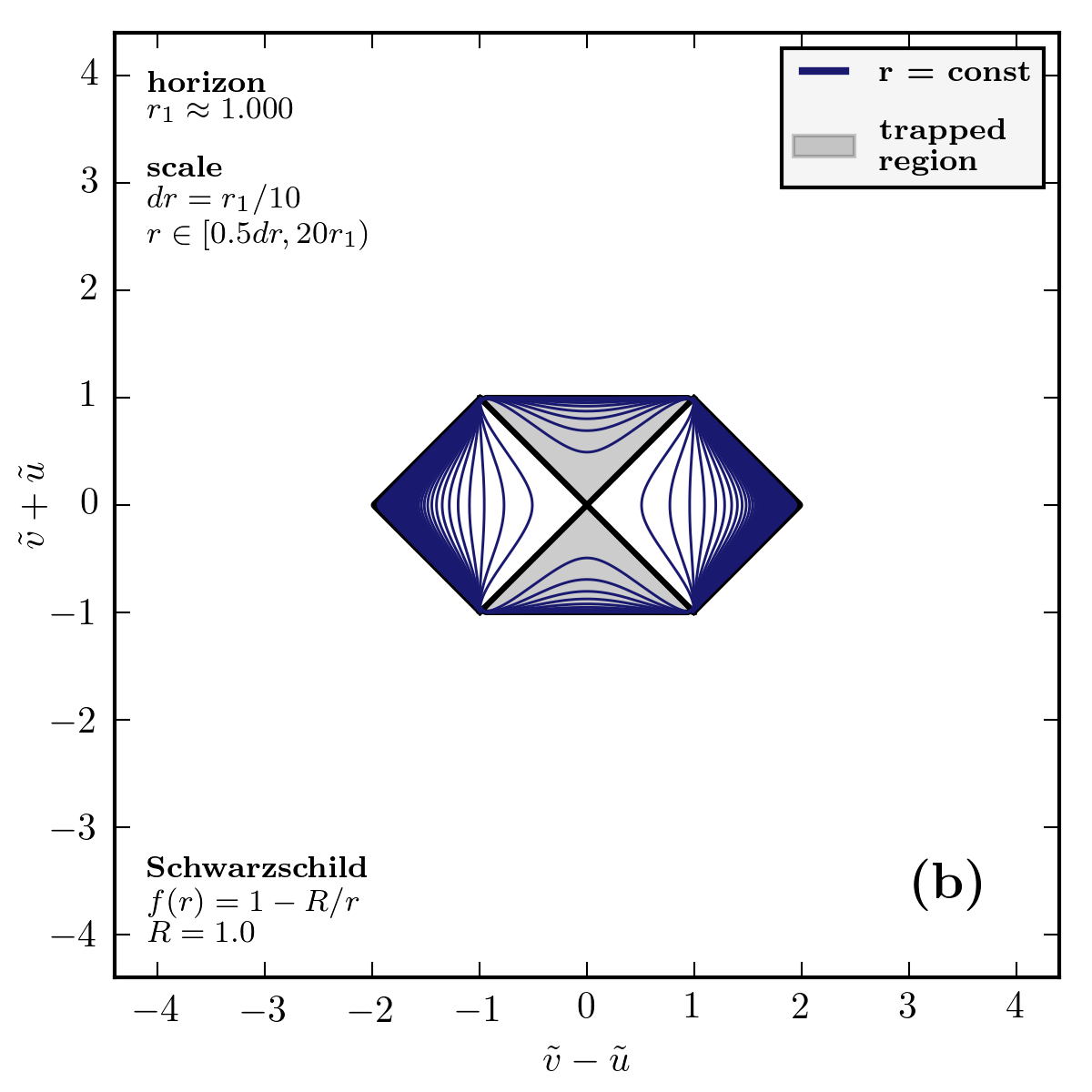} 
\\
\includegraphics[width=\plotwidth]{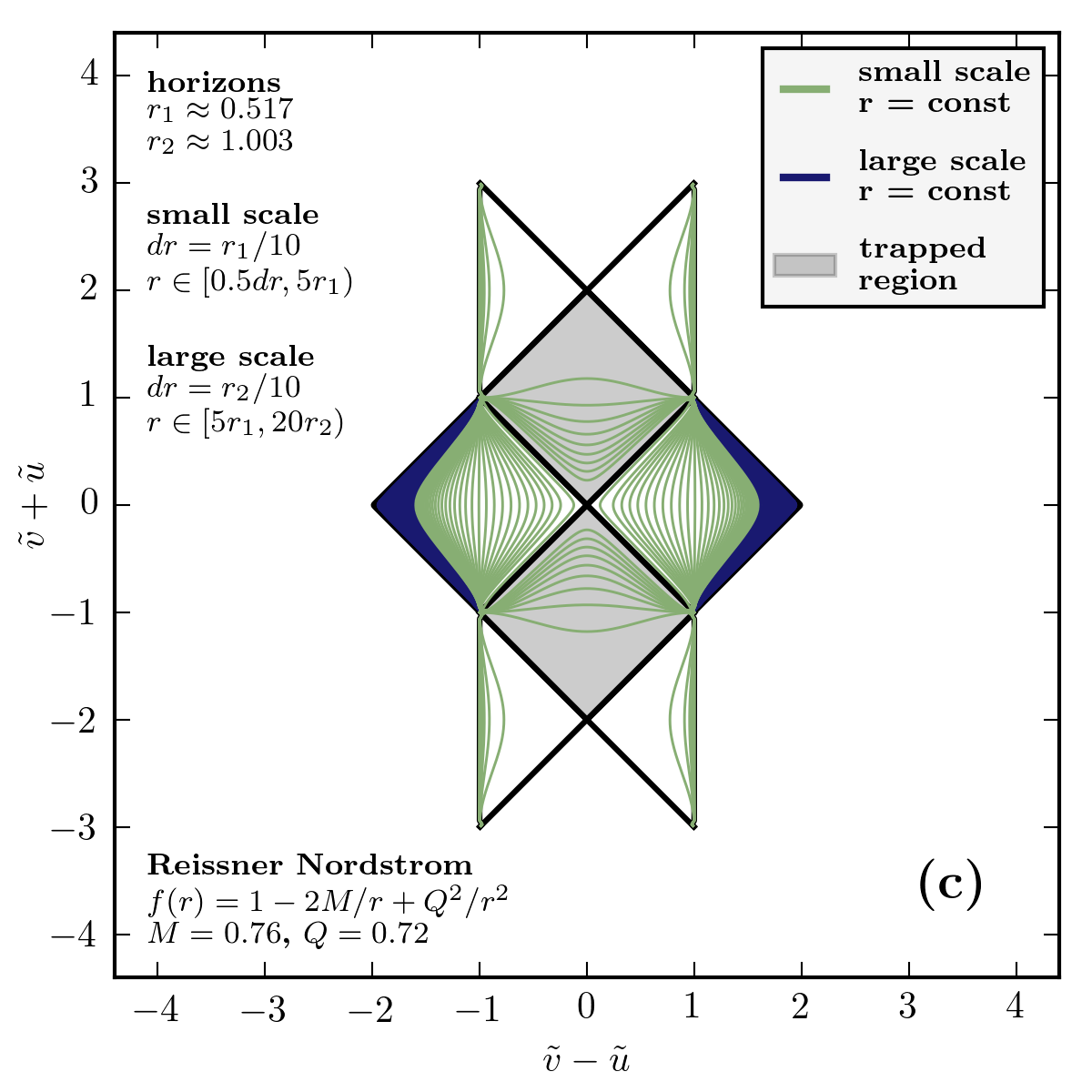} 
&
\includegraphics[width=\plotwidth]{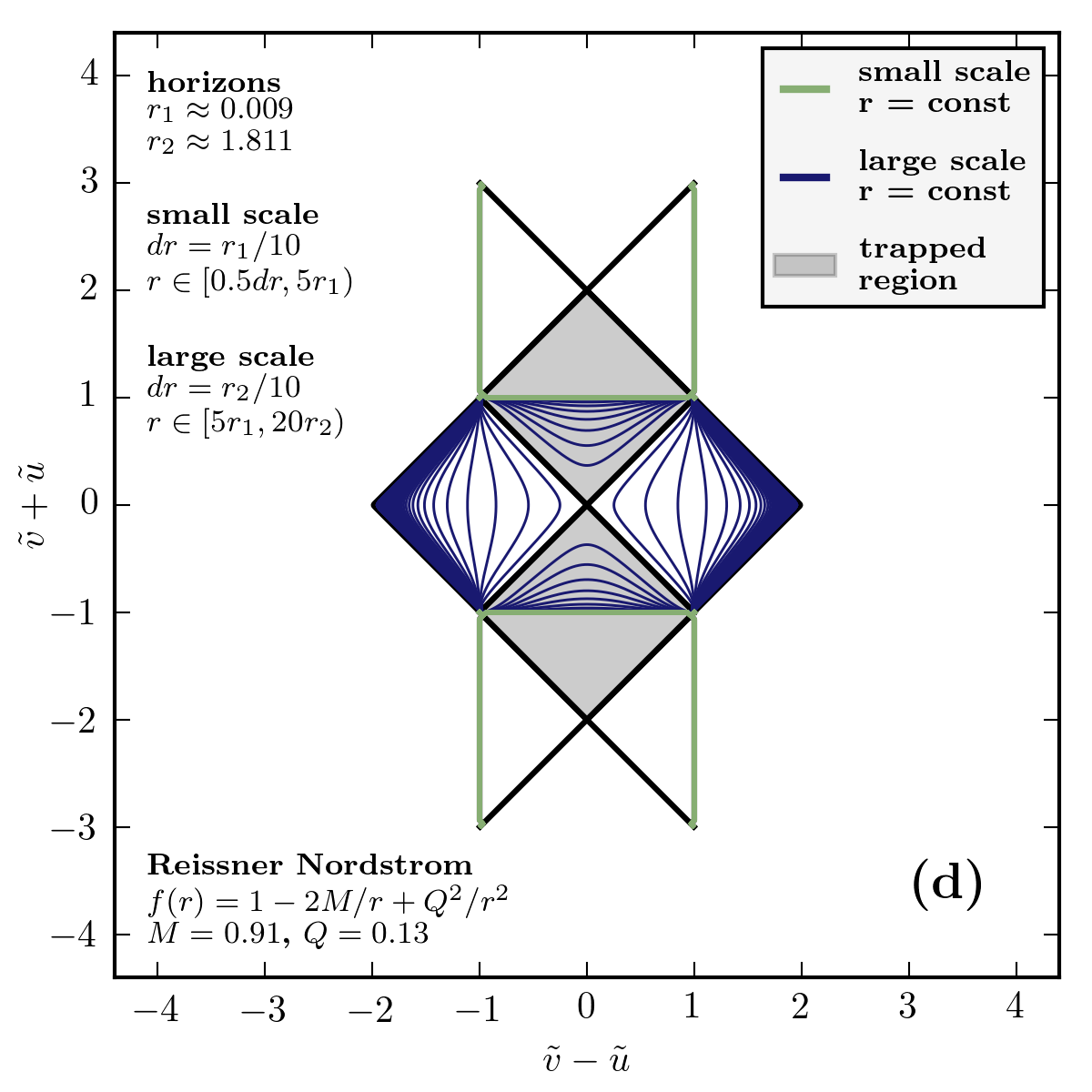} 
\\
\includegraphics[width=\plotwidth]{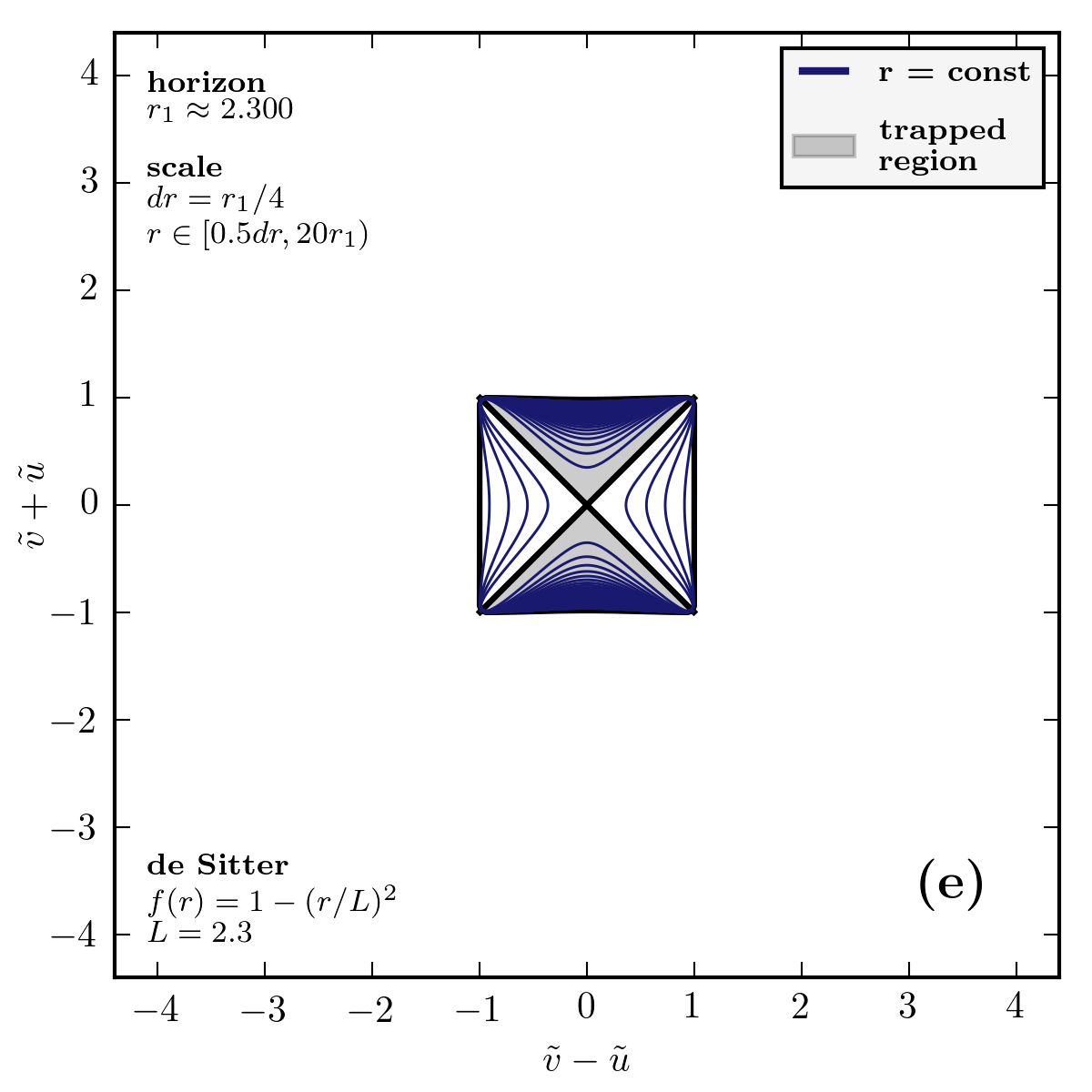} 
&
\includegraphics[width=\plotwidth]{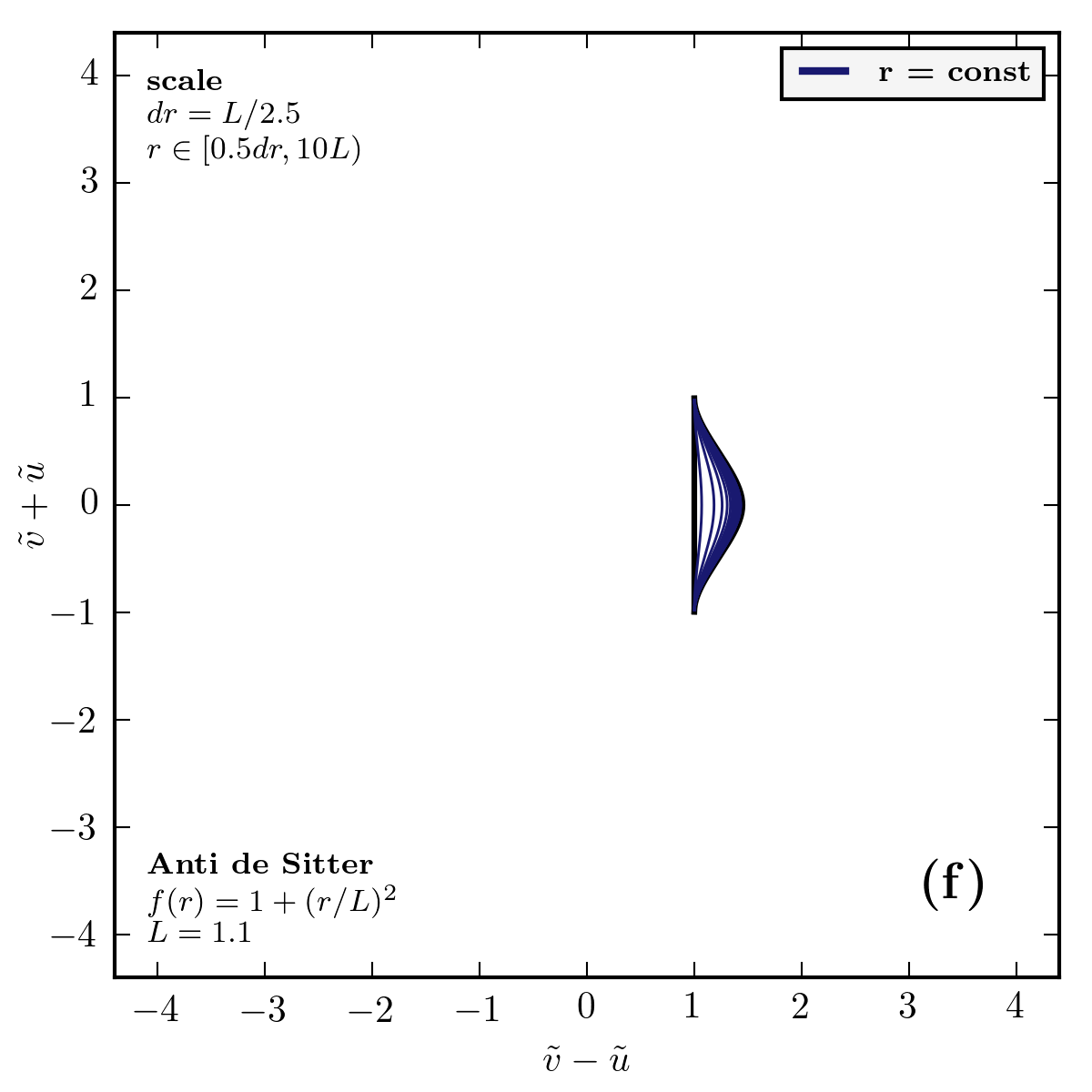} 
\end{tabular}
\caption[Example SSS diagrams]{ 
\label{fig:example1}
(Color online). Example SSS diagrams generated by an implementation of the methods described in this article. For each diagram, lines of constant radius are given at various length scales. For parameters and line spacing scales see diagram annotations. All these diagrams utilize the global diagram constants $c=0$ and $s_0=10$ (see section \ref{sec:alg1}). (a) Minkowski spacetime. (b) Schwarzschild black hole. (c) Highly charged (nearly extremal) Reissner-Nordstrom black hole. (d) Reissner-Nordstrom black hole with small charge. (e) de Sitter spacetime. (f) Anti de Sitter spacetime.
}
\end{figure}

\begin{figure}[ht]
\centering
\begin{tabular}{cc}
\textbf{\footnotesize (a) Lines of $\rstar=const$}
&
\textbf{\footnotesize (b) Undefined radius point}
\\
\includegraphics[scale=.625, trim={0 0 2.5mm 0}, clip]{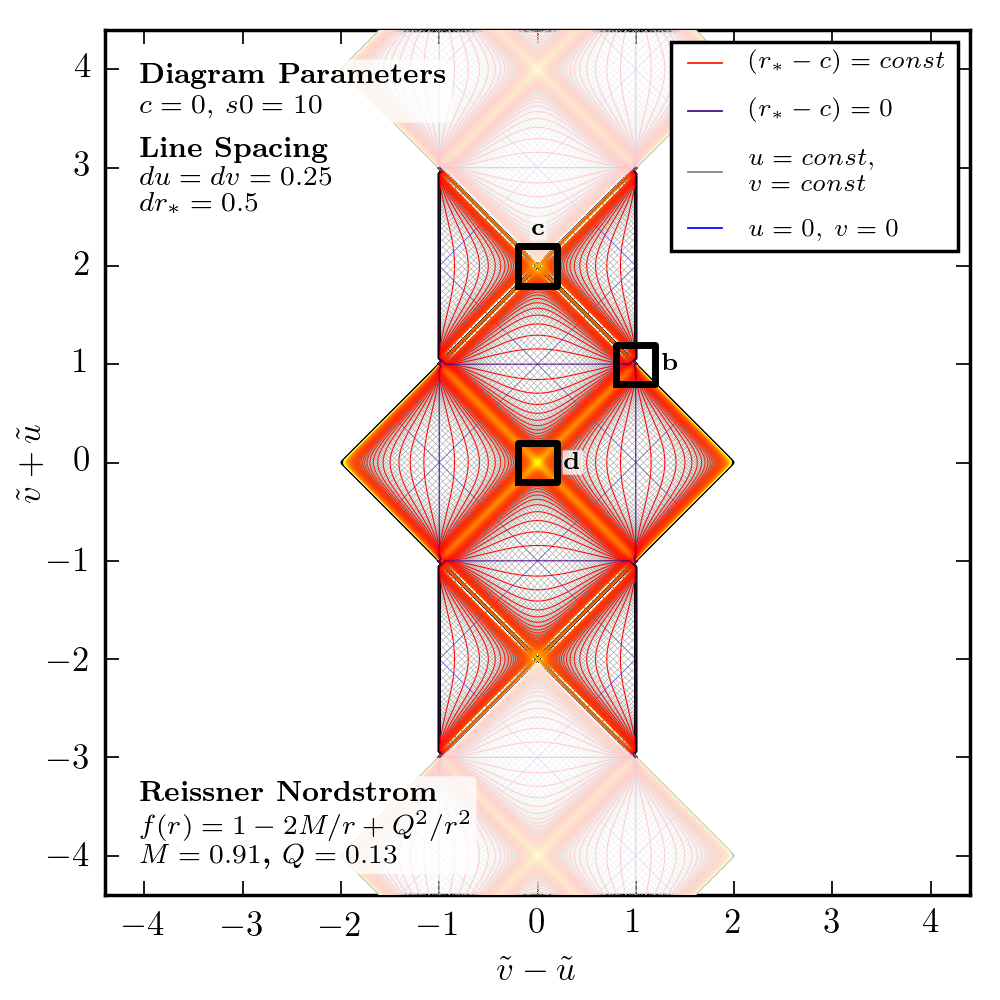}
\includegraphics[scale=.625]{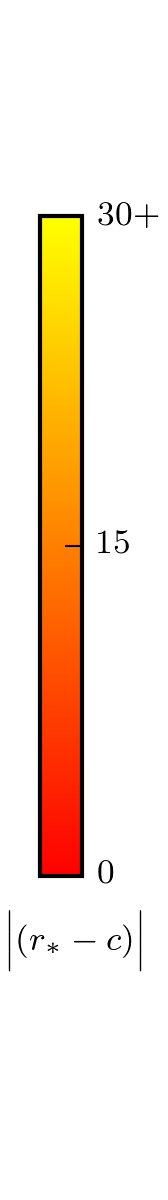}
&
\includegraphics[scale=1]{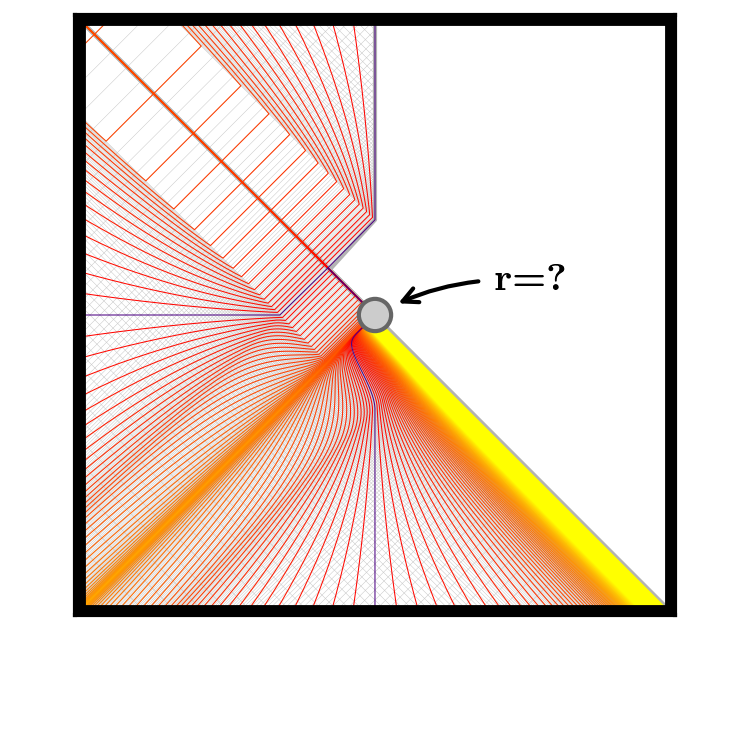}
\\
\textbf{\footnotesize (c) Inner horizon vertex  }
&
\textbf{\footnotesize (d) Outer horizon vertex }
\\
\includegraphics[scale=1, trim={0 0.25in 0 0}, clip]{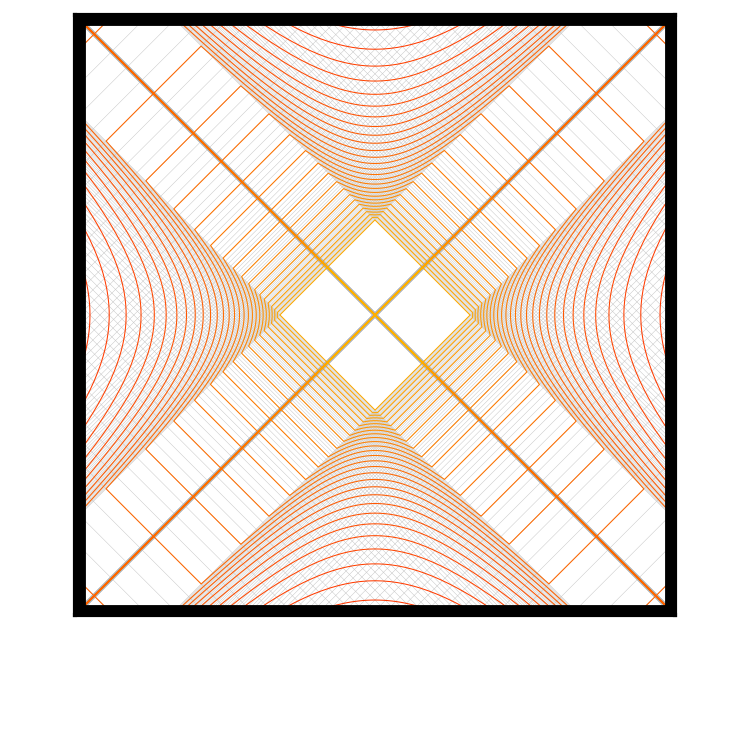} 
&
\includegraphics[scale=1, trim={0 0.25in 0 0}, clip]{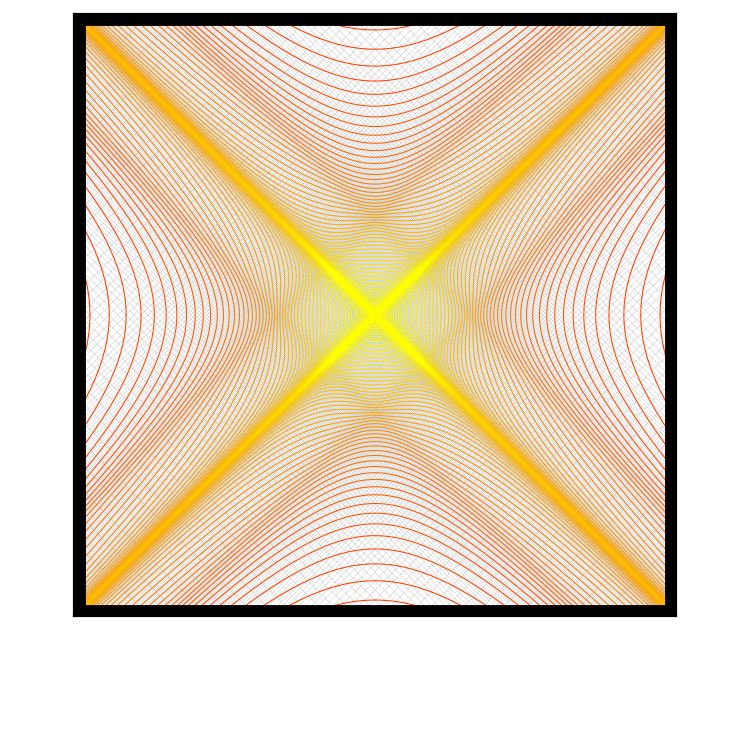}
\end{tabular}
\caption[Diagram detail view]{ 
\label{fig:detail}
(Color online).
Diagram and detail views of the same Reissner-Nordstrom spacetime from figure \ref{fig:example1}d above. Evenly spaced lines of constant tortoise coordinate $\rstar=const$ (orange color scale) provide a different perspective from the previously shown lines of constant radius. Each of the panels (b,c,d) provides a zoom view of the corresponding labeled box in (a). A periodic continuation of the central region is shown in faded color in panel (a), but in full color in the zoom panel (c). Some curves in the detail views may appear to have discontinuities or numerical precision jumps, but this is not the case. In fact, all the visible curves are once-differentiable, numerically accurate, and numerically well-resolved. Apparent kinks and discontinuities are due to turn-on of exponential behavior in the piecewise function $h(s)$ (see section \ref{subsec:dnpen}); these turn-ons occur when $|u/2|>s_0$ and when $|v/2|>s_0$. Near the inner horizon in panel (c), all reasonably spaced lines of $\rstar=const$ in the exponential region are squished against the horizons and into the corner, due to the large magnitude of the slope $k_i = f'(r_i)$ at the inner horizon radius. More gradual deformations near the outer horizon in panel (d) correspond to a less extreme value of the slope. Near the undefined radius point in panel (b), the Penrose coordinate metric is discontinuous across the horizon within the diamond defined by $|u/2|>s_0$ and $|v/2|>s_0$ (see section \ref{subsec:undefined_rad}). Changing the parameter $s_0$ moves the location of the exponential turn-ons. As $s_0 \to \infty$, these features are pushed arbitrarily far against the horizons and into the corners.
}
\end{figure}

\begin{figure}[ht]
\centering
\newcommand{\plotscale}{0.6}
\begin{tabular}{|c|c|}
\hline
\textbf{\footnotesize(a) Lines of constant tortoise coordinate}
&
\textbf{\footnotesize(b) Lines of constant radius }
\\
\includegraphics[scale=\plotscale]{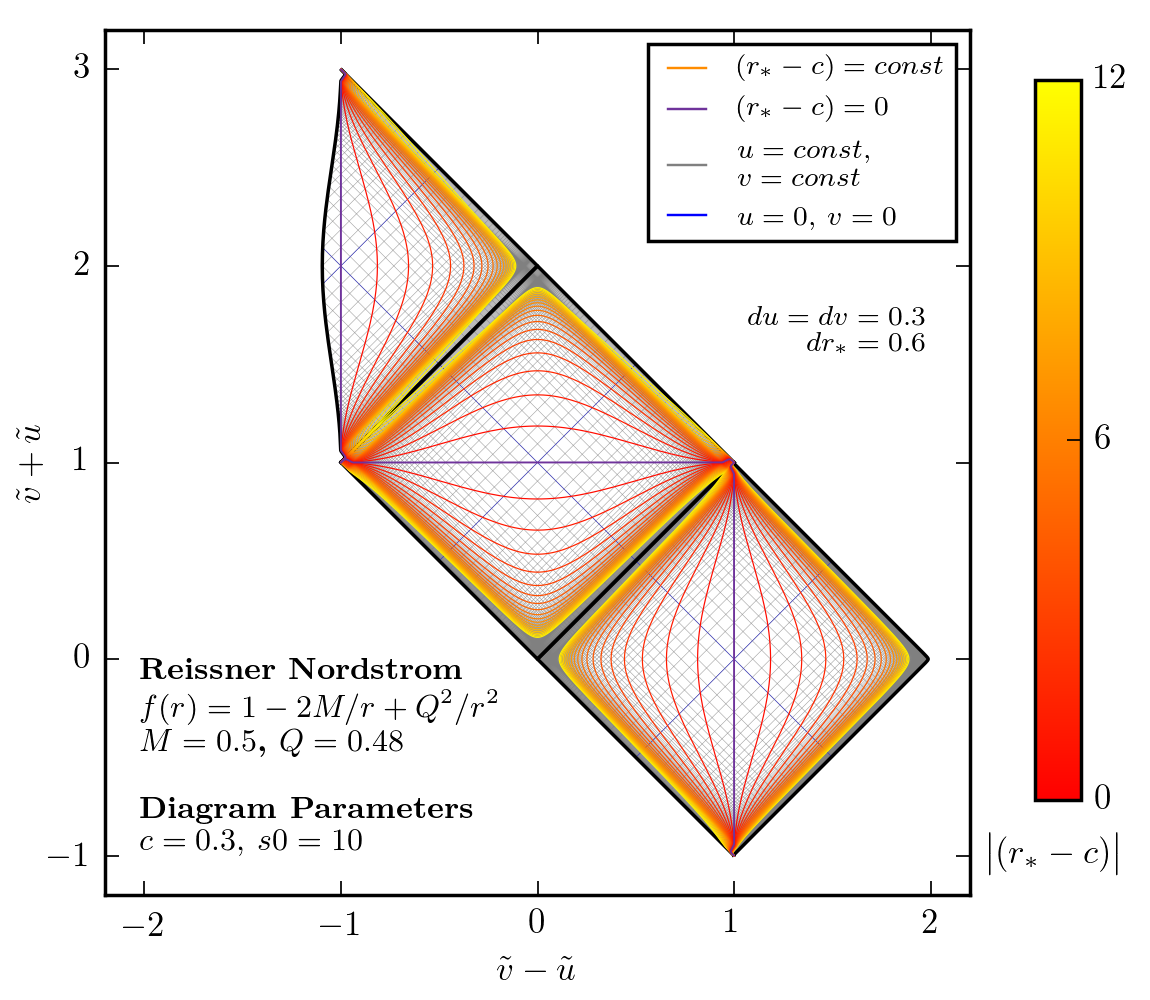}  
&
\includegraphics[scale=\plotscale]{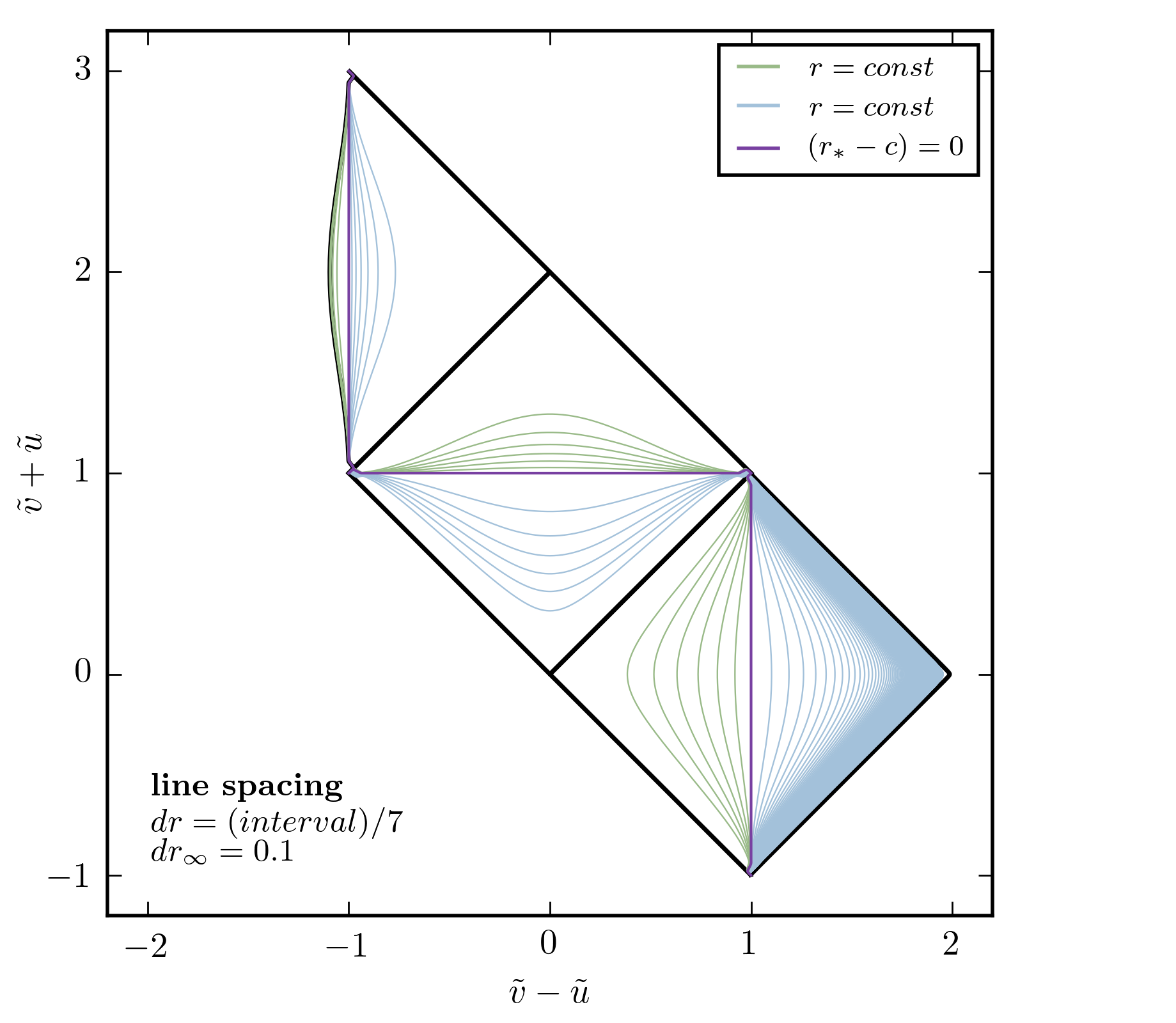}  
\\
\hline
\textbf{\footnotesize (c) Lines of constant tortoise coordinate}
&
\textbf{\footnotesize(d) Lines of constant radius }
\\
\includegraphics[scale=\plotscale]{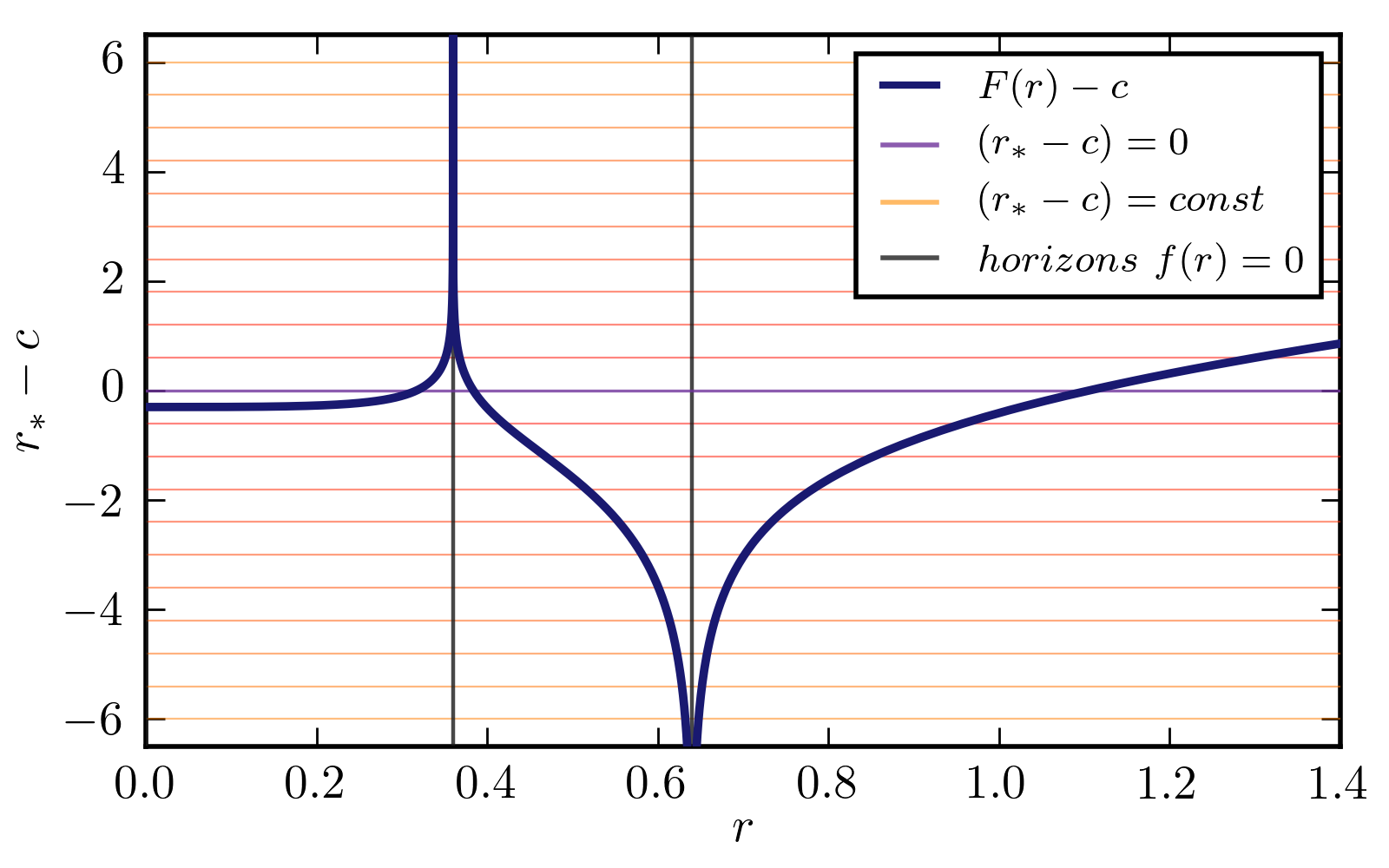}  
&
\includegraphics[scale=\plotscale]{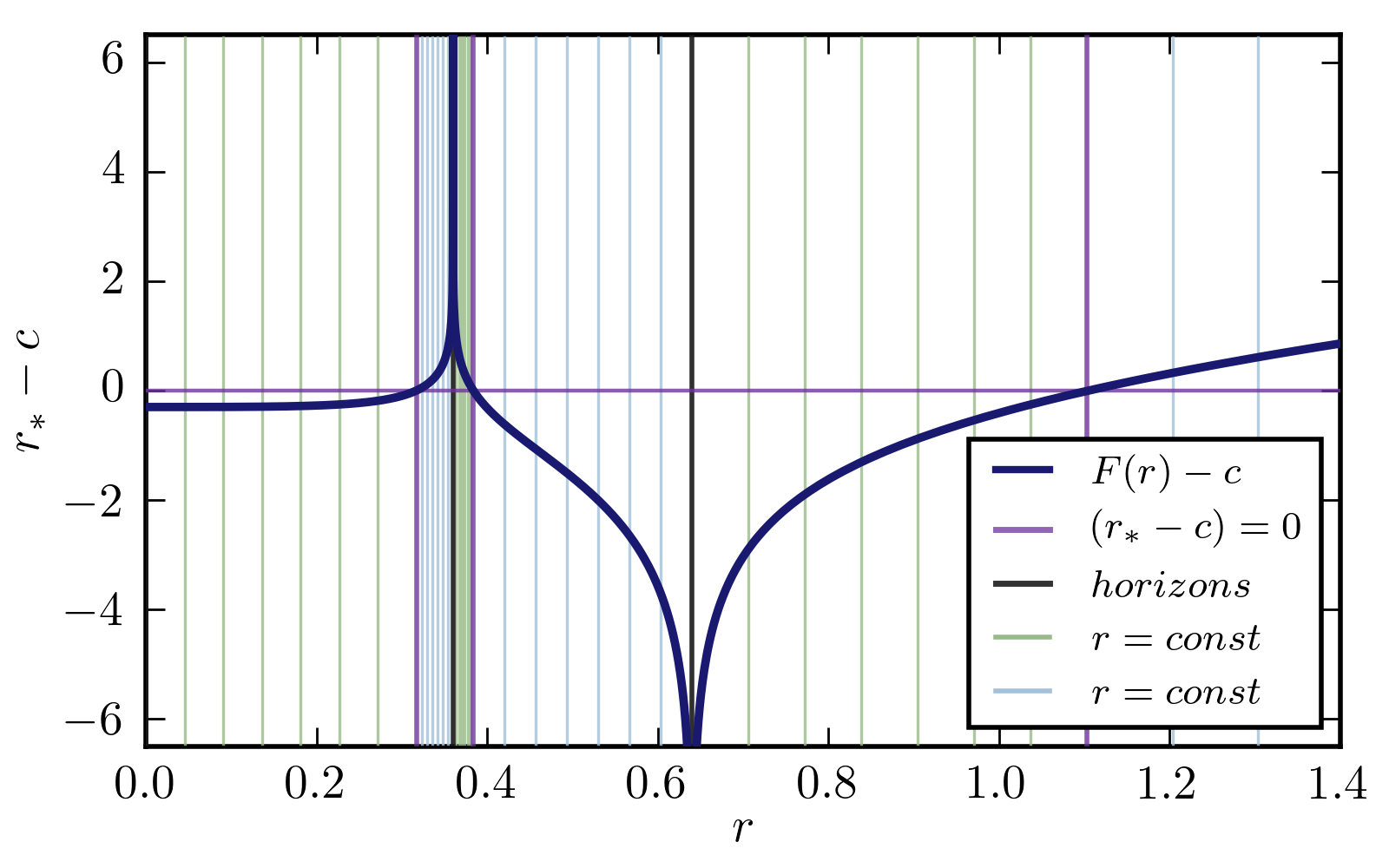}  
\\ \hline
\end{tabular}
\caption[Understanding generic diagram features]{ 
\label{fig:genericfeatures}
(Color online).
Generic features of these diagrams can be understood by inspecting the relationship between lines of constant radius $r$ and lines of constant tortoise coordinate $\rstar=F(r)$. In this example, all four panels correspond to an EF patch of a Reissner Nordstrom spacetime, with parameters as given in panel (a). Panels (a) and (b) respectively show lines of constant $\rstar$ and $r$ in the diagram. Panels (c) and (d) depict the same lines against graphs of $F(r)$. The interval boundaries in (b,d) occur wherever either $f(r)=0$ or $F(r)-c=0$, and there are six equally spaced lines of constant radius in each interval (except in the last interval where $dr=0.1$ as $r\to\infty$). Observing panel (a) highlights some features which are generic to the method: the line $F(r)-c=0$ always runs straight through the middle of each block, and the lines of $\rstar-c=const$ have a regular and predictable spacing in the diagram. These may be used as a regularly spaced reference for sketching block boundaries and lines of constant radius. Comparing to the graph of $F(r)$ in panel (c), one observes that within each interval, most radius values lie in a small range of $\rstar$ values, with most of the range of $\rstar$ lying arbitrarily close to a horizon. This property is generic when horizons are present, and leads to a ``bunching up" of lines of constant radius, since sampled values of radius generally lie in a small interval of $\rstar$. This bunching is visible in panels (b,d), where lines of constant radius are evenly spaced within each individual interval, but remain bunched in the diagram nonetheless. Less carefully sampled radius values will generally be even more bunched than these, and bunching is stronger near horizons with large $|k_i|$. Horizons with large $|k_i|$ thereby ``repel'' lines of $r=const$ and take up a large effective space in the diagram. Changing the parameter $c$ moves all bundles simultaneously while maintaining relative positions. Note also the location of the block boundary at $r=0$ (top left boundary). Since $F(0)=0$ always, when $c=0$ this boundary will be a straight line at $F(r)-c=0$. In the present case $c>0$, and the boundary is pushed out to the left of vertical.
}
\end{figure}

\begin{figure}[ht]
\centering
\newcommand{\plotscale}{0.625}
\begin{tabular}{cc}
\includegraphics[scale=\plotscale, trim={0 3pt 0 3pt}, clip]
{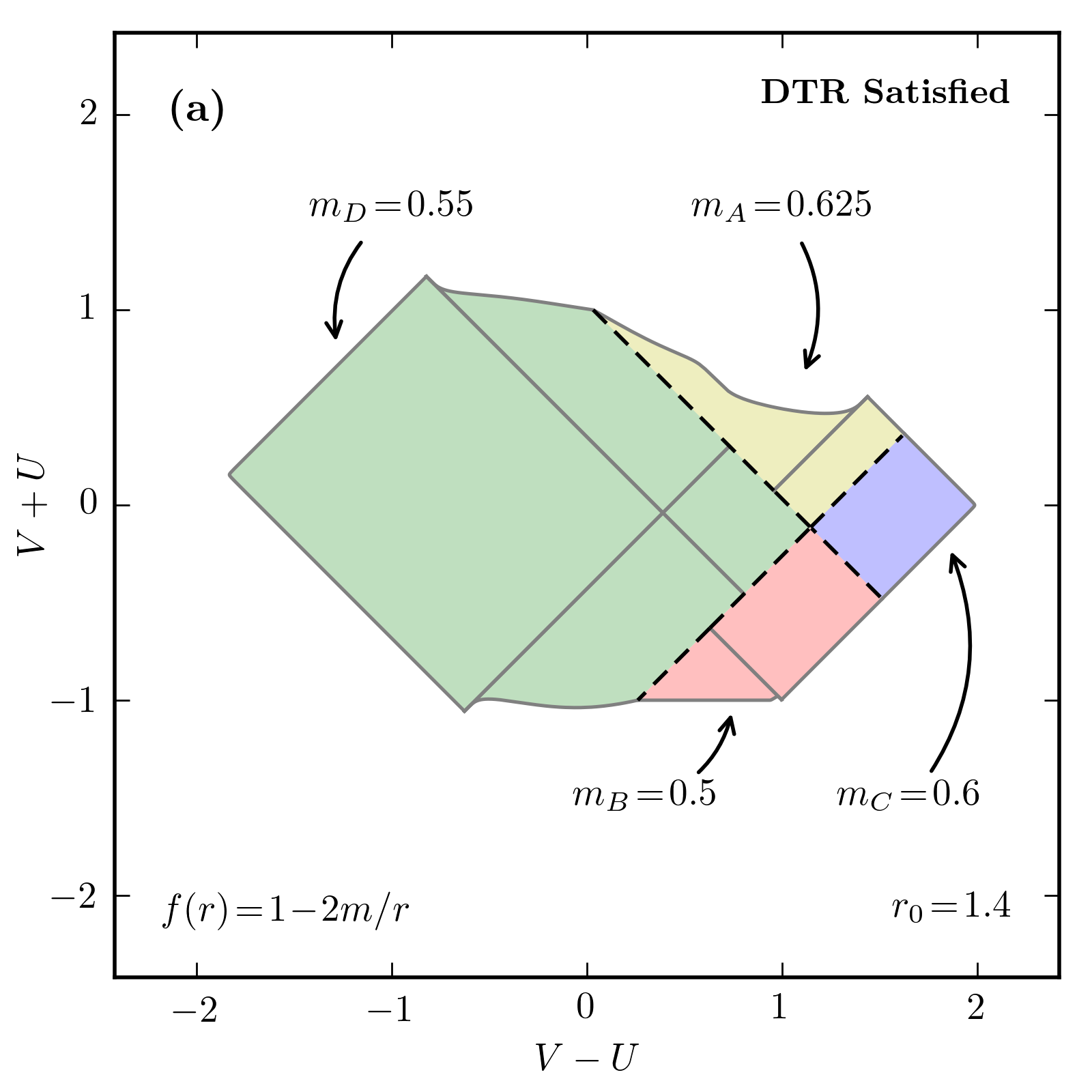}
&
\includegraphics[scale=\plotscale, trim={0 3pt 0 3pt}, clip]
{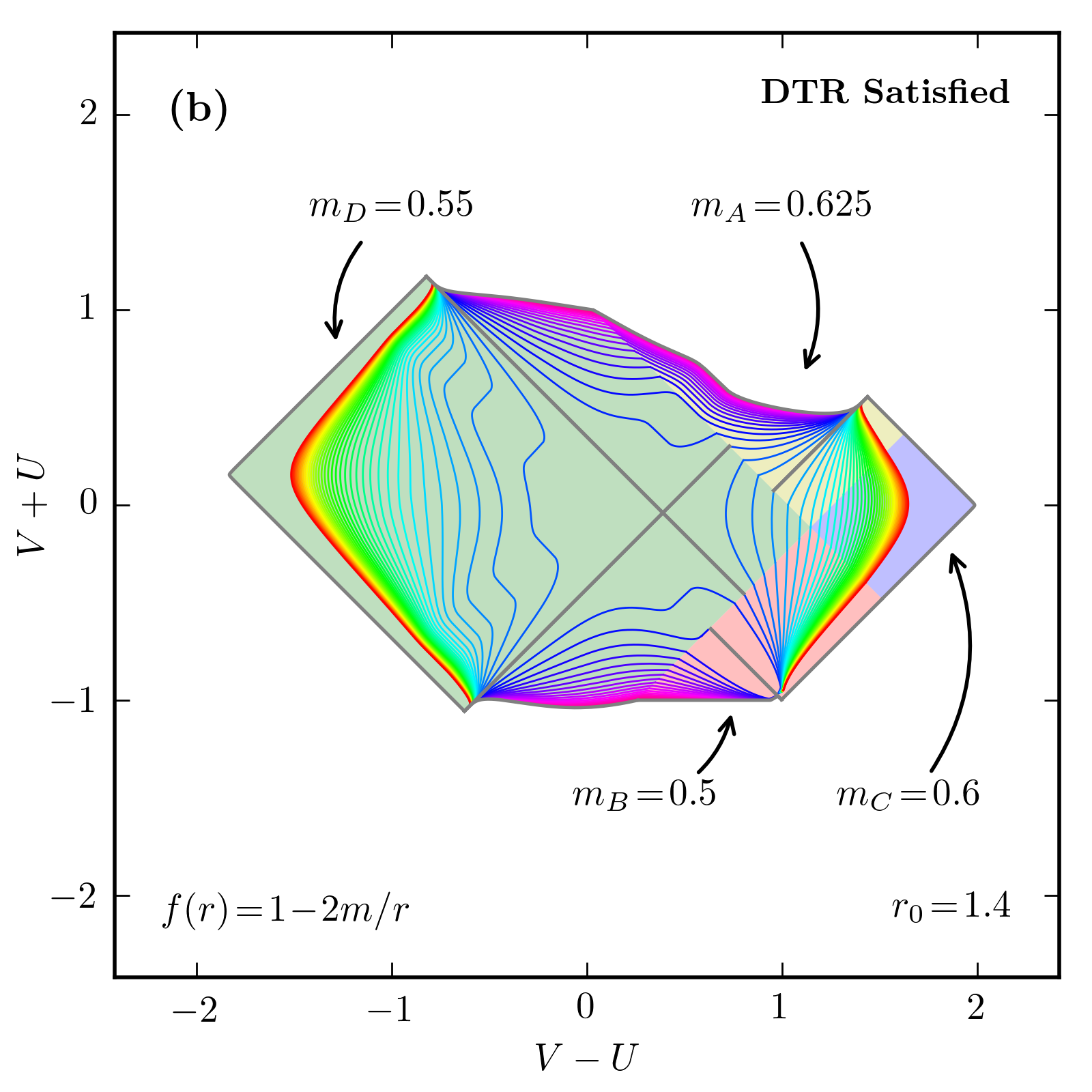}
\includegraphics[scale=\plotscale, trim={2mm 3pt 2mm 3pt}, clip]
{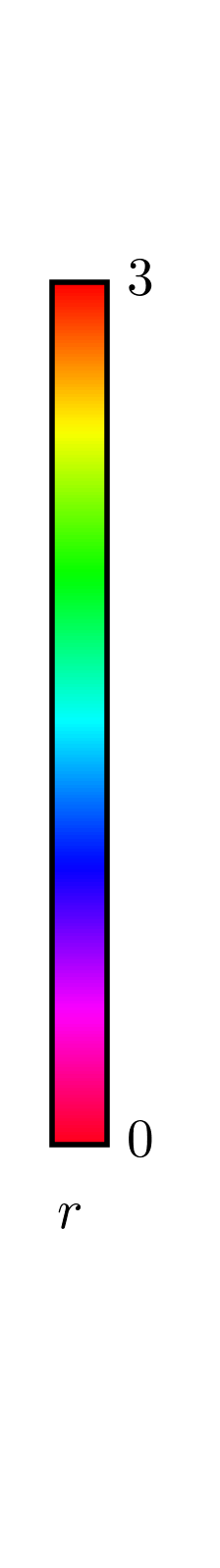}
\\
\includegraphics[scale=\plotscale, trim={0 3pt 0 3pt}, clip]
{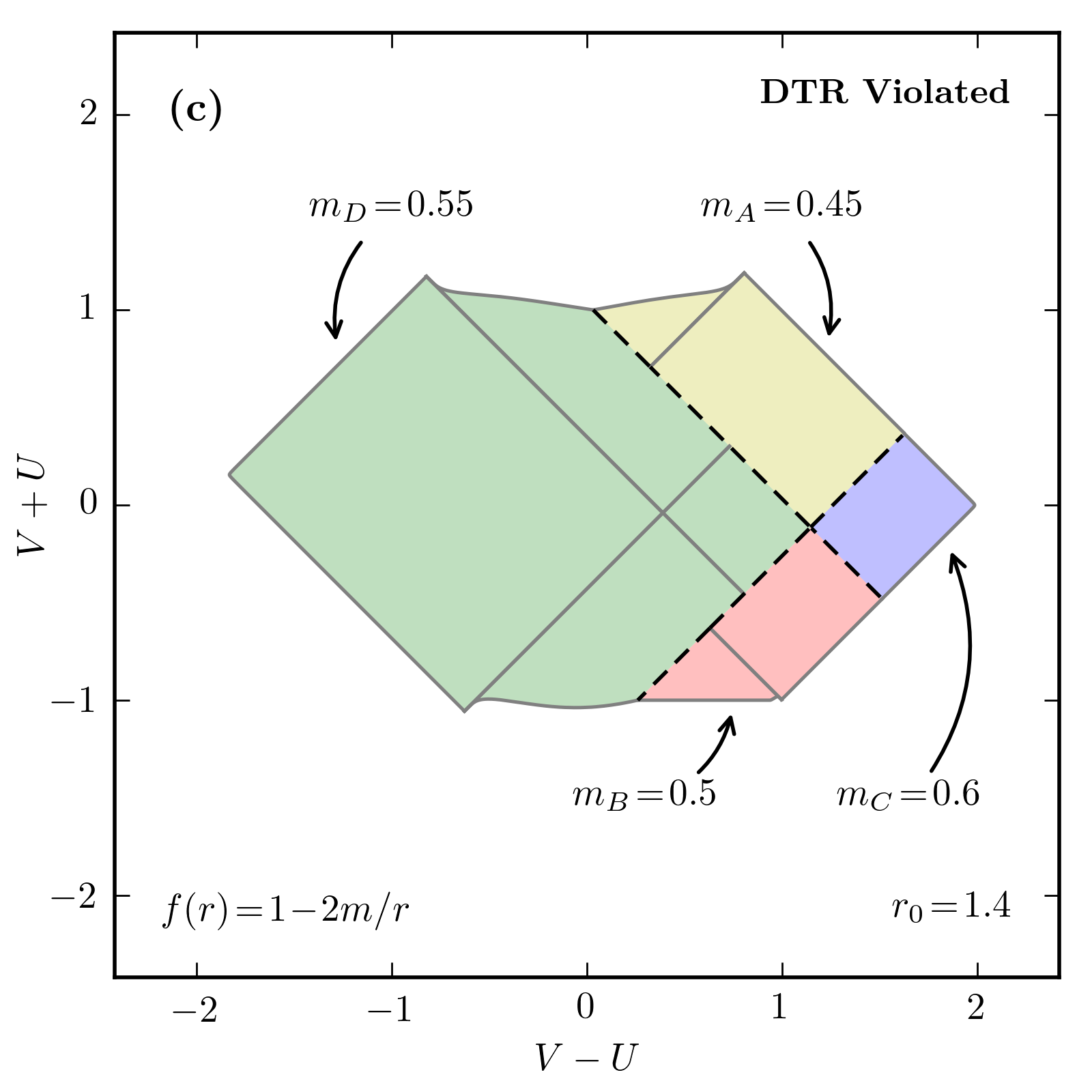}
&
\includegraphics[scale=\plotscale, trim={0 3pt 0 3pt}, clip]
{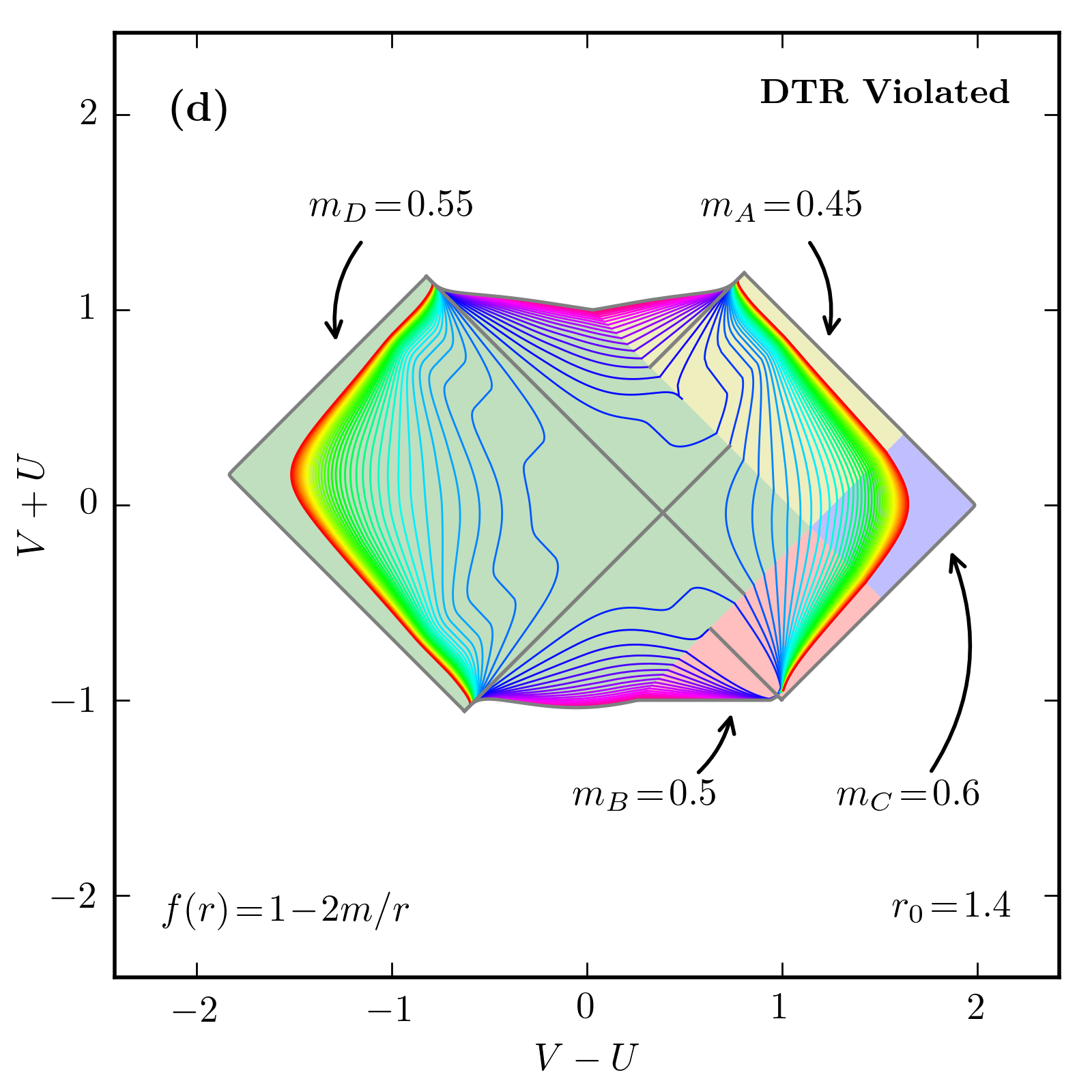}
\includegraphics[scale=\plotscale, trim={2mm 3pt 2mm 3pt}, clip]
{fig18_cbar.png}
\\
\includegraphics[scale=\plotscale, trim={0 3pt 0 3pt}, clip]
{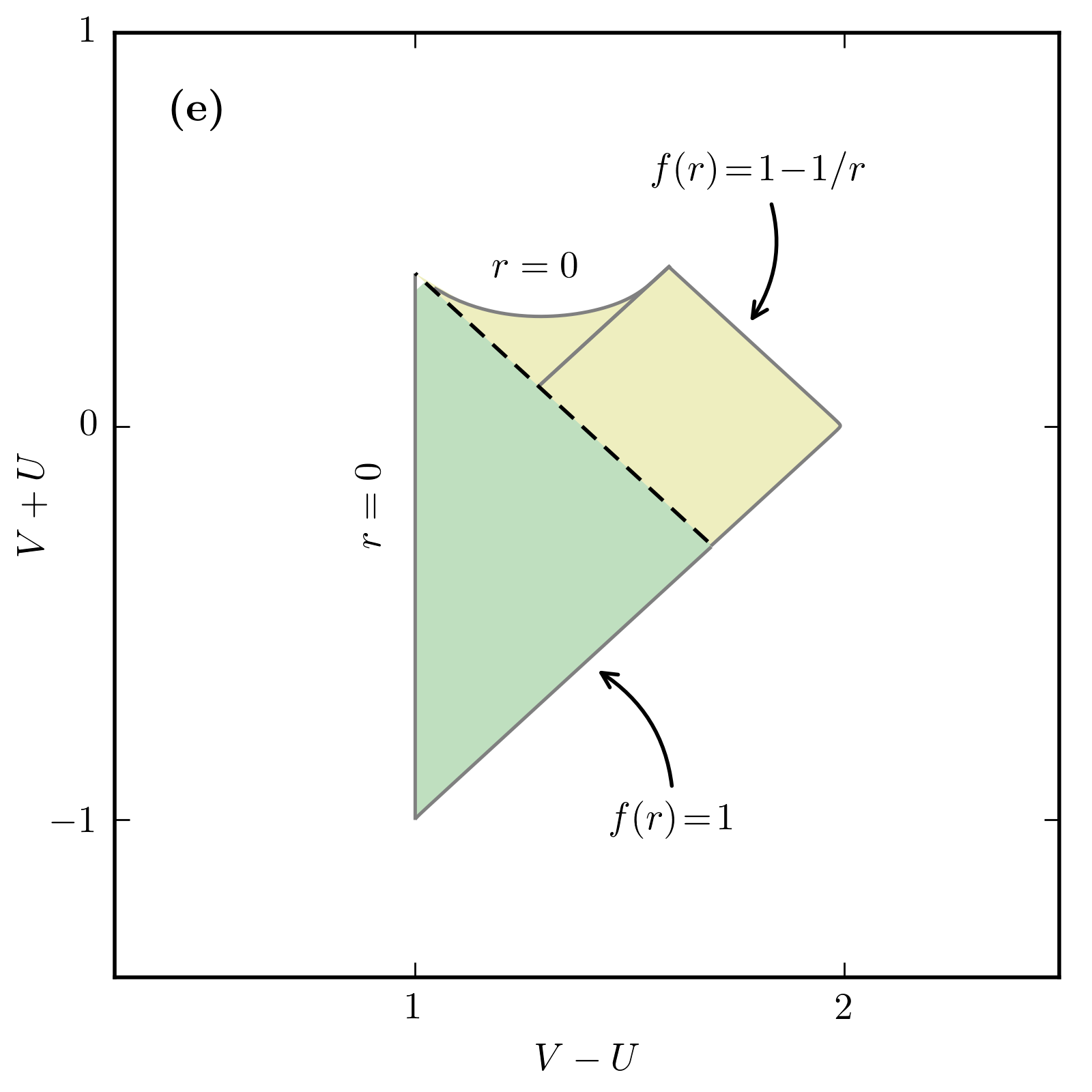}
&
\includegraphics[scale=\plotscale, trim={0 3pt 0 3pt}, clip]
{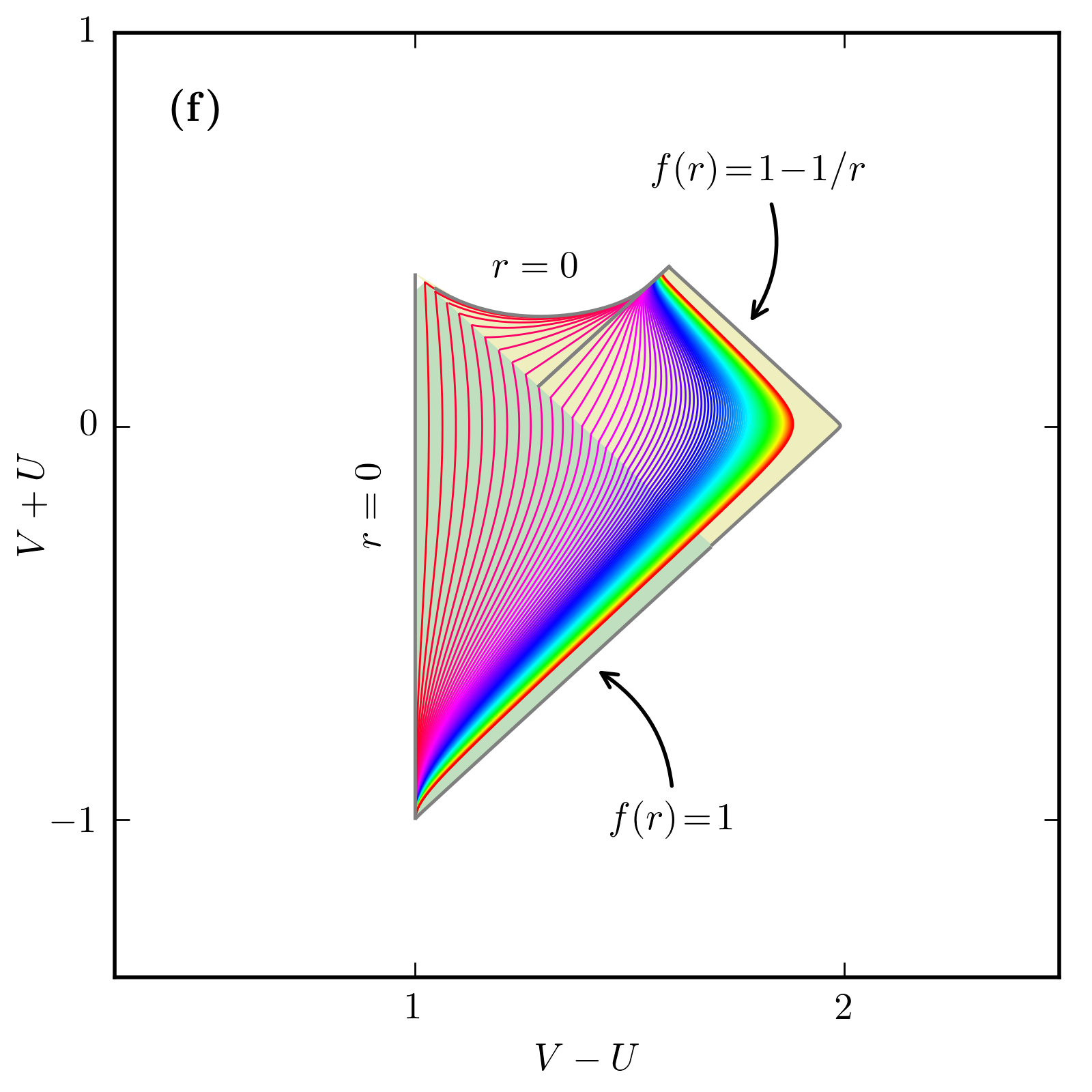}
\includegraphics[scale=\plotscale, trim={2mm 3pt 2mm 3pt}, clip]
{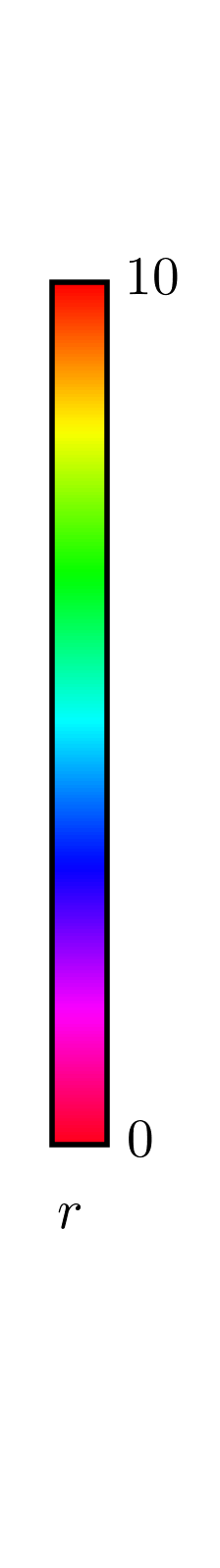}
\end{tabular}
\vspace*{-5pt}
\caption[Example piecewise-SSS diagrams]{ 
\label{fig:example2}
(Color online). Example piecewise-SSS diagrams generated by an implementation of the methods described in this article. Diagram constants set to $c=0$ and $s_0=10$. In each row, left and right panels show the same example with different features. Left panels show conformal boundaries (gray lines), horizons (gray lines), and junction hypersurfaces (black dashed lines). Right panels show lines of constant radius (color scale). Each line of constant radius is continuous at all junctions, and obtains an unusual wiggly appearance from the junction matching transformations. \mbox{Panels (a,b)} show four SSS regions joined at an energy-conserving corner junction. Panels (c,d) are similar to (a,b), but with DTR violated (energy not conserved). Panels (e,f) show a Schwarzschild black hole forming from shell collapse in Minkowski space.
}
\end{figure}

Examples generated by the implementation are given in figures \ref{fig:example1}--\ref{fig:example2} (located after the end of the text but before the appendix). Figure \ref{fig:example1} depicts example diagrams for SSS spacetimes. Figure \ref{fig:detail} shows a detailed zoom view of a particular diagram. Figure \ref{fig:genericfeatures} helps elucidate generic features of the diagrams resulting from these methods. And figure \ref{fig:example2} gives example diagrams for piecewise-SSS spacetimes resulting from null shell junctions.

Features of the examples and implementation are described in captions of the example figures, since they are best understood in context of the results. Just a few further comments are in order here.

The first comment regards the SSS diagrams. Comparison between the extended Schwarzschild diagram in figure \ref{fig:example1}b and the Reissner-Nordstrom (R-N) diagram in figure \ref{fig:example1}d is immediately striking: for all lines of constant radius on the length scale of the outer horizon radius, the two diagrams are nearly identical. They differ only at length scales on the order of the R-N inner horizon radius. The R-N diagram appears, in fact, as a Schwarzschild diagram with the $r=0$ singularity ``rigidly pulled up" to become timelike. This is consistent with the generic effects of ``bunching" and ``repelling" described in figure \ref{fig:genericfeatures}. Due to the large value of $|k_i|$ at the inner horizon, the upper square (between green bunches) of the R-N diagram is almost all located at the inner horizon radius plus or minus ``epsilon", while all reasonably spaced lines of constant radius are bunched at the edges. When the R-N spacetime becomes highly charged (nearly extremal), as in figure \ref{fig:example1}c, these effects are mitigated.

The second regards energy conservation in the piecewise-SSS diagrams. Notably, there is no obviously visible difference between the energy conserving junction in figure \ref{fig:example2}a--b and the energy non-conserving junction in figure \ref{fig:example2}c--d. The DTR relation must be independently verified. Moreover, energy conservation by the DTR relation is not always intuitive; even in the energy conserving Schwarzschild example of figure \ref{fig:example2}a--b, the total incoming and outgoing shell masses do not add up in the naive way. Heuristically, this is because the DTR formula must take gravitational potential energy into account \cite{barrabes2003}.

For more general remarks, see the captions of figures \ref{fig:example1}--\ref{fig:example2}.

\section{Concluding remarks}
\label{sec:disc}

We have given a complete analysis of the theory of Penrose diagrams as applied to strongly spherically symmetric spacetimes and their piecewise-SSS cousins. Having set down the rules, these methods may be used to analyze causal structure in a broad class of spacetimes. In a forthcoming publication, these methods will be applied to the case of a black hole which forms from stellar collapse and subsequently evaporates by emitting Hawking radiation.

\section*{Acknowledgments}
The authors would like to thank A. Kuttner for many useful and relevant discussions. This research was supported by the Foundational Questions Institute (FQXi.org), of which AA is Associate Director, and by the Faggin Presidential Chair Fund.

%%%%%%%%%%%%%%%%%%%%%%%%%%%%

\clearpage

%%%%%%%%%%%%%%%%%%%%%%%%%%%%

%%%%%%%%%%%%%%%%%%%%%%%%%%%%%%%%%%%%%%%

\clearpage

%%%%%%%%%%%%%%%%%%%%%%%%%%%%%%%%%%%%%%%%%

%%%%%%%%%%%%appendix%%%%%%%%%%%%%%%%%%
\appendix

\section[{\hspace{18mm} Comparison to existing methods}]
{Comparison to existing methods}
\label{sec:comp}

The method described in this article generalizes and unifies techniques from various sources. In order to provide the proper context, we conduct here a brief review of the methods historically used to create similar diagrams.

The problem of generating causal diagrams for SSS spacetimes has previously been tackled in two fundamental ways: the method of block diagrams and the method of global Penrose coordinates. The block diagram method provides explicit global Penrose coordinates only when the metric function has one or zero horizon radii (locations where $f(r)=0$). When many horizons are present, the block diagram method allows qualitative analysis by the identification of overlapping blocks in neighboring quad-block regions (see section \ref{sec:ssss}), but does not define a self-consistent system of global coordinates in the area of overlap.

The more well known of the two methods is that of block diagrams, in which the causal structure of spacetime is pieced together from quad-block units. This method was first applied to the special case of Schwarzschild spacetime by \mbox{Kruskal \cite{kruskal60}}. Maximally extended Schwarzschild spacetime, with its one horizon radius, consists of a single quad-block, and so in this case global Penrose coordinates were also achieved. Kruskal's method was then generalized by Walker to the method of block \mbox{diagrams \cite{walker70}}. In Walker's treatment, quad-blocks could be constructed so long as the metric function took a special form, with the tortoise function defined by an indefinite integral (although definite integrals were implemented to generate the tortoise function within individual blocks). Later, however, Brill and Hayward \cite{hayward94} used a definite integral formulation to extend Walker's method to the case of an arbitrary metric function, in a manner equivalent to the results of section \ref{subsec:inout}. In this way, the question of constructing arbitrary block diagrams (but not Penrose diagrams) for SSS spacetimes has long since been settled.

\begin{figure}[t]
\centering
\begin{tabular}{cc}
\includegraphics[width=2.5in]{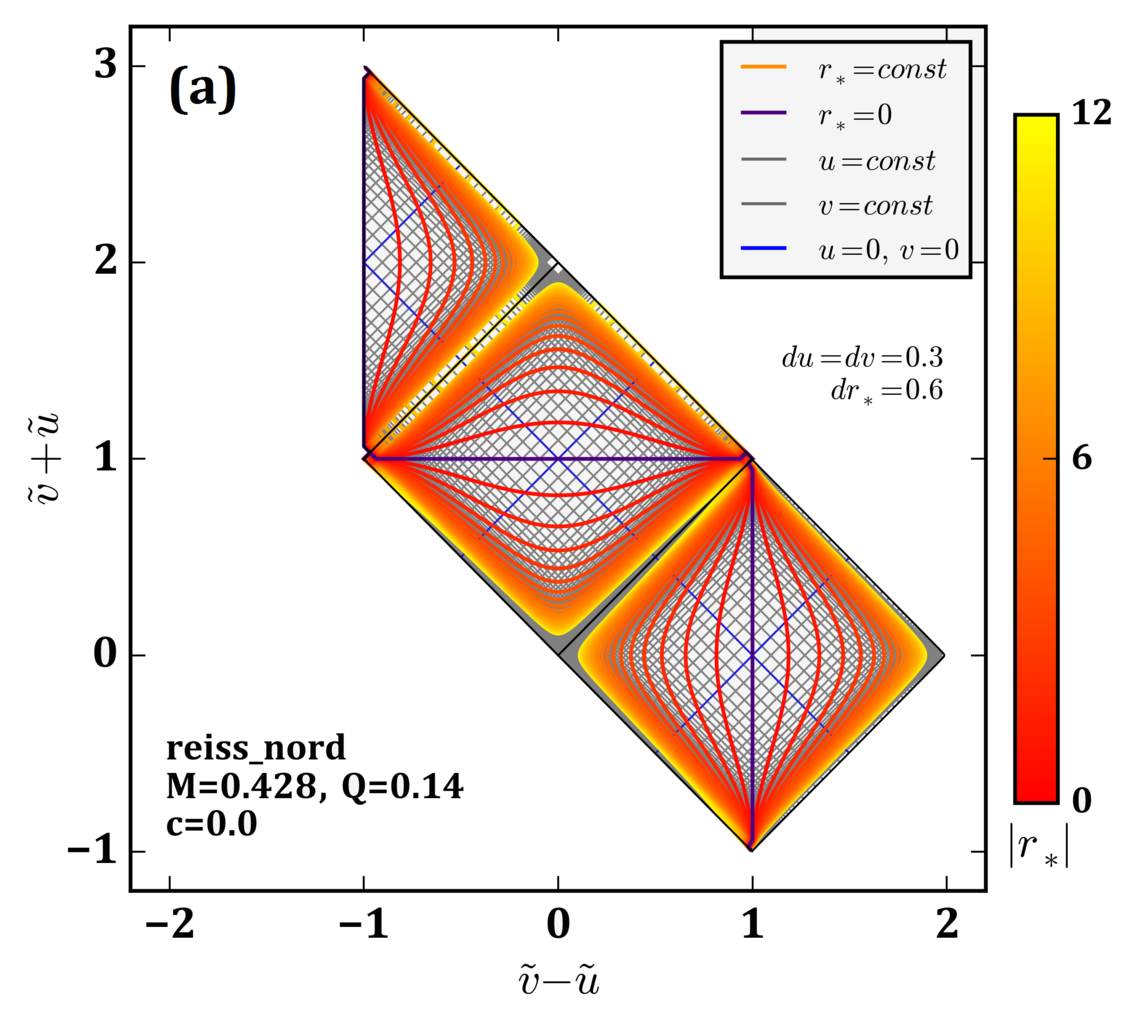} 
&
\includegraphics[width=2.5in]{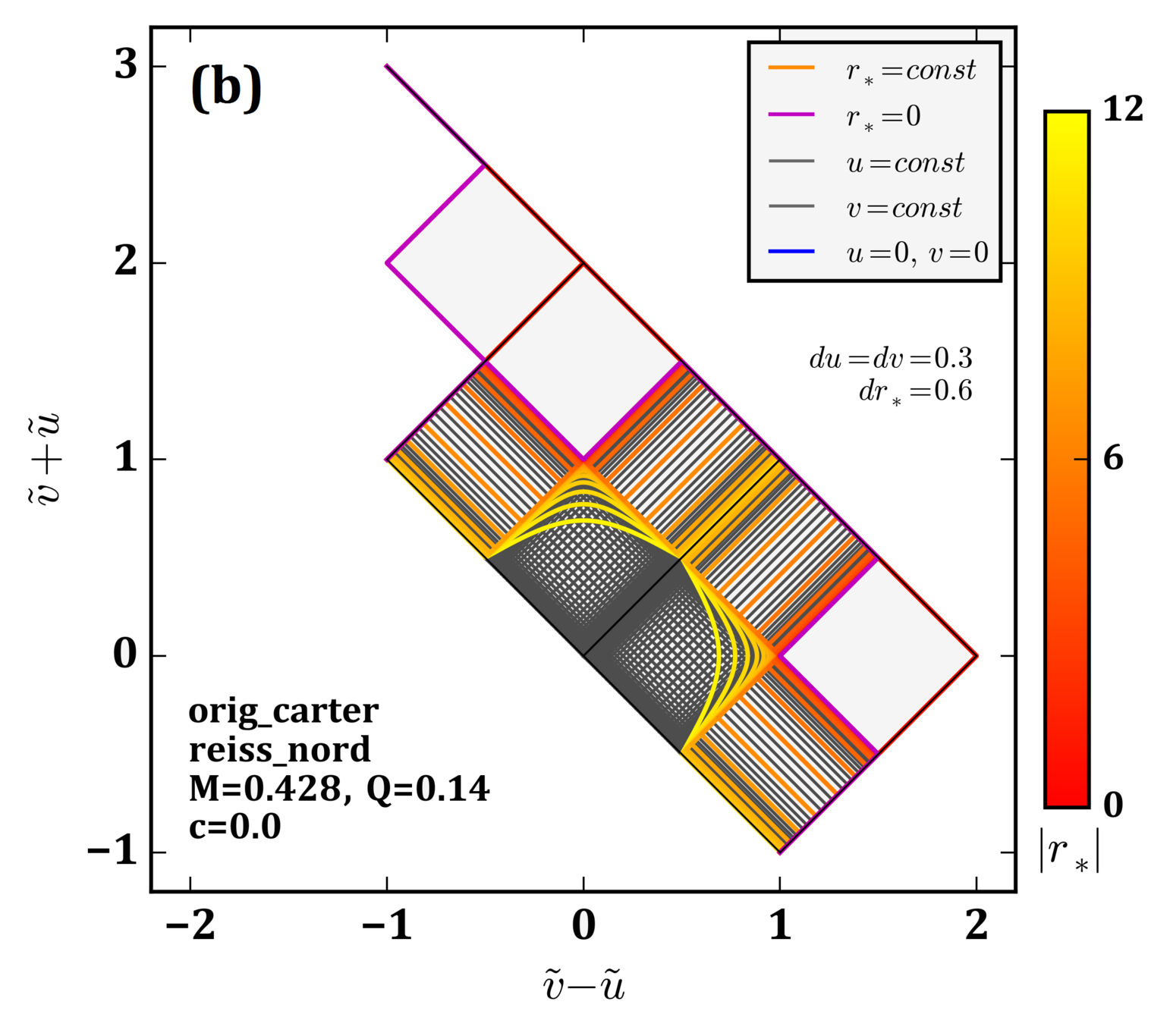} 
\\
\includegraphics[width=2.5in]{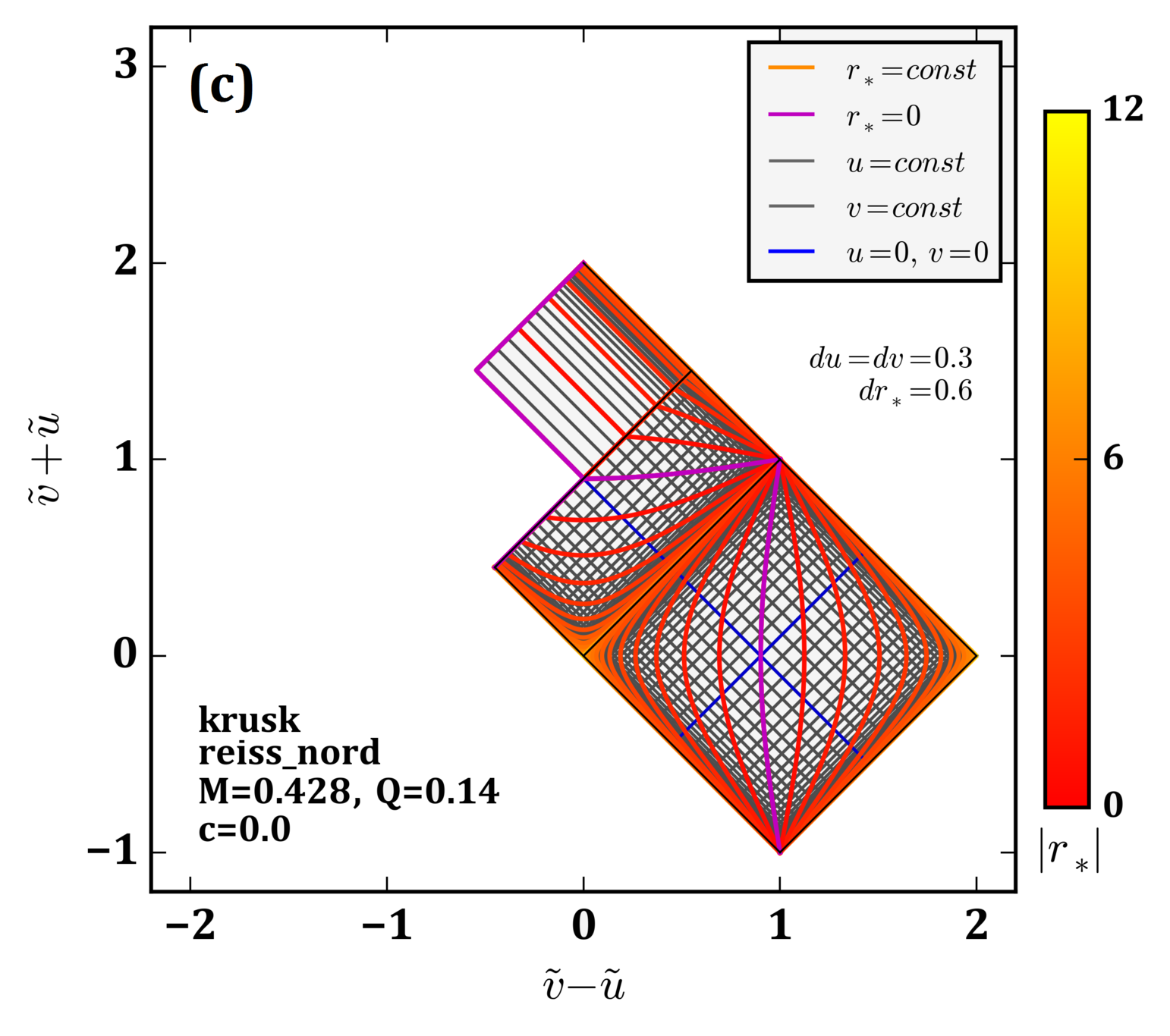}  
&
\includegraphics[width=2.5in]{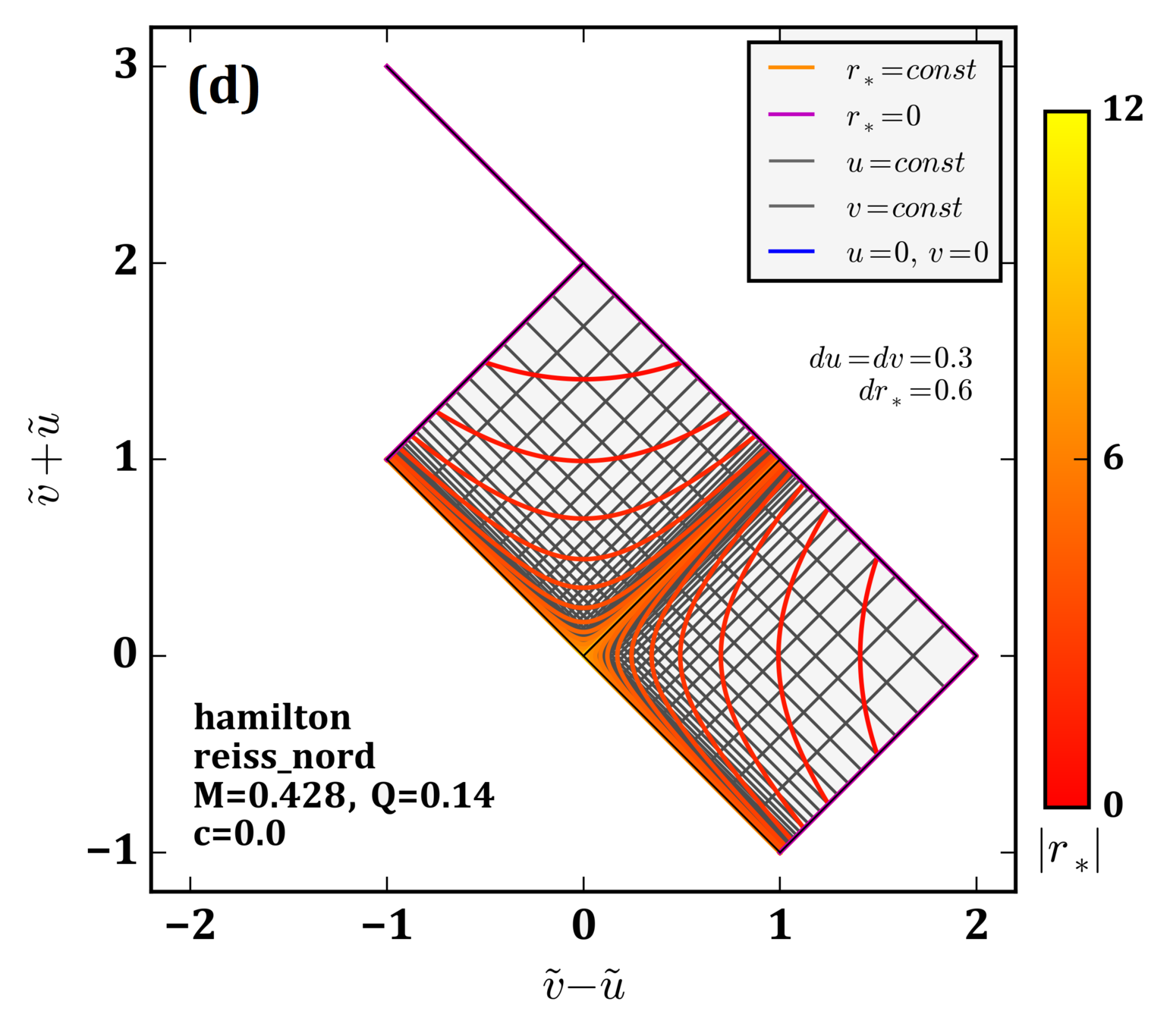} 
\end{tabular}
\caption[Comparison to existing two-horizon methods]{
\label{fig:comparison}
(Color online).
An EF region for a Reissner-Nordstrom black hole, generated according to four different methods. In order to emphasize the effects of each transformation, lines of constant tortoise coordinate $\rstar$, rather than constant radius, are shown. \mbox{(a) The} algorithm of section \ref{sec:alg1}, with $s_0=10$. (b) The method described by Carter in 1966 \cite{carter66}. (c) A method which is adapted from the Kruskal method to allow two horizons. \mbox{(d) A} method due to Hamilton \cite{hamiltonbook}. In all cases the preliminary transformation to double null coordinates $(u,v)$ utilizes the constant $c=0$.
}
\end{figure}

The less well known set of methods is that in which global Penrose coordinates are explicitly constructed. The most important progress in this direction was made by Carter \cite{carter66}, who found global Penrose coordinates for special cases with two horizons. Although Carter's method relied on the ability to determine the tortoise function by an indefinite integral, and assumed the existence of exactly two horizons, it is in fact easily generalizable, and was the most important contribution to the present methods. In fact, when the indefinite integral form of the tortoise function is known, Carter's two-horizon method is essentially equivalent to the method of section \ref{sec:alg1}, but with the substitution $\hks(s) = e^{2 k_+ s} - e^{-2 k_- s} $ in every block. The original Carter method does have several drawbacks. One is that the exponential form of $h(s)$ yields very strange-looking lines of constant radius (see figure \ref{fig:comparison}). Additionally, applied to a case with more than two horizons, this method would suffer from major problems near the undefined radius points (see section \ref{subsec:undefined_rad}), and need to be modified. The results of section \ref{sec:alg1} can be thought of as a generalization of the methods of Carter; the key new additions are the introduction of the global tortoise function and the new form of the function $\hks(s)$. These additions allow the method to be extended to an arbitrary metric function with an arbitrary number of horizons, while simultaneously rendering the diagram more readable.

In order to understand the advantages of the present method, it is useful to compare the known methods for constructing Penrose diagrams in the case of two horizons. Such a comparison is depicted in figure \ref{fig:comparison}. Once the tortoise function $F(r)$ is set and the double-null coordinates $(u,v)$ are defined in each block, there are several known methods of obtaining the diagram coordinates. First, of course, there is the method of section \ref{sec:alg1}. Second, there is the original method of Carter \cite{carter66}. Both have been described above. A third method has been used by Hamilton \cite{hamiltonbook}; this method could potentially also be generalized, but is less theoretically appealing than Carter's method. Finally, it is possible to define a two-horizon extension of the Kruskal method such that the diagram matches the standard Kruskal diagram outside of some inner limit; this two-horizon Kruskal-like method cannot be extended to more horizons, however, and so although it is shown for comparison, we omit its exact formulation. In figure \ref{fig:comparison}, diagrams featuring lines of constant tortoise coordinate $\rstar$ are shown for each of these four methods. Showing the tortoise coordinate instead of radius isolates the effect of the different coordinates systems, eliminating distortion due \mbox{to $F(r)$}.

\section[{\hspace{18mm} Unit conventions}]
{Unit conventions}
\label{sec:unitconventions}
A major goal of the techniques described in this article is to realize Penrose diagrams of physically relevant spacetimes. As such, we must clarify the status of units in these calculations. First of all, note that we utilize standard conventions for geometrized units, such that $G=c=1$, in accordance with \cite[(Appendix F)]{wald84}. The tricky issue of units in the coordinates, metric components, and Penrose diagrams is discussed below.

In GR the units of the line element are necessarily 
$[ds^2]=[\textrm{length}]^2$. Units for other geometric quantities depend on a choice of convention. Among the possible conventions, there is a unique simplest choice: all metric components $g_{\mu\nu}$ should be unitless, while all coordinates carry units of length. This convention has several benefits. All components of a tensor have the same units. The components of an even-rank tensor have the same units as its scalar contractions. And for any given tensor, the covariant, contravariant, and mixed varieties all have the same units. In particular the Ricci curvature components and Ricci scalar obey $[{R^{\mu}}_{\nu}]=[{R^{}}_{\mu\nu}]=[{R^{\mu\nu}}]=[R]=[\textrm{length}]^{-2}$, and after conversion to SI units by $\bar{T}_{\mu\nu}= (c^4/G) \, T_{\mu\nu} = (c^4 / 8\pi G) \, G_{\mu\nu}$ all components of the stress-energy tensor have units $[{\bar{T}_{\mu\nu}}] = [\textrm{energy}] \, [\textrm{length}]^{-(D-1)}$ (where $D$ is the number of spacetime dimensions). There are other standard unit conventions in which these need not hold. For example in the standard Euclidean spherical coordinates $(\bar{r},\theta,\phi)$ where $[\bar{r}]=[\textrm{length}]$ and $[\theta]=[\phi]=[1]$, the components $T_{\mu\nu}$ do not even all have the same units. 

We achieve this simplest convention while maintaining notational simplicity by factoring out and suppressing a universal length scale. Let the physical line element $d\bar{s}^2 = \bar{g}_{\mu\nu} \, d\bar{x}^\mu d\bar{x}^\nu$ have any arbitrary unit convention such that $[d\bar{s} ^2] = [\textrm{length}]^{2}$. Let $l$ be an arbitrary length scale. It is always possible to nondimensionalize the coordinates and metric components to obtain  $d\bar{s}^2 = l^2 \, ds^2$ with $ds^2 = g_{\mu\nu} \, dx^\mu \, dx^\nu$, where $[ds^2]=[g_{\mu\nu}]=[dx^\mu]=[x^\mu]=[1]$. Then the physical line element can always be expressed as $d\bar{s}^2 = g_{\mu\nu} \, d(lx^\mu) \, d(lx^\nu)$ in terms of the unitful coordinates $(lx^\mu)$ and the unitless metric. This form of the physical line element satisfies the desired convention.

We will always write the unitless line element $ds^2$  in the unitless coordinates $x^\mu$ as a shorthand for the true line element $d\bar{s}^2 = l^2 \, ds^2 = g_{\mu\nu} \, d(lx^\mu) \, d(lx^\nu)$ in the unitful coordinates $(lx^\mu)$, never making the arbitrary length scale explicit. Note that derivatives $\partial_{lx^\mu} = l^{-1} \, \partial_{x^\mu}$ pick up an extra factor of the length scale when calculating in the shorthand. By keeping track of these and similar factors, one can easily relate geometric quantities in the unitful geometry to those in the unitless geometry. 

Additionally, all parameters in the metric components are by convention unitless. Any physical, unitful, parameters must be obtained by studying the resultant geometry of the physical metric $d\bar{s}^2$.

Consider the demonstrative example of three dimensional Euclidean flat space in spherical coordinates. We write this geometry in the unitless shorthand coordinates $(r,\theta,\phi)$, with line element $ds^2 = dr^2 + r^2 \, ( d\theta^2 + \sin^2\theta \, d\phi^2)$. This serves as shorthand for the the unitful coordinates $\bar{x}^\mu = (lr, l\theta, l\phi)$, and the unitful metric 
$d\bar{s}^2 = d(lr)^2 + (lr/l)^2 \, ( d(l\theta)^2 + \sin^2(l\theta/l) \, d(l\phi)^2 )$, with $l\theta \in (0,\pi l)$ and $l\phi \in (0,2\pi l)$.

Moreover, to see how parameters in the metric correspond to physical parameters, consider for example the Schwarzschild metric 
$ds^2 = -(1-R/r) \, dt^2 + (1-R/r)^{-1} \, dr^2 + r^2 \, d\Omega^2$, as expressed here in the unitless shorthand coordinates. The parameter $R$ is unitless. By inspection of the geometry $d\bar{s}^2$, one finds that the physical Schwarzschild radius is $\bar{R} = Rl$, and the physical Schwarzschild mass in SI units is $\bar{M}= Rlc^2 / 2G$.

The Penrose diagrams we construct are for the unitless metric $ds^2$. The Penrose coordinates are unitless, and in all examples no absolute length scale appears in the metric components. Consequently, the appearance of the Penrose diagrams never depends on the arbitrary length scale factor. To return to units, one simply notes that if $(u,v)$ are Penrose coordinates for $ds^2$, then $(lu,lv)$ are Penrose coordinates for $d\bar{s}^2 = l^2 ds^2$. Once the length scale $l$ is set, all lengths are defined in units, and other geometrical quantities must be calculated in units as described above.

\section[{\hspace{18mm} Spherical symmetry with a fixed origin}]
{Spherical symmetry with a fixed origin}
\label{sec:spherical}

To properly describe SSS spacetimes requires the concept of spherical symmetry about a fixed origin. Since manifolds need not contain their symmetry axes, this requires a bit of extra work.

To set the stage, let us recall some facts about the symmetry of psuedo-Riemannian manifolds (i.e. metric manifolds of arbitrary signature). Such a $D$-dimensional manifold $M$ usually has a continuous (Lie) group $\mathcal{G}$ of isometries. Continuous isometries are generated by flow along Killing vector fields, which satisfy $\del_{(a} k_{b)} =0$. The set of Killing vector fields, under the Lie bracket operation, form a Lie algebra equivalent to that of $\mathcal{G}$. The maximum number of independent Killing vector fields (maximum dimension of $\mathcal{G}$) is always $D\,(D+1)/2$. Spaces saturating this maximum are maximally symmetric, have constant curvature, and are homogenous and isotropic, at least locally \cite{thirring92}. In particular the $k$-sphere $\mathbb{S}^k$, with metric induced from Euclidean space, is a maximally symmetric space of positive curvature. The isometry group of $\mathbb{S}^{k-1}$ is the orthogonal group $O(k)$, defined as the Lie group of real orthogonal $k\times k$ matrices. The corresponding Lie algebra $\mathfrak{so}(k)$, defined as the real algebra of real antisymmetric $k\times k$ matrices, has dimension $k\,(k-1)/2$. It is by analogy with this symmetry algebra that we define spherical symmetry in spacetime.

Let $M$ be a spacetime of dimension $D$. $M$ is called \textit{$n$-spherically symmetric} if \mbox{(i) the} Lie algebra of Killing vector fields of $M$ has a spacelike subalgebra $\sigma$ isomorphic to $\mathfrak{so}(n+1)$, such that (ii) the local dimension of $\sigma$ as a subspace of $T_p M$ is either $0$ or $n$ at each point $p \in M$. In general there may be many such subalgebras $\sigma$, each corresponding to a different symmetry axis. Therefore we will call a choice of one specific $\sigma$ a \textit{choice  of  the  origin  of  spherical  symmetry},  and  the  pair  $(M,\sigma)$  an \textit{$n$-spherically symmetric  spacetime  with  fixed  origin} (but we will abuse notation and simply refer to $M$, with $\sigma$ implied).

Now suppose $M$ is $n$-spherically symmetric with origin fixed by $\sigma$. Points where the local dimension of $\sigma$ is 0 are fixed points of the spherical symmetry; we call the set of all such points the \textit{axis} $A_{\sigma}$. The axis may be empty or nonempty. When nonempty, it is typically a $D-(n+1)$ dimensional submanifold of $M$. The complement of the axis we shall denote $B_{\sigma}$. By assumption, $\sigma$ has local dimension $n$ on $B_\sigma$. Thus by Frobenius's theorem \cite{schutz80}, $B_\sigma$ is foliated by $n$-dimensional integral submanifolds of $\sigma$ corresponding to isometric flows along $\sigma$. Each point $p \in B_\sigma$ is contained in exactly one such submanifold, called its \textit{orbit} and denoted $\orb_\sigma(p)$. 

We wish to show that each of these orbits has the intrinsic geometry of a sphere (its intrinsic geometry being induced by the metric on $M$). We do so by determining Killing vectors of the orbits. Each Killing flow generates an isometry of $M$, which in turn induces an isometry on the invariant subspace $\orb_\sigma(p)$. Since additionally $\sigma$ is tangent to the orbits, $\bar{\sigma} = \sigma \, \big|_{\textrm{\footnotesize Orb}_\sigma(p)}$ is an algebra of Killing vector fields on the orbit, and the vector field commutators of $\sigma$ are preserved by the restriction. One could imagine, however, that when restricted to the orbit, $\sigma$ may no longer have the same number of independent generators (i.e. if two linearly independent fields, when restricted, became linearly dependent). This possibility is mostly (i.e. for $n\neq 3$) ruled out by the group theoretic considerations of \cite{montgomery42}. A more concise proof follows from the algebraic approach, as follows. The restriction $\sigma \to \bar{\sigma}$, being both linear and commutator-preserving, defines a Lie algebra homomorphism.  The kernel of the homomorphism must be an ideal. But since simple Lie algebras have no nontrivial proper ideals, and since our assumption on the local dimension of $\sigma$ implies that $\dim{\bar{\sigma}}\geq n$, simplicity of $\sigma$ implies that $\dim{\bar{\sigma}} = \dim \sigma$. Thus, $\dim{\bar{\sigma}}= n (n+1)/2$ whenever $\mathfrak{so}(n+1)$ is simple, leaving only the cases $n=1$ and $n=3$ \cite{hall03}. The same dimensionality holds for the case $n=1$, since in that case $n(n+1)/2 = n$ already. Let us assume the case $n=3$ causes no problems.

The above reasoning shows that $\sigma$ restricts to a spacelike $n(n+1)/2$ dimensional algebra of Killing vector fields on the $n$-dimensional orbits, so it follows that each orbit is a maximally symmetric Riemannian manifold with symmetry algebra isomorphic to $\mathfrak{so}(n+1)$. This implies that each orbit must be locally isometric to the sphere $\mathbb{S}^n$ with metric $ds^2 = (r_{{\orb}})^2 \; d\Omega^2$. It is therefore justified to define the \textit{areal radius (relative to $\sigma$)} at each point $p\in B_\sigma$ by $r_{\sigma}(p) = r_{{\orb_\sigma(p)}}$. Correspondingly, for points $p \in A_\sigma$ on the symmetry axis we define $r_\sigma(p)=0$. In this way the areal radius is defined for all points in $M$.

At every point not on the symmetry axis, there exists a local coordinate system exhibiting the foliation by spheres. Let $p \in B_\sigma \subset M$. Then $p$ has a neighborhood with coordinates $(a^\lambda,\Omega^k)$, in which the metric reads
\begin{equation} \label{eqn:spheremetric}
ds^2 = h_{\mu\nu}(a^\lambda) \, da^\mu \, da^\nu + r(a^\lambda)^2 \; d\Omega^2 \; ,
\end{equation}
where $h_{\mu\nu}(a^\lambda)$ is a $(D-n)$-dimensional Lorentzian metric depending only on $a^\lambda$, the coordinates $\Omega^k$ parameterize a $\sigma$-orbit for fixed $a^\lambda$, and $r(a^\lambda)$ is the areal radius of such an orbit. For each point $q$ with coordinates $(a^\lambda,\Omega^k)$ in this patch, $r(a^\lambda)=r_\sigma(q)$. That such a coordinate system exists can be shown by construction of surface-orthogonal geodesic coordinates \cite{carrol03}; another proof is given by \cite[(s. 13-5)]{weinberg72}.

Let it be emphasized: once an origin of spherical symmetry is fixed by choosing $\sigma$, the areal radius $r=r_\sigma(p)$ is an intrinsic property of each point $p \in M$, independent of coordinate system. $M$ is foliated by spheres (lying tangent to $\sigma$) with intrinsic metric $ds^2=r_\sigma(p)^2 \, d\Omega^2$, except on the axis of symmetry where $r_\sigma(p)=0$ by definition. Everywhere except on the axis, there is a local coordinate system respecting this foliation, in terms of which the metric is (\ref{eqn:spheremetric}).

\section[{\hspace{18mm} Proof of regularity at the horizons}]
{Proof of regularity at the horizons}

Here we prove that the Penrose-coordinate metric coefficients are continuous and nonzero at all horizons and horizon vertices in Penrose diagrams obtained by the methods of section \ref{sec:alg1}. The only exception is that the metric is sometimes discontinuous at horizon points in a small neighborhood near $|t|\to\infty$, in non-vertex corners of the diagram, as described in section \ref{subsec:undefined_rad}. Before proceeding, properties of the global tortoise function of section \ref{sec:alg1} are proved, and some useful identities are developed. 

\subsection*{Properties of the Global Tortoise Function}
\label{subsec:tortproof}

The identities (\ref{eqn:analytic}) and (\ref{eqn:cont}) are proved by showing that for each zero $r_i$ of $f(r)$ there exists a cut punctured disk $D_{cut}$ about the origin in the complex plane, and a function $D(z)$ analytic at $z=0$, such that for $z=x$ on the real axis in $D_{cut}$,
\begin{equation} \label{eqn:logmag}
F(r_i+x) = k_i^{-1} \ln |x| + D(x) \, .
\end{equation}
It follows that the complex function 
$F(r_i+z) -  k_i^{-1} \ln |z| = D(z)$
is analytic at $z=0$, which, as seen below, quickly implies the desired results.

The proof of (\ref{eqn:logmag}) runs as follows. Denote $F(r)=\Re F_c(r)$, where $F_c(r)$ is the defining integral in (\ref{eqn:globtort}) with the real part sign removed. Evaluation of the semicircular contours in the small $\epsilon$ limit shows that $\Im F_c(r)= \pi \, \sum_{i=1}^{N} k_{i}^{-1} \, \Theta(r-r_i)$. By assumption, $f(r_i+z)$ is complex analytic at $z=0$ with $f(r_i)=0$ and $f'(r_i)=k_i$. Power series arguments show that there exists a $D_{cut}$ (as above) on which 
\begin{equation} \label{eqn:fprop}
f(r_i+z) = k_i \, z \, (1 + A(z)) \, ,
\end{equation}
with $A(z)$ analytic at $z=0$ and $A(0)=0$, and on which $f(r_i + z)^{-1}$ is analytic with analytic antiderivative
\begin{equation}
\tilde{F}(r_i+z) = k_i^{-1} \log(z) + B(z) + C \, ,
\end{equation}
where $B(z)$ is also analytic at $z=0$. $F_c(r_i+z)$, which can be written as a constant plus an integral contained in $D_{cut}$, analytically extends to the antiderivative  $\tilde{F}(r_i+z)$ for a suitable constant. Comparing imaginary parts in $F_c(r_i+z)=\tilde{F}(r_i+z)$ for $z=x$ on the real axis shows that $B(z)+C = D(z) + i\pi (k_1^{-1}+\ldots + k_i^{-1}) $ with  $\Im D(x) = 0$. Thus $F(r_i+x) = \Re F_c(r_i+x) = k_i^{-1} \, \ln |x| + \Re D(x) = k_i^{-1} \, \ln |x| + D(x) $.

From (\ref{eqn:logmag}), property (\ref{eqn:cont}) follows directly. Then, using (\ref{eqn:logmag}) and (\ref{eqn:fprop}), one finds
\begin{equation} \label{eqn:analytic2}
|f(r_i+z)| \; e^{- k_i F(r_i+z)} = |k_i| (1 + A(z)) \, e^{- k_i D(z)} \, .
\end{equation}
Near $z=0$, we define this formula by its right hand side, which is analytic and positive at $z=0$, and obtain the property (\ref{eqn:analytic}) by restricting to the real axis. This yields the limit $|f(r_i)| \; e^{- k_i F(r_i)} = |k_i| \, e^{- k_i D(0)} $ at $z=0$.

\subsection*{Useful identities}
\label{subsec:ident}

Here we collect some useful identities for the functions $h(s)$ and $H_k(s)$, and the coordinate transformations $\udl(u)$ and $\vdl(v)$, of section \ref{subsec:dnpen}. The identities below include some inverse functions, which are useful for determining radii of points in the diagram, and some derivative identities, which are useful for analysis of the metric.

First of all, there are the inverse functions
\begin{equation}
{H_k}^{\! -1}(s) = \left\lbrace
\begin{tabular}{ll}
$ \frac{1}{k} \ln(1+ks)   $           & $k \neq 0$ \\
$ s $                      & $k = 0$
\end{tabular}
\right. ,
\end{equation}
\begin{equation} 
(\hks)^{-1}(s) = \left\lbrace
\begin{tabular}{lll}
$-s_0 + {H_{k_-}}^{\!\!\!\!\! -1}(s+s_0)$, & & $s <  - s_0$ \\
$s$, & & $|s| \leq s_0$ \\
$s_0 + {H_{k_+}}^{\!\!\!\!\! -1}(s-s_0)$, & & $s > s_0$
\end{tabular}
\right. \; ,
\end{equation}
which must be used to determine radii of points in the diagram.

Meanwhile, the derivatives we want are
\begin{equation}
{H_k}'(s) = \left\lbrace
\begin{tabular}{ll}
$e^{ks}$           & $k \neq 0$ \\
$1$                      & $k = 0$
\end{tabular}
\right. \; ,
\end{equation}
\begin{equation} 
(\hks)'(s) = \left\lbrace
\begin{tabular}{lll}
${H_{k_-}}'(s+s_0)$, & & $s <  - s_0$ \\
$1$, & & $|s| \leq s_0$ \\
${H_{k_+}}'(s-s_0)$, & & $s > s_0$
\end{tabular}
\right. \; .
\end{equation}
From the derivatives one observes that ${H_k}'(s) = 1 + k H_k(s)$.

Now denote $h(s)=\hks(s)$. Due to the piecewise nature of this function, it is useful to also define the piecewise constant
\begin{equation}
\label{eqn:kappa}
\kappa(s) \equiv \left\lbrace
\begin{tabular}{lll}
$k_-$, & & $s <  - s_0$ \\
$0$, & & $|s| \leq s_0$ \\
$k_+$, & & $s > s_0$
\end{tabular}
\right. \; .
\end{equation}
In this notation (and suppressing the argument), one can write the simple expressions
\begin{equation}
\label{eqn:kappah}
\kappa \, h(s) = -1 + |\kappa| \, s_0 + e^{-|\kappa| s_0} \;  e^{\kappa s}  \; ,
\end{equation}
\begin{equation}
\label{eqn:hprime1a}
h'(s) = e^{-|\kappa| s_0} \;  e^{\kappa s}  \; ,
\end{equation}
\begin{equation}
\label{eqn:hprime1b}
h'(s) = 1 - |\kappa| \, s_0 + \kappa \, h(s) \; .
\end{equation}

Let $\xdl(x)$ be a joint notation for $\udl(u)$ and $\vdl(v)$. For both $u$ and $v$, the transformation from block to Penrose coordinates has the general form 
\begin{equation}
\label{eqn:xdl}
\tan \, \pi \, (\xdl - \cdl) = a \,  h( b x)  \; ,
\end{equation}
for real constants $a,b,c$, where $h(s)$ is monotonic increasing. It follows that $\xdl(x)$ is monotonic and invertible, and the derivative is
\begin{equation} 
\label{eqn:dxdl}
\frac{dx}{d\xdl} = 
\left(\frac{\pi}{ab} \right) 
\left( \frac{1+a^2 \, h(bx)^2}{h'(bx)}\right) \; .
\end{equation}

Applying (\ref{eqn:hprime1a}) and (\ref{eqn:hprime1b}) with (\ref{eqn:xdl}) and (\ref{eqn:dxdl}), one obtains
\begin{equation} 
\label{eqn:dxdl2}
\frac{dx}{d\xdl} = 
\left(\frac{\pi}{ab} \right) 
\left(\frac{e^{-|\kappa|s_0}}{e^{-\kappa bx}} \right)
 \frac{1+\cot^2 \pi (\xdl-\cdl)}
{\left[\kappa \, a^{-1}  + (1 - |\kappa|s_0  )\cot \pi(\xdl-\cdl) \right]^{\,2}} \; 
\end{equation}
with $\kappa = \kappa(bx)$, wherever $\xdl-\cdl \neq 0$ (and again, $x$ represents either $u$ or $v$).

All necessary identities are now in hand.

\subsection*{Proof of regularity}
\label{subsec:proofan}
We shall demonstrate that, in the Penrose coordinates $(\udl,\vdl)$, the metric coefficients are continuous and nonzero in the coordinate basis at horizons and horizon vertices, except possibly in arbitrarily small corners of the diagram where $|u/2|>s_0$, $|v/2|>s_0$, and $|t|\to \infty$. The proof proceeds by direct calculation of the metric components in a neighborhood of horizon points.

As usual (see section \ref{subsec:coordnames}), suppose that $(x,y)$ is a shorthand representing either $(u,v)$ or $(v,u)$. Begin with metric coefficients written in the form (\ref{eqn:penmetric}). In order to evaluate the metric at $r=r_i$, one sets the free parameter $k=k_i$. The function (\ref{eqn:analytic}), which carries the explicit radial dependence, was already shown to be analytic (and thus continuous) and positive at $r_i$. All that is left is to evaluate the functions $G_x(x,k)$ and $G_y(y,k)$ in the appropriate limits. The details run as follows.

In light of (\ref{eqn:dxdl2}), the metric coefficients (\ref{eqn:Gu}-\ref{eqn:Gv}) in each block can be rewritten
\begin{equation}
\label{eqn:Gu2}
G_u(u,k) = 
\left(\frac{e^{-|\kappa_u|s_0}}{e^{(k-\kappa_u)u/2}} \right)
 \frac{1+\cot^2 \pi (\udl-\cdl_u)}
{\left[\kappa_u    + \epsilon_u \, (1 - |\kappa_u|s_0  )\cot \pi(\udl-\cdl_u) \right]^{\,2}} \; ,
\end{equation}
\vspace{1mm}
\begin{equation}
\label{eqn:Gv2}
G_v(v,k) = 
\left(\frac{e^{-|\kappa_v|s_0}}{e^{-(k-\kappa_v)v/2}} \right)
 \frac{1+\cot^2 \pi (\vdl-\cdl_v)}
{\left[\kappa_v    + \epsilon_v \, (1 - |\kappa_v|s_0  )\cot \pi(\vdl-\cdl_v) \right]^{\,2}} \; ,
\end{equation}
where $\kappa_u = \kappa(u/2)$ and $\kappa_v=\kappa(-v/2)$, utilizing the notation (\ref{eqn:kappa}).

Now suppose there are multiple blocks, all matched to their neighbors according to the prescription of section \ref{subsec:joinpen}. Further, suppose without loss of generality that one of the blocks has the parameters $\cdl_u=\cdl_v=0$. Since all blocks must have been shifted by integer values of $\Delta \cdl$, it follows from the periodicity $\cot (x+\pi) = \cot x$ that in \textit{every} block the metric coefficients are
\begin{equation}
\label{eqn:Gu3}
G_u(u,k) = 
\left(\frac{e^{-|\kappa_u|s_0}}{e^{(k-\kappa_u)u/2}} \right)
 \frac{1+\cot^2 \pi\udl}
{\left[\kappa_u    + \epsilon_u \, (1 - |\kappa_u|s_0  )\cot \pi\udl \, \right]^{\,2}} \; ,
\end{equation}
\vspace{1mm}
\begin{equation}
\label{eqn:Gv3}
G_v(v,k) = 
\left(\frac{e^{-|\kappa_v|s_0}}{e^{-(k-\kappa_v)v/2}} \right)
 \frac{1+\cot^2 \pi\vdl}
{\left[\kappa_v    + \epsilon_v \, (1 - |\kappa_v|s_0  )\cot \pi\vdl \, \right]^{\,2}} \; .
\end{equation}
All metric dependence on the shift parameters has been eliminated.

It remains to implement the limit $r \to r_i$. In any block bordering a horizon radius $r=r_i$, it is easy to see from their definitions that the coordinate limits have a standard pattern, depending on the sign of $k_i=f'(r_i)$. Suppose that $\sgn(k_i)=\pm 1$. Then in any block bordering $r_i$, one finds $F(r)\to \mp \infty$ and $u \to \pm \infty$ and $v \to \mp \infty$. As a result, one finds the important result that as $r \to r_i$, either  $\kappa(u/2)=k_i$, or $\kappa(-v/2)=k_i$, or both. By using the free parameter $k=k_i$, the limiting value can now be evaluated.

Suppose that two blocks $B_A$ and $B_B$ are joined along a line of constant $\ydl$, as in section \ref{subsec:joinpen}. In both blocks, $|y| \to \infty$ near the horizon, and the limits of the preceding paragraph are such that $\kappa_y = k_i$. Then letting $k=k_i$,  one finds
\begin{equation}
\label{eqn:Gy}
G_y(y,k_i) = 
\left(e^{-|k_i|s_0} \right) \,
 \frac{1+\cot^2 \pi\ydl}
{\left[k_i    + \epsilon_y \, (1 - |k_i|s_0  )\cot \pi\ydl \, \right]^{\,2}} \; ,
\end{equation}
which obtains the limit
\begin{equation}
\label{eqn:horlimit}
G_y(y,k_i) = k_i^{-2} \;\; e^{-|k_i|\, s_0}
\end{equation}
when $r = r_i$, since $\cot^2 \pi y =0$ there. The limit (\ref{eqn:horlimit}) is independent of block parameters, and thus automatically equal in both blocks, ensuring that $G_y(y,k_i)$ is positive and continuous across the horizon.

Meanwhile, at horizon points, the coordinate $x$ parameterizing the horizon remains finite. On the horizon, the corresponding metric factor becomes
\begin{equation}
\label{eqn:Gx}
G_x(x,k_i) = 
\left(\frac{e^{-|k_i|s_0}}{e^{\pm(k_i-\kappa_x)x/2}} \right)
 \frac{1+\cot^2 \pi\xdl}
{\left[\kappa_x    + \epsilon_x \, (1 - |\kappa_x|s_0  )\cot \pi\xdl \, \right]^{\,2}} \; ,
\end{equation}
where $\pm$ corresponds to $x=u$ and $x=v$ respectively. This expression holds on both sides of the horizon, in both blocks $B_A$ and $B_B$, and is strictly positive. Since the matching conditions ensure that $\epsilon_x^A = \epsilon_x^B$, the function (\ref{eqn:Gx}) is equal on both sides of the horizon for all $x$ such that $\xdl_A(x)=\xdl_B(x)$ (which also implies $\kappa_x^A = \kappa_x^B$). Near horizon vertices, where all transformations $h(s)$ have the same exponential factor, and in the bulk of blocks, where $h(s)=s$, this matching is guaranteed. However, when both $|u/2|>s_0$ and $|v/2|>s_0$ near $|t|\to\infty$, then $h(s)$ may be different in the neighboring blocks for the relevant direction, yielding a discontinuity. By choosing $s_0$ large, the neighborhood affected by this is exception may be made arbitrarily small. Thus at all horizon points, except near  $|t|\to\infty$, the function $G_x(x,k_i)$ is positive and continuous across the horizon. At a horizon vertex point, the limit (\ref{eqn:horlimit}) applies to both $x$ and $y$, for all surrounding blocks, and the metric is, again, continuous.

To summarize, utilizing (\ref{eqn:Gy}) and (\ref{eqn:Gx}), the metric in a neighborhood of any horizon point or vertex point at $r=r_i$ can be written
\begin{equation}
ds^2 = -  \left( \frac{4 \pi^2}{e^{-k_i c}}  \right) \frac{|f(r)|}{e^{k_i F(r)} } \;\, G_y(y,k_i) \; G_x(x,k_i) \;  d\udl \, d\vdl + r^2 \; d\Omega^2 \; .
\end{equation}
Every factor in the coefficient of $-d\udl d\vdl$ has by been shown to be positive and continuous as a function of $(\xdl,\ydl)$ in this neighborhood, except in arbitrarily small neighborhoods where $|t| \to \infty$. Thus, except for the polar coordinate singularity at $r=0$, and except for the above-noted $|t| \to \infty$ exception, the global metric is continuous and strictly positive everywhere. Our goal of obtaining non-degenerate continuous global Penrose coordinates extending across an arbitrary number of horizons has been achieved.

Given the analyticity of the radial dependence in the metric (see (\ref{eqn:analytic})), one might hope to give a coordinate system which extends the full metric analytically across horizons. Indeed, this is possible by simply choosing a suitable $h(s)$. However, a trade-off has to be made. If analyticity at the horizons is retained, either differentiability of the interior metric, simplicity of $h(s)$, or simplicity of the diagram appearance must be sacrificed. An example of the latter case may be seen in figure \ref{fig:comparison}(b). Since there is no physically motivated benefit to retaining analyticity, we give it up.

\section[{\hspace{18mm} More properties of SSS spacetimes}]
{More properties of SSS spacetimes}
\label{sec:moreprops}
For convenience, we collect here some geometrical formulae for a four-dimensional SSS spacetime. Let the metric function be written
\begin{equation}
f(r)=1 - \frac{2 \, m(r)}{r}    \; .
\end{equation}
We imply no restriction whatsoever on the function $m(r)$. This is simply a very useful way to write the metric function. It is only sometimes appropriate to interpret $m(r)$ as the total mass inside radius $r$. Useful derived quantities include the shell mass
\begin{equation} \label{eqn:mprime}
\mu(r) = m'(r)
\end{equation}
(see (\ref{eqn:density}) and below) which determines the Einstein tensor, and the function
\begin{equation}
\eta(r)= \left( \frac{2 \, m(r)}{r} - \frac{4 \, m'(r)}{3} +
\frac{r \, m''(r)}{3} \right) \; ,
\end{equation}
which controls the Weyl tensor.

\subsection*{Orthonormal basis}

We define an orthonormal basis $\ehat_a$ in each block. Where $f(r)>0$ define $\ehat_0= \sqrt{f(r)^{-1}} \, \partial_t$ and $\ehat_1= \sqrt{f(r)} \, \partial_r$. Where $f(r)<0$ define $\ehat_0= \sqrt{-f(r)} \, \partial_r$ and $\ehat_1= \sqrt{-f(r)^{-1}} \, \partial_t$. And, everywhere, define $\ehat_2= r^{-1} \, \partial_\theta$ and $\ehat_3= (r \, \sin \theta )^{-1} \, \partial_\phi$. In this construction $\ehat_0$ is always timelike. Both $\ehat_0$ and $\ehat_1$ can be continuously extended across horizons, which is why the Einstein tensor in this basis (below) will have no discontinuities associated with the piecewise definition. However the basis cannot be continuously extended, as the extensions of $\ehat_0$ and $\ehat_1$ would coincide at the horizon. Despite this shortcoming, we often choose to work in this basis for its conceptual and calculational simplicity.

\subsection*{Curvature components in coordinate bases}

These quantities were computed with help from the \textit{Mathematica} package \textit{RGTC} \cite{bonanos03}. Sign conventions for the curvature tensors are equivalent to those found in Wald \cite{wald84}.

In the $(t,r,\theta,\phi)$ coordinate basis, the nonzero Christoffel symbols are 
$\G{t}{rt}=\G{t}{tr}=-\G{r}{rr} = \frac{f'}{2 f}$ 
and 
$\G{r}{tt}= \frac{f' f}{2}$ 
and  
$\G{r}{\theta\theta}=(\sin\theta)^{-2} \; \G{r}{\phi\phi} = -r f$ 
and
$\G{\theta}{r\theta} = \G{\theta}{\theta r} = \G{\phi}{r\phi} = \G{\phi}{\phi r} = \frac{1}{r}$
and
$\G{\theta}{\phi\phi} = -\sin\theta \cos \theta$
and
$\G{\phi}{\theta\phi} = \G{\phi}{\phi\theta} = \cot \theta$.

The Riemann tensor has the symmetries $R_{abcd}=-R_{abdc}=-R_{bacd}=R_{badc}$ and $R_{abcd}=R_{cdab}$ and $R_{abcd}+R_{acdb}+R_{adbc}=0$ \cite{griffiths09}. Using the symmetries, all nonzero components of $R_{abcd}$ are generated in the coordinate basis by
$R_{trtr}=\frac{f''}{2}$ 
and
$R_{t\Omega t\Omega} =  \frac{rff'}{2}$
and
$R_{r\Omega r\Omega} = - \frac{rf'}{2f}$
and
$ R_{\theta\phi\theta\phi} = r^2 \, \sin^2 \theta \,(1-f)$,
where
$R_{a\theta b\theta} = R_{a\Omega b\Omega}$
and
$ R_{a \phi b \phi} = (\sin^2 \theta) ( R_{a\Omega b\Omega})$.

The Weyl tensor has the same symmetries as the Riemann tensor above. Using the symmetries, all nonzero components of $C_{abcd}$ are generated in the coordinate basis by
$C_{trtr} = - \frac{\eta}{r^2}$ 
and
$C_{t\Omega t\Omega} =  \frac{\eta f}{2}$
and
$C_{r\Omega r\Omega} = - \frac{\eta}{2f} $
and
$ C_{\theta\phi\theta\phi} =  \eta \, r^2 \sin^2 \theta $, 
where
$C_{a\theta b\theta} = C_{a\Omega b\Omega}$
and
$ C_{a \phi b \phi} = (\sin^2 \theta) ( C_{a\Omega b\Omega})$.

For calculations involving null curves and surfaces, it is often useful to also know the covariant derivative in null coordinate systems.

In double null coordinates with metric $ds^2 = -f(r) \, du dv + r^2 \, d\Omega^2$, the relevant Christoffel symbols can be evaluated by the chain rule on $r$ and $f(r)$. The nonzero components are
$\G{u}{uu}=-\G{v}{vv}=-\frac{f'}{2}$
and
\mbox{$\G{u}{\theta\theta} = - \G{v}{\theta\theta} = 
(\sin\theta)^{-2} \; \G{u}{\phi\phi} = - (\sin\theta)^{-2} \; \G{v}{\phi\phi} = 
r $}
and
$ \G{\theta}{u\theta} = \G{\theta}{\theta u} = 
\G{\phi}{u\phi} = \G{\phi}{\phi u} = 
-\frac{f}{2r}$
and 
$ \G{\theta}{v\theta} = \G{\theta}{\theta v} = 
\G{\phi}{v\phi} = \G{\phi}{\phi v} = 
\frac{f}{2r}$
and
$\G{\theta}{\phi\phi} = -\sin \theta \cos \theta$
and
$\G{\phi}{\theta\phi}=\G{\phi}{\phi \theta} = \cot \theta$.

Meanwhile, in the EF coordinate basis (with metric (\ref{eqn:efmet})), the nonzero Christoffel symbols are, by direct computation,
$\G{r}{rw}=\G{r}{wr}=-\G{w}{ww} = \frac{f'}{2 \epsilon}$ 
and  
$\G{w}{\theta\theta}=(\sin\theta)^{-2} \; \G{w}{\phi\phi} = \frac{r}{\epsilon}$ 
and 
$\G{r}{ww}= \frac{f' f}{2}$ 
and  
$\G{r}{\theta\theta}=(\sin\theta)^{-2} \; \G{r}{\phi\phi} = -r f$ 
and
$\G{\theta}{r\theta} = \G{\theta}{\theta r} = \G{\phi}{r\phi} = \G{\phi}{\phi r} = \frac{1}{r}$
and
$\G{\theta}{\phi\phi} = -\sin\theta \cos \theta$
and
$\G{\phi}{\theta\phi} = \G{\phi}{\phi\theta} = \cot \theta$.

\subsection*{Matter content, curvature scalars, and energy conditions}
 
The Einstein tensor in the orthonormal basis $\ehat_a$ (defined above) is given by
\begin{equation}
\label{eqn:stress}
{G^a}_b = 
8\pi 
\left(
\begin{tabular}{cccc}
   $-\rho$   &   $0$   &   $0$   &   $0$   \\
   $0$   &   $p_1$   &   $0$   &   $0$   \\
   $0$   &   $0$   &   $p_2$   &   $0$   \\
   $0$   &   $0$   &   $0$   &   $p_3$   
\end{tabular}
\right)
=
8\pi 
\left(
\begin{tabular}{cccc}
   $-\rho$   &   $0$   &   $0$   &   $0$   \\
   $0$   &   $-\rho$   &   $0$   &   $0$   \\
   $0$   &   $0$   &   $p_{\Omega}$   &   $0$   \\
   $0$   &   $0$   &   $0$   &   $p_{\Omega}$   
\end{tabular}
\right) \; ,
\end{equation}
where
\begin{equation} \label{eqn:density}
\begin{tabular}{ccc}
$ \rho =   \mu / 4 \pi r^2 $ 
&
\hspace{2cm}
& 
$p_\Omega = - \mu' / 8 \pi r $ .
\end{tabular}
\end{equation}
Thus, $\mu = m'(r)$ may always be interpreted as proportional to the mass of a thin shell at radius $r$. When $\mu$ is constant, density diffuses naturally as $1/r^2$, and there is no transverse pressure.

The curvature scalars are
\begin{equation}
K_0 \equiv R = 16 \pi \, (\rho - p_\Omega)
\end{equation}
\begin{equation}
K_1 \equiv R_{ab}R^{ab} = 128 \pi^2 \, (\rho^2 + p_\Omega^2)
\end{equation}
\begin{equation}
K_2 \equiv C_{abcd} C^{abcd} =  12 \, \eta^2 / r^4
\end{equation}
\begin{equation}
K_3 \equiv R_{abcd} R^{abcd} = K_2 + 2 \, K_1 - (1/3) \, K_0^2 \; .
\end{equation}

The basis $\ehat_a$ diagonalizes ${G^a}_b$, with one timelike and three spacelike eigenvectors. The corresponding eigenvalues are the negative of the density $-\rho$ (timelike) and the principal pressures $p_i$ (spacelike).  It is well known that in this case, the classical energy conditions take on simple forms \cite{hawking73}. In particular, the null energy condition amounts to $\rho+p_i \geq 0$. For us, $\rho + p_1 = 0$ always, and the remaining NEC constraint is equivalent to $2\mu \geq r \mu'$. The weak energy condition amounts to the NEC plus the constraint $\rho \geq 0$, which for us becomes $\mu \geq 0$. The flux energy condition of \cite{visser13} amounts to $\rho^2 - p_i^2 \geq 0$ in the diagonalized case. Again, the SSS case trivially saturates $\rho^2 - p_1^2 = 0$, and the remaining constraint is equivalent to $4\mu^2 \geq r^2 \mu'^{\,2}$.

\subsection*{Trapped surfaces}

A closed spacelike surface $S$ is called \textit{future trapped} (\textit{past trapped}) if its mean curvature vector \mbox{$H^a \equiv -  \; \theta^{+} \, (k_{-})^a -  \, \theta^{-} \, (k_{+})^a$} is everywhere future-directed timelike (past-directed timelike) on $S$ \cite{senovilla11b}. In this expression, the $k_{\pm}$ are future-directed null vectors, each orthogonal to $S$, normalized by \mbox{$(k_+)_a (k_-)^a = -1$}. The $\theta^{\pm}$ are the corresponding future null expansions; they can be defined by 
\mbox{
$\theta^{\pm} \equiv - \, \gamma^{AB} (k_\pm)_\mu \, e^\nu_A \, \ddel_\nu \, e^\mu_B  $
},
where $e^\mu_A$ is a basis on the tangent space $T_p S$ of $S$, and $\gamma^{AB}$ is the inverse of the induced spatial metric \mbox{on $S$}.

In an SSS spacetime, the sphere $(t,r,\theta,\phi)=(t_0,r_0,\theta,\phi)$, contained within a block $I_j$, is a closed trapped surface if and only if $f(r_0)<0$. This is proved by direct calculation of the mean curvature vector of the sphere at SSS block points. It is simplest to work in double-null block coordinates $(u,v,\theta,\phi)$, in terms of which the metric is (\ref{eqn:dnmetric}). Since $S$ is a surface of constant coordinates, its tangent space is spanned by the coordinate basis $e_\theta^\mu = \delta_\theta^\mu$ and $e_\phi^\mu = \delta_\phi^\mu$, and the induced metric $\gamma_{AB}$ has line element $ds^2 = r_0^2 \; d\theta^2 + r_0^2 \, \sin^2 \theta \; d\phi^2$. A pair of mutually normalized null vectors orthogonal to $S$ is given by 
$(k_+)^\mu = (|f|/2)^{-1/2} \;  \delta^\mu_v$ 
and 
$(k_-)^\mu = \sgn(f) \, (|f|/2)^{-1/2} \;  \delta^\mu_u$. 
With this setup, the expansions simplify to 
$\theta^{\pm}
 = 
- (k_\pm)_\mu \left( \gamma^{\theta\theta} e^\nu_\theta \ddel_\nu e^\mu_\theta + \gamma^{\phi\phi} e^\nu_\phi \ddel_\nu e^\mu_\phi \right)
=
- (k_\pm)_\mu \left( \gamma^{\theta\theta} \, \G{\mu}{\theta\theta} + \gamma^{\phi\phi} \, \G{\mu}{\phi\phi} \right) $.
Thus we obtain 
$\theta^+ = \sgn(f) \sqrt{2 |f|/r_0^2}  $ 
and 
$\theta^- =  - \sqrt{2 |f|/r_0^2}  $.
The mean curvature $H^a$ is timelike if and only if $\sgn(\theta^+)=\sgn(\theta^-) \neq 0$, which implies that $S$ is trapped if and only if $f(r_0)<0$. Additionally, it is always true that $\theta^- \leq 0$. Thus when $S$ is trapped, $S$ is future-trapped (past-trapped) if and only if $k_\pm$ as defined above are both future-directed (past-directed). This completes the proof of the desired result.

\subsection*{Cartesian coordinates near the origin}

In a neighborhood of $r=0$ (contained in $I_0$), one can naively make the transformation from $(t,r,\theta,\phi)$ into the cartesian coordinates $(t,\vex)$, according to the standard spherical coordinate transformation, such that $r^2=\vex \cdot \vex$. The metric becomes
\begin{equation}
ds^2 = - f(r) \; dt^2 + d\vex^{\,2}   + 
\left( \frac{ 2 \, m(r) }{ r^3 \,f(r) } \right)  (\vex \cdot d\vex)^2 
\end{equation}
which can be used to study the limit as $r \to 0$, as seen below.

\subsection*{Singularity}

The metric may be either \textit{singular} or \textit{nonsingular} at the origin. If $m(r)$ has a Laurent series expansion about $r=0$ then the following are equivalent: 
\begin{enumerate}
\item $M$ is called \textit{nonsingular} at the origin;
\item $m(r) = \sum_{k=0}^{\infty} c_k \, r^{\,k+3} \;\; $  as $r\to 0$;
\item $f(r) = 1 + \mathcal{O}(r^2) \; $ as $r\to 0$;
\item Curvature scalars $K_0$, $K_1$, $K_2$, $K_3$ all finite at origin;
\item Cartesian metric has a finite limit as $r \to 0$;
\item $M$ contains the $r=0$ axis as a set of points, and is geodesically complete there 
\\(i.e. all geodesics terminating there can be extended);
\end{enumerate}
{\small \textit{Proof.} By definition $(i)\lra(ii)$, and trivially $(ii)\lra(iii)$. Direct calculation shows $(ii)\lra(iv)$ and $(ii)\lra(v)$. Then $(v)\rightarrow(vi)$ and $(vi)\rightarrow(iv)$ by existence of normal neighborhoods \cite{oneill83}, assuming sufficient differentiability.} \vspace{1mm}

Evidently if an SSS spacetime is nonsingular at the origin, then $f(0)=1$, the axis $r=0$ is a timelike curve, and to quadratic order the dominant behavior as $r\to 0$ is either flat, de Sitter, or Anti de Sitter.

%%%%%%%%%%%%%%%%%%%%%%%%%%%%%%%%%%%%%%

%%%%%%%%%%%%%%biblio%%%%%%%%%%%%%%%%%%
\section*{References}
\bibliographystyle{unsrt}
\bibliography{biblio}
%%%%%%%%%%%%%%%%%%%%%%%%%%%%%%%%%%%%%%

\end{document}